\documentclass[final,12pt,a4paper]{article}

\usepackage[T1]{fontenc}

\usepackage{graphicx}
\usepackage{subfigure}

\usepackage{amsmath}
\usepackage{amsgen}
\usepackage{amssymb}
\usepackage{amsfonts}
\usepackage{amsbsy}

\usepackage{exscale}
\usepackage{a4wide}

\usepackage[numbers]{natbib}

\numberwithin{equation}{section}

\renewcommand{\Im}[0]{\mathrm{Im}\ }

\DeclareMathAlphabet{\bi}{OML}{cmm}{b}{it}

\newfont{\tensy}{cmsy10}
\newcommand{\chem}[1]{{$\fontdimen16\tensy=3.0pt
    \fontdimen17\tensy=3.0pt \mathrm{#1}$}}

\newcommand{\etal}{{\rm et.~al.~}}
\newcommand{\ie}[0]{{\rm i.e.}\ }
\newcommand{\eg}[0]{{\rm e.g.}\ }
\newcommand{\Tc}[0]{T_{\mathrm{C}}}
\newcommand{\up}[0]{\uparrow}
\newcommand{\Nup}[0]{\downarrow}
\newcommand{\en}[0]{\epsilon}
\newcommand{\ek}[0]{\epsilon_{\bi{k}}}
\newcommand{\om}[0]{\omega}

\newcommand{\Si}[0]{\Sigma}
\newcommand{\si}[0]{\sigma}
\newcommand{\las}[0]{\langle}
\newcommand{\ras}[0]{\rangle}
\newcommand{\la}[0]{\left\las}

\newcommand{\llas}[0]{\las\las}
\newcommand{\ra}[0]{\right\ras}

\newcommand{\rras}[0]{\ras\ras}
\newcommand{\Tr}[0]{\mathrm{Tr}}

\newcommand{\ket}[1]{\left|#1\ra}

\newcommand{\com}[2]{[#1,#2]}

\newcommand{\res}[0]{m$\Omega$cm~}
\newcommand{\con}[0]{(m$\Omega$cm)$^{-1}$~}

\newcommand{\rme}{\mathrm{e}}
\newcommand{\rmi}{\mathrm{i}}
\newcommand{\rmd}{\mathrm{d}}

\newcommand{\bsigma}{\boldsymbol{\sigma}}

\begin{document}


\title{Ferromagnetism and electron-phonon\\coupling in the manganites}
\author{D M Edwards\\Department of
    Mathematics\\Imperial College\\London SW7 2BZ, UK\\
    \ttfamily{d.edwards@ic.ac.uk}}
\maketitle
\begin{abstract}
  The physics of ferromagnetic doped manganites, such as
  \chem{La_{1-x}Ca_xMnO_3} with $x\approx0.2$--0.4, is reviewed. The concept
  of double exchange is discussed within the general framework of itinerant
  electron magnetism. The new feature in this context is the coupling of
  electrons to local phonon modes. Emphasis is placed on the quantum nature
  of the phonons and the link with polaron physics. However it is stressed
  that the manganites fall in an intermediate coupling regime where standard
  small-polaron theory does not apply. The recently-developed many-body
  coherent potential approximation is able to deal with this situation and
  Green's recent application to the Holstein double-exchange model is
  described. Issues addressed include the nature of the basic electronic
  structure, the metal-insulator transition, a unification of colossal
  magnetoresistance, pressure effects and the isotope effect, pseudogaps in
  spectroscopy and the effect of electron-phonon coupling on spin waves.
\end{abstract}

\newpage
\tableofcontents
\newpage


\section{Introduction}\label{sec:introduction}

Since the discovery of high temperature superconductivity about fifteen years
ago it has become clear that a metallic state obtained by doping an
antiferromagnetic insulating oxide is substantially different from that of
ordinary metals. Despite enormous experimental and theoretical effort
neither the superconductivity nor the normal state of this unusual type of
metal is at all well understood. There is no generally accepted view about
the mechanism of superconductivity in the cuprates. To gain a wider
perspective it is clearly important to study other oxides which become
metallic upon doping. In fact interest has recently revived in the manganites
\chem{A_{1-x}A'_xMnO_3} (A=trivalent rare-earth ion, A$'$=divalent alkaline
earth ion), which were first studied fifty years ago. An excellent history of
this early work is given in the introduction of an extensive review of
mixed-valence manganites by Coey \etal \cite{Coey99}.
These compounds are ferromagnetic in the metallic state, typically for
$x\approx0.2$--0.4, and the origin of this metallic ferromagnetism is much
less problematic than that of superconductivity in the cuprates. However
above the Curie temperature $\Tc$ many of the manganites become insulating,
and this metal-insulator transition is unlike anything that occurs in
ordinary ferromagnetic metals such as Fe, Co, Ni or Gd. There are competing
theories of this behaviour, and of the detailed exchange mechanism
responsible for the ferromagnetism. Also the importance of electron-phonon
coupling, and the possible existence of small polarons, is hotly
debated. These are some of the issues which are focussed upon in this review.

Two of the most-studied manganites are \chem{La_{1-x}Sr_xMnO_3} (LSMO) and
\chem{La_{1-x}Ca_xMnO_3} (LCMO). The parent compound \chem{LaMnO_3}, with
nominal valence \chem{La^{3+}Mn^{3+}O_3^{2-}}, is an antiferromagnetic
insulator. \chem{Mn^{3+}} ions have four d electrons and in the cubic
environment the \chem{t_{2g}} states, with d wavefunctions of the $xy$, $yz$, $zx$
type, lie lower in energy than the \chem{e_g} ($x^2-y^2$, $3z^2-r^2$) ones. The
spins of the four d electrons are aligned, by Hund's rule, so the
\chem{Mn^{3+}} ion adopts a \chem{t_{2g}^3e_g^1} configuration with spin
2. The effect of doping, which removes $x$ electrons per Mn atom, depends on
whether \chem{LaMnO_3} is a Mott-Hubbard or charge-transfer insulator, in the
terminology of Zaanen \etal \cite{ZaSaAl85}. In the latter case holes are
created in the oxygen 2p band and, since the observed ferromagnetic
saturation moment is less than 4$\mu_{\rm B}$ per Mn atom, their spins must
be antiparallel to the $S=2$ Mn ions. In the former case electrons are
removed from a narrow \chem{e_g} band and the three \chem{t_{2g}} electrons
may be regarded as forming a local spin with $S=3/2$. This is the traditional
view \cite{Ze51} which is still most widely held. The \chem{e_g} band
contains $n=1-x$ electrons per atom with the possibility of metallic
behaviour. The electrons are strongly correlated, avoiding double occupation
of a Mn site, so that for $n=1$ ($x=0$) the system is a Mott insulator. The
itinerant \chem{e_g} electrons are completely spin-polarized and align the
local spins via Zener's \cite{Ze51} double-exchange mechanism. The validity
of this general picture is thoroughly discussed in
sections~\ref{sec:electronic-structure} and~\ref{sec:itin-electr-ferr}.

\begin{figure}[htbp] 
  \centering
  \includegraphics[width=0.3\textwidth]{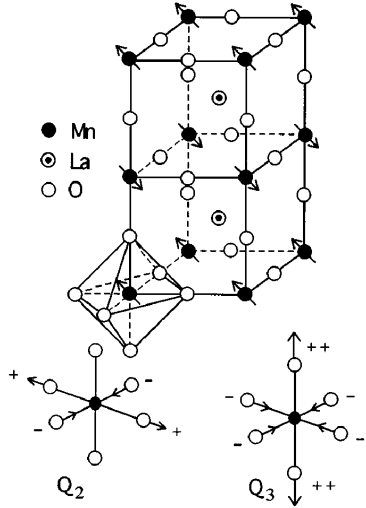}
  \caption{\label{fig:fig1}%
   The ideal cubic perovskite structure, showing the AF type A order for
   \chem{LaMnO_3}. Also shown are the breathing mode ($Q_1$), the
   basal-plane distortion mode ($Q_2$), and the octahedral stretching mode
   ($Q_3$) which are present in the actual structure. (from reference
   \cite{SaPoVu96})}
\end{figure}

The structure of manganites such as \chem{LaMnO_3} is based on the ideal
cubic perovskite (\chem{CaTiO_3}) structure shown in
figure~\ref{fig:fig1}. However the oxygen octahedra surrounding each
Mn atom undergo two types of distortion. One is a rotation and tilting due
to the \chem{La^{3+}} ion being smaller than the \chem{O^{2-}} and the other
is a tetragonal distortion due to the Jahn-Teller (JT) effect. Since the
octahedra are connected the rotations and tilting alternate in sense so that
the final orthorombic structure has a unit cell whose volume corresponds to
four perovskite unit cells. The JT distortion is predominantly of the $Q_2$
type (see figure~\ref{fig:fig1}) in which each octahedron is
stretched along one of the axes in the basal plane. Clearly, the locally
stretched axis alternates in direction by $90^\circ$. The JT distortion
arises from the degenerate \chem{e_g^1} configuration; it splits the
\chem{e_g} level so as to lower the energy of the Mn $3z^2-r^2$ orbital, with
local $z$ axis along the long Mn--O bonds. The JT distortion in
\chem{LaMnO_3} is a large effect, with an octahedral aspect ratio of 1.12
compared with 1.004 in \chem{CaTiO_3} \cite{SaPoVu96}. It encourages orbital
ordering in the basal plane with local $3z^2-r^2$ orbitals pointing in
alternate directions. This has been observed \cite{MuHiGi98}. The magnetic
ordering of \chem{LaMnO_3}, as indicated
in figure~\ref{fig:fig1}, is A-type antiferromagnetic, with
ferromagnetic basal planes arranged antiferromagnetically. In a doped system
such as \chem{La_{0.7}Sr_{0.3}MnO_3}, which is a ferromagnetic metal, the
crystal structure is much more nearly cubic \cite{MaEn96}. The slight
distortion seems to be predominantly due to octahedral rotation and there is
no evidence of a static JT effect \cite{SaPoVu96}. However a dynamic JT
effect may occur as the \chem{e_g} electron moves from site to site and this
constitutes a JT polaron. An important parameter in all manganites is the
Mn--O--Mn bond angle which is less than $180^\circ$ in a distorted structure.
This reduced angle leads to reduced Mn--Mn hopping and hence to a narrowing
of the \chem{e_g} band.

\begin{figure}[htbp]
  \centering
  \subfigure[(from reference \cite{ToAsKu96})]{%
  \begin{minipage}[b]{0.3\textwidth}
    \centering \includegraphics[width=\textwidth]{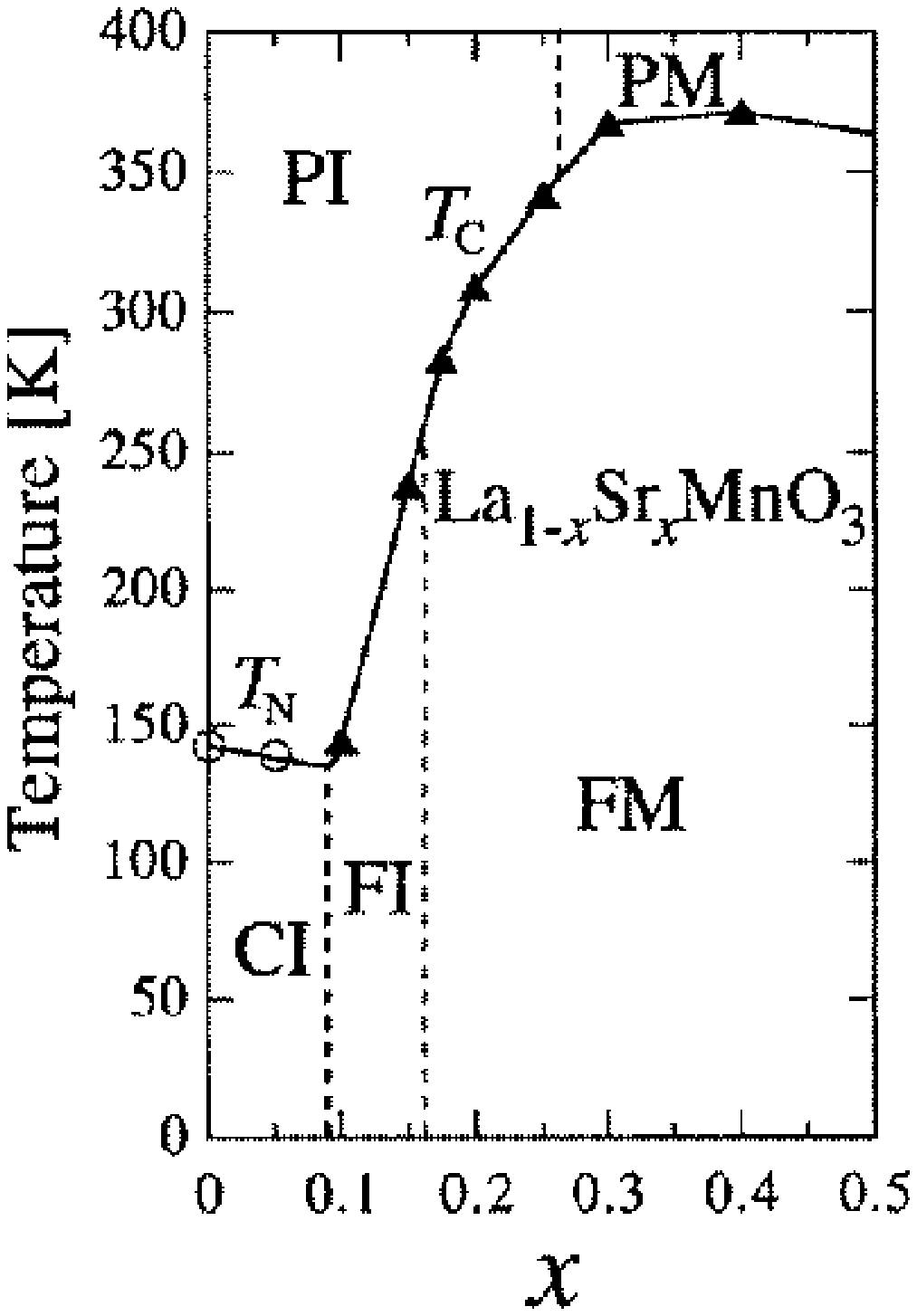}
  \end{minipage}}%
\subfigure[(from reference \cite{ScRaBaCh95})]{%
  \begin{minipage}[b]{0.6\textwidth}
    \centering \includegraphics[width=\textwidth]{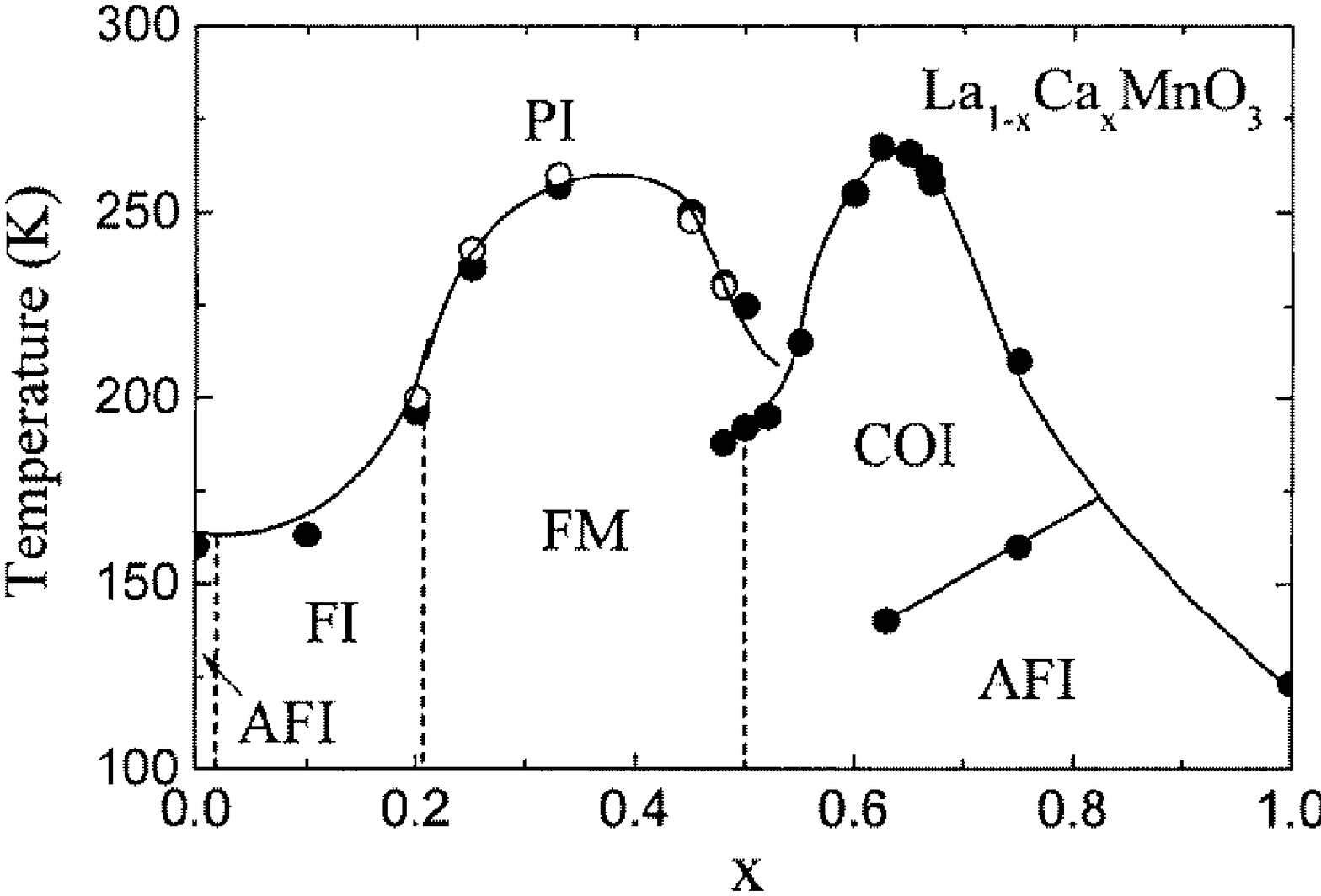}
  \end{minipage}}  
  \caption{\label{fig:fig2}%
    Phase diagram for \chem{La_{1-x}Sr_xMnO_3} (a) and
    \chem{La_{1-x}Ca_xMnO_3} (b). The various states are:
    paramagnetic insulating (PI), paramagnetic metal (PM), canted insulating
    (CI), ferromagnetic insulating (FI), ferromagnetic metal (FM),
    antiferromagnetic insulating (AFI) and charge-ordered insulating (COI). $\Tc$
    and $T_{\rm N}$ are Curie and Ne\'{e}l temperatures, respectively.}
\end{figure}

Phase diagrams in the $x$-$T$ plane of LSMO and LCMO are shown in
figure~\ref{fig:fig2}. For $x<0.5$ the two diagrams appear to be quite
similar with a maximum Curie temperature $\Tc$ at $x\approx0.33$. However the
maximum $\Tc$ in LCMO is considerably lower than in LSMO and the paramagnetic
state above $\Tc$ is insulating in LCMO and a poor metal in LSMO. In LCMO
charge-ordering sets in at $x\approx0.5$ and in the ferromagnetic insulator
regimes at $x\lesssim0.2$, in both LSMO and LCMO, substitutional disorder
must play an important role. Much of the theoretical work reviewed here is
concentrated on $x\approx0.3$, where neither charge order nor orbital order
is expected, and it is assumed that the system is homogeneous. However, it
has been argued by Moreo \etal \cite{MoreoYuDa99} and Nagaev \cite{Na96} that
charge inhomogeneity is widespread in the manganites with a tendency for
holes, produced by doping, to segregate in ferromagnetic
clusters. Unfortunately most quantitative work along these lines has employed
models in which long range Coulomb forces are neglected. This leads to
unrealistic predictions of macroscopic phase separation \cite{MoreoYuDa99}.

\begin{figure}[htbp]
  \centering
  \includegraphics[width=0.5\textwidth]{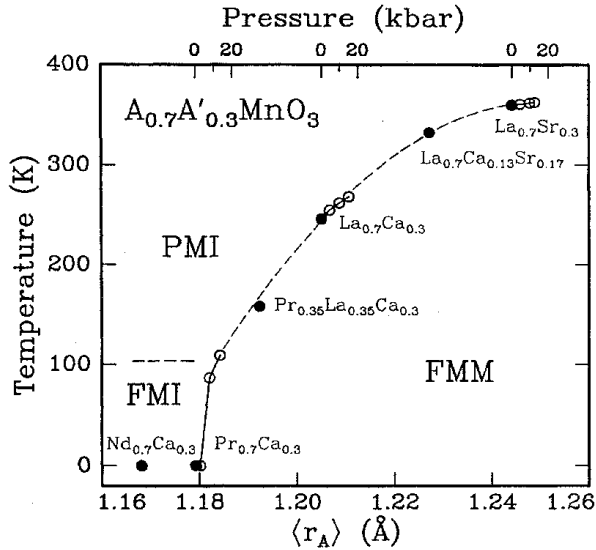}
  \caption{\label{fig:fig3}%
    The phase diagram of \chem{A_{0.7}A'_{0.3}MnO_3} as a function of the
    bandwidth. Closed circles represent variations of
    the bandwidth due to internal pressure (variations of the average A-site
    ionic radius $\las r_A\ras$), and open circles indicate variations
    due to externally applied pressure. (from reference \cite{Hw95})}
\end{figure}

\begin{figure}[htbp] 
  \centering
  \includegraphics[width=0.5\textwidth]{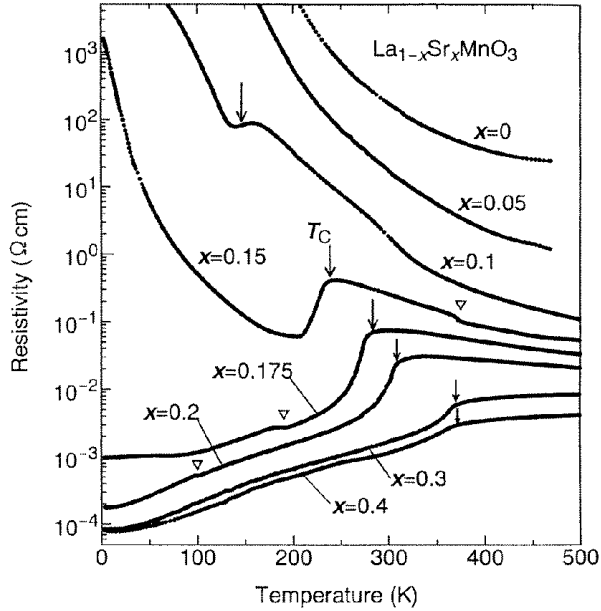}
  \caption{\label{fig:fig4}%
    Resistivity versus temperature for \chem{La_{1-x}Sr_xMnO_3}
    crystals. Arrows indicate the Curie temperature $\Tc$ for the
    ferromagnetic phase transition. (from reference \cite{UrMoAr95})}
\end{figure}

\begin{figure}[htbp] 
  \centering
  \includegraphics[width=0.5\textwidth]{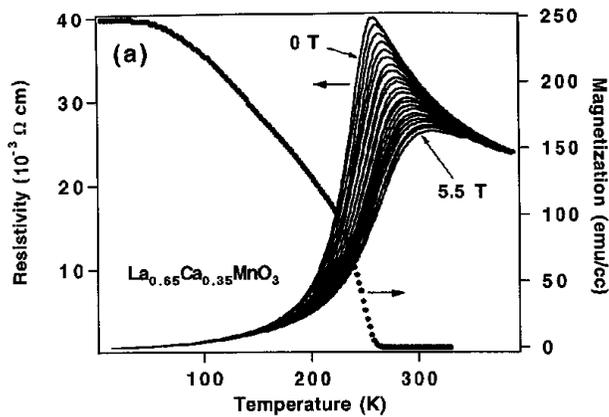}
  \caption{\label{fig:fig5}%
    Resistivity (solid lines) of \chem{La_{0.65}Ca_{0.35}MnO_3}, acquired in
    applied fields ranging from 0 to 5.5 T, and the bulk magnetization in an
    in-plane applied field of 50 G ($\bullet$), both as a function of
    temperature. (from reference \cite{McWaZh96})}
\end{figure}

The importance of the Mn--O--Mn bond angle was mentioned above. This can be
varied in the system \chem{A_{0.7}A'_{0.3}MnO_3}, with a fixed number of
electrons $n=0.7$ in the \chem{e_g} conduction band, by varying either the
average A-site ionic radius $\las r_{\rm A}\ras$ or by applied pressure.
Rodriguez-Martinez and Attfield \cite{RoAt96} have also stressed the role of
the variance in $r_{\rm A}$. An increase of either $\las r_{\rm A}\ras$ or
applied pressure tends to push the Mn--O--Mn bond angle closer to
$180^\circ$, thereby increasing the Mn--Mn hopping parameter and consequently
the width of the \chem{e_g} band. The sensitivity of the Curie temperature
$\Tc$ to such changes is shown in figure~\ref{fig:fig3}. The effect on the
transport properties is even more striking. The contrast between the
temperature-dependence of the resistivity $\rho(T)$ for LSMO and LCMO, with
$x\approx0.3$, is seen in figures~\ref{fig:fig4} and~\ref{fig:fig5}. In LSMO
the resistivity above $\Tc$ ($\sim370$ K) continues to rise, as in a poor
metal, whereas in LCMO the resistivity peaks at $\Tc$ ($\sim260$ K) and
decreases as the temperature is raised above $\Tc$. This indicates a
metal-insulator transition at $\Tc$. Furthermore in LSMO, above $\Tc$,
$\rho\sim4$ \res whereas the resistivity of LCMO peaks at about 40 \res. A
satisfactory theory of the manganites must be able to account for this huge
difference in behaviour between two apparently very similar materials. The
effect of applied magnetic fields of magnitude 0--5.5 T on $\rho(T)$ in LCMO
is shown in figure~\ref{fig:fig5}. The large change in resistance near $\Tc$
is termed `colossal magnetoresistance' (CMR) and it is this phenomenon which
has inspired much of the recent research on the manganites. However the large
fields required make it unlikely that this intrinsic property will be used in
a sophisticated device for sensing magnetic fields. Low-field
magnetoresistance observed in polycrystalline \chem{La_{2/3}Sr_{1/3}MnO_3} is
attributed to spin-polarized tunneling between `half-metallic' grains
\cite{Hw96,MaBuIs97}. The half-metallic property of manganites is discussed
in the next section on electronic structure.


\section{Electronic structure}\label{sec:electronic-structure}

Before discussing the origin of metallic ferromagnetism in the doped
manganites it is necessary to understand their electronic structure. The most
important question is whether doping \chem{LaMnO_3} introduces holes
predominantly into a Mn d band or into an O p band. This depends of course on
which occupied band lies higher in energy. The question must be answered by a
combination of theory and experiment. The usual starting-point is a band
calculation based on the local spin-density approximation (LSDA). In this
approximation all electrons of a given spin experience the same crystal
potential and therefore all d orbitals of that spin, apart from a modest
crystal field splitting between \chem{t_{2g}} and \chem{e_g}, have the same
on-site energy. In order to obtain a \chem{d^n} configuration, with $n\neq0$,
5 or 10, it is inevitable that the d band must overlap the Fermi level.
Consequently in LSDA calculations for \chem{LaMnO_3}, with a Mn \chem{d^4}
configuration, the d band lies above the filled O p band
\cite{PiSi96,SaPoVu96,SaShBa95}.  The JT lattice distortion of $Q_2$ type is
required to open a small gap in the \chem{e_g} band so that the system
becomes an insulator with orbital ordering. Also, the JT distortion
stabilizes the observed antiferromagnetic structure over the ferromagnetic
one \cite{SaPoVu96}. The calculated Mn magnetic moment is 3.4$\mu_{\rm B}$
compared with 3.7$\mu_{\rm B}$ found by neutron scattering \cite{SaShBa95}.
Thus the LSDA calculations are compatible with the observed magnetic and
structural details in \chem{LaMnO_3}.

\begin{figure}[htbp] 
  \centering
  \includegraphics[width=0.5\textwidth]{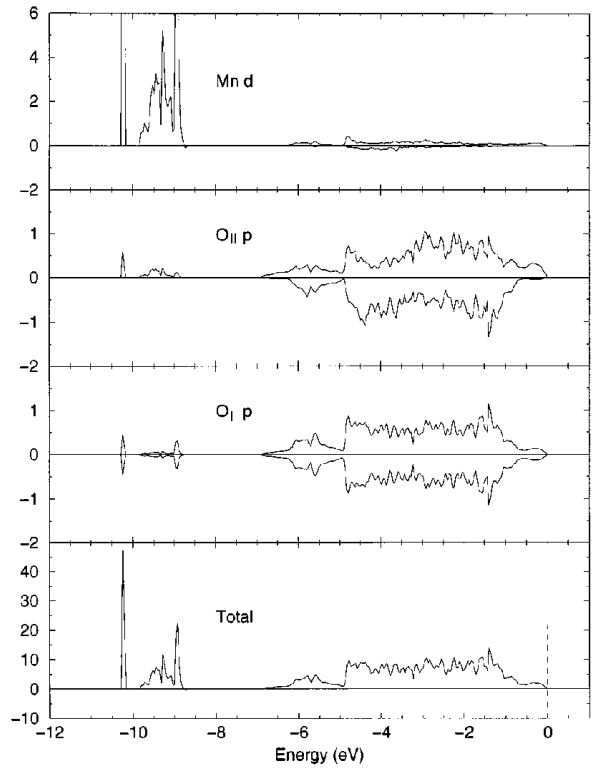}
  \caption{\label{fig:fig6}%
    The (projected) DOS of \chem{LaMnO_3}, with AAF ordering. Positive and
    negative DOS are for up- and down-spin states, respectively. Energies are
    relative to the top of the valence band. The projected Mn-d and
    \chem{O_{II}}-p DOS are for Mn and O on an up-spin basal plane. (from
    reference \cite{SuKaMaHa00})}
\end{figure}

However it is well-known that in some oxides, like NiO \cite{ZaSaAl86}, the p
band lies above the d band and the LSDA fails \cite{TeWiOgKu84}. This is the
case of a charge-transfer insulator where excitations across the gap are O p
$\rightarrow$ Ni d. To describe such a system within band theory one must use
extensions of the LSDA, such as LDA+U or the self-interaction correction
(SIC), or a Hartree-Fock method. In these approaches different d orbitals may
see different potentials. For example one $\up$ spin \chem{e_g} orbital could
lie well below the Fermi energy and the other $\up$ spin \chem{e_g} orbital
could lie above it. Since a particular \chem{e_g} orbital must be picked out
for occupation on a given site cubic symmetry is broken and orbital ordering
occurs. The splitting between the \chem{e_g} levels would be the Hubbard
on-site Coulomb interaction $U$. The p band can then lie above the occupied
\chem{e_g} band. The density of states (DOS) obtained by Su \etal
\cite{SuKaMaHa00} in a Hartree-Fock calculation for \chem{LaMnO_3} is shown
in figure~\ref{fig:fig6}. Here the occupied majority spin d states
(\chem{t_{2g}^3e_g^1}) lie well below the oxygen band and the Mn moment is
close to 4$\mu_{\rm B}$. The DOS sketched by Satpathy \etal \cite{SaPoVu96}
to illustrate the results of a LDA+U calculation for \chem{LaMnO_3} is
slightly less extreme. The majority spin d band lies just within the bottom
of the oxygen p band. \chem{LaMnO_3} orders magnetically as an A-type
antiferromagnet (AAF) with ferromagnetic ordering within the basal planes. In
figure~\ref{fig:fig6} it is seen that the projected \chem{O_{II}-p} DOS,
corresponding to oxygen in the basal planes (see figure~\ref{fig:fig1}), has
an exchange splitting of about 0.75 eV at the top of the band. If the system
were ferromagnetically ordered one would expect a similar, or slightly
greater, exchange splitting. Thus, according to this picture, the
ferromagnetic metal obtained by doping \chem{LaMnO_3} sufficiently with Sr,
for example, would contain $S=2$ Mn ions aligned by itinerant p holes of
opposite spin.

\begin{figure}[htbp]
  \centering
  \subfigure[]{%
  \begin{minipage}[b]{0.35\textwidth}
    \centering \includegraphics[width=\textwidth]{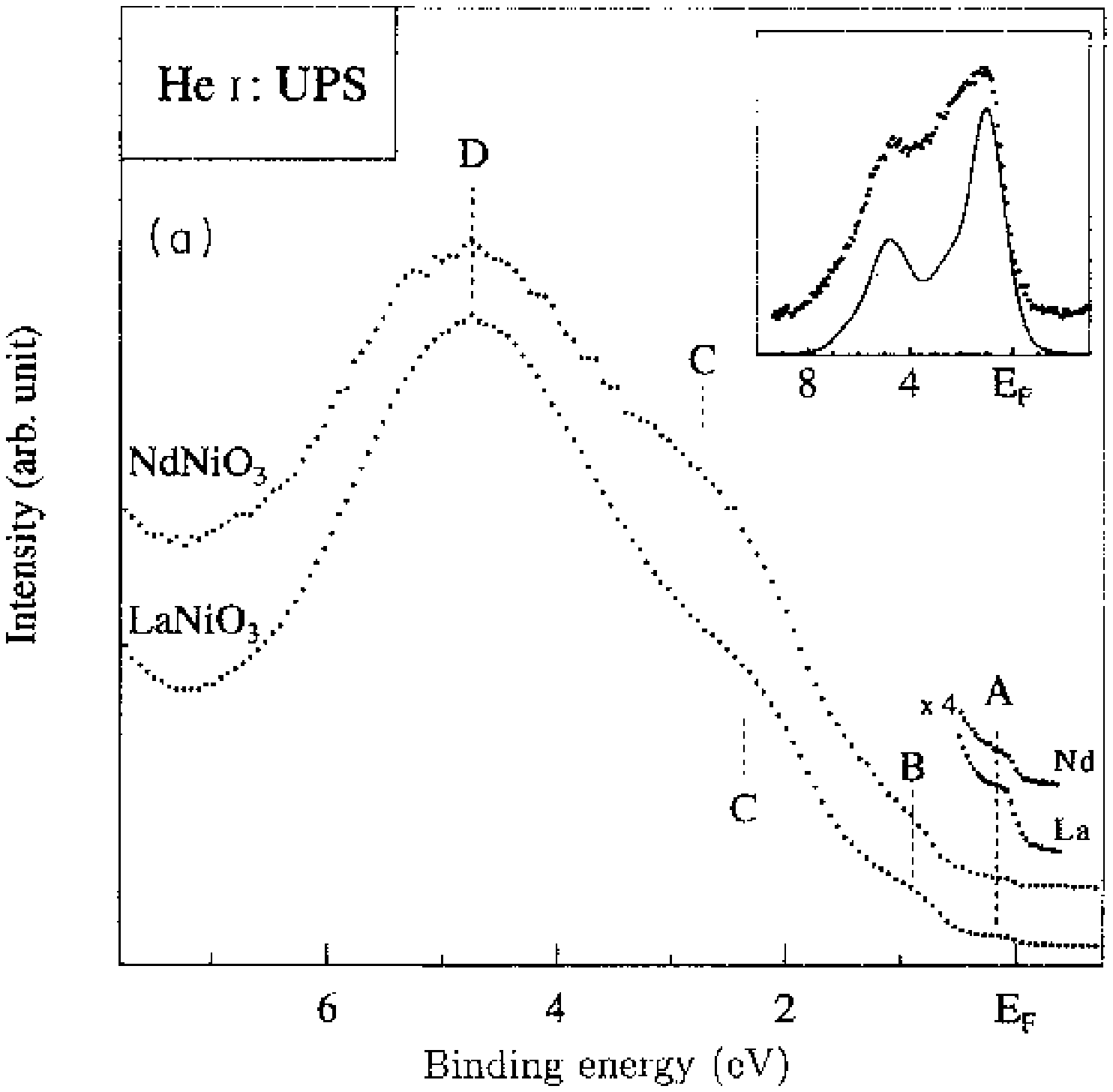}
  \end{minipage}}%
\subfigure[]{%
  \begin{minipage}[b]{0.55\textwidth}
    \centering \includegraphics[width=\textwidth]{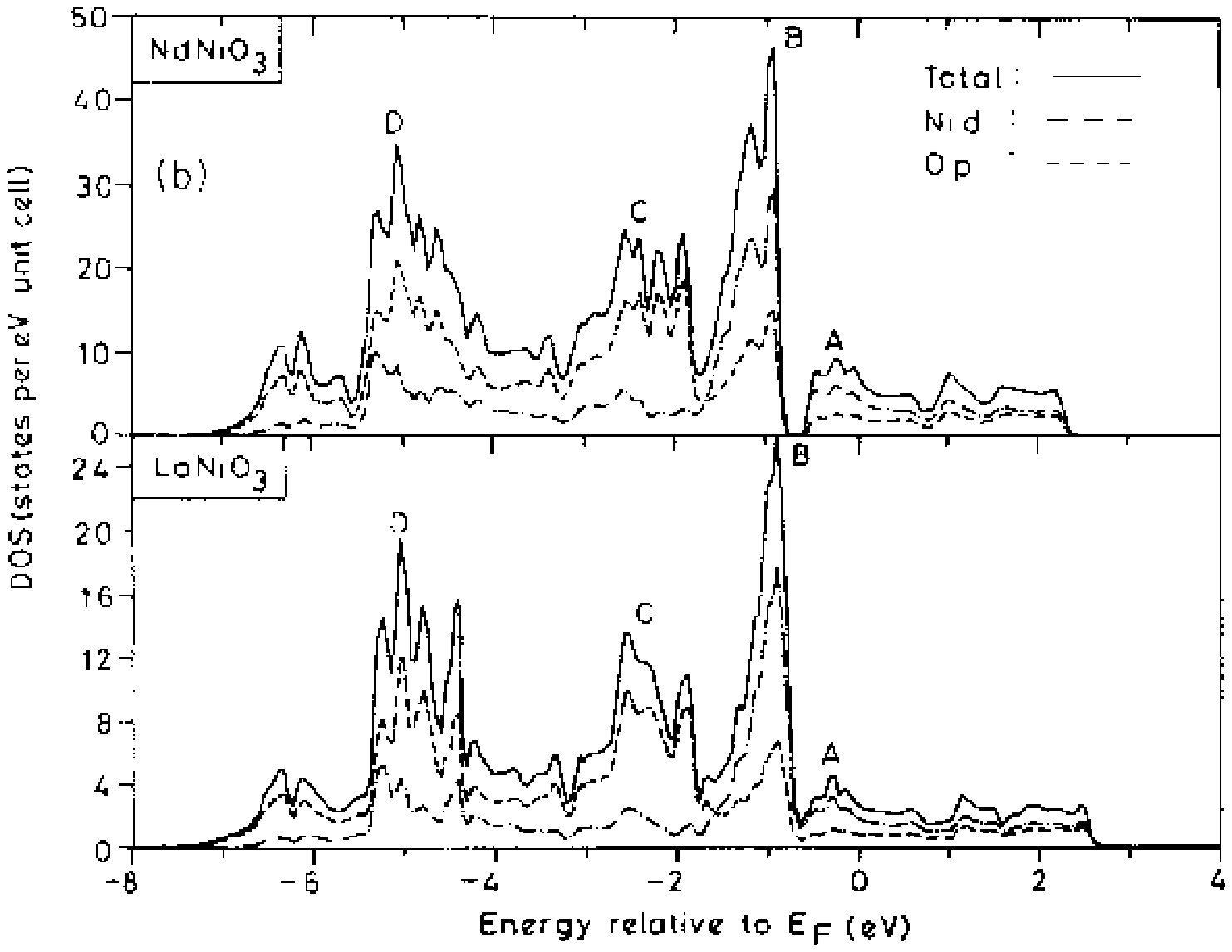}
  \end{minipage}}  
  \caption{\label{fig:fig7}%
    (a) He I UP spectra of \chem{LaNiO_3} and \chem{NdNiO_3}. The
    experimental (dotted line) and calculated (solid line) XP spectra of
    \chem{LaNiO_3} are shown in the inset. (b) The total DOS (solid line), Ni
    d (dot-dashed line), and O p (dashed line) partial DOS for \chem{LaNiO_3}
    and \chem{NdNiO_3} calculated within LMTO-ASA \cite{BaChSa94}.}
\end{figure}

\begin{figure}[htbp] 
  \centering
  \includegraphics[width=0.5\textwidth]{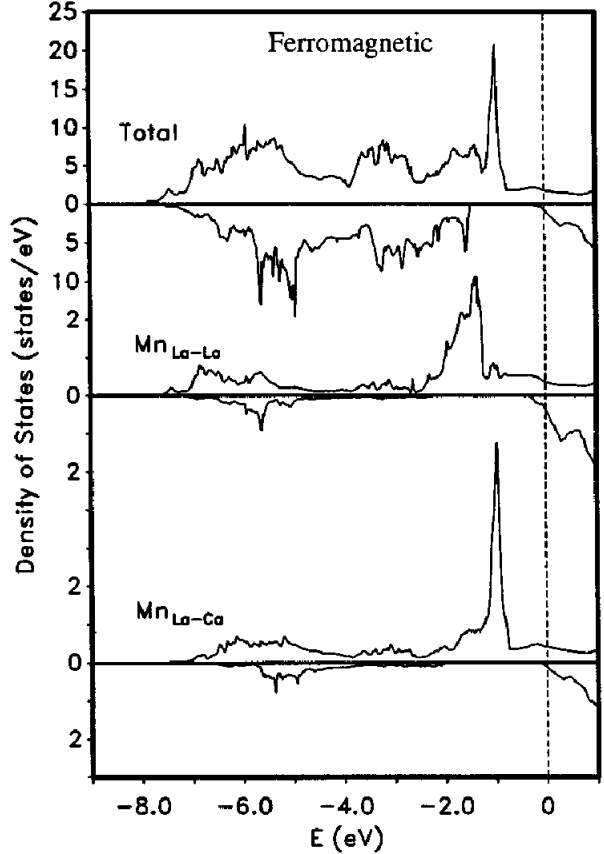}
  \caption{\label{fig:fig8}%
    The total DOS and
    local DOS on each inequivalent Mn ion for
    ordered \chem{La_{2/3}Ca_{1/3}MnO_3} with FM spin ordering \cite{PiSi96}.
    The subscripts denote the types of cation
    planes sandwiching that layer of Mn ions.}
\end{figure}

Both of the very different theoretical pictures presented above can explain
the magnetic and structural properties of \chem{LaMnO_3}. To distinguish
between them we must examine some spectroscopic data. Barman \etal
\cite{BaChSa94} give a very clear picture of the situation in the related
material \chem{LaNiO_3}. This is a non-magnetic metal with the \chem{Ni^{3+}}
ion in a nominal \chem{t_{2g}^6e_g^1} configuration, that is
\chem{t_{2g}^3e_g^{0.5}} for each spin. Thus the band structure should be
very similar to the majority spin band in ferromagnetic
\chem{La_{0.5}Sr_{0.5}MnO_3}. Indeed the calculated DOS in reference
\cite{BaChSa94}, shown in figure~\ref{fig:fig7}, resembles in detail majority
spin bands found in LSDA calculations for \chem{LaMnO_3} \cite{SaPoVu96},
\chem{La_{2/3}Ca_{1/3}MnO_3} \cite{PiSi96} and \chem{La_{2/3}Ba_{1/3}MnO_3}
\cite{SiPi98}. Comparison should be made with the total DOS in
figure~\ref{fig:fig8}, due to Pickett and Singh \cite{PiSi96}. Their
calculation is for ordered \chem{La_{2/3}Ca_{1/3}MnO_3} with planes of La and
Ca in a regular sequence. The feature of the DOS marked A in
figure~\ref{fig:fig7}(b) is an \chem{e_g} band and B corresponds to the
\chem{t_{2g}} band. C and D are two features of the broad oxygen p band. In
the inset of figure~\ref{fig:fig7}(a) a calculated x-ray photoemission (XP)
spectrum, including the effects of matrix elements as well as resolution and
lifetime broadenings, is compared with the experimental spectrum. The peak
near the Fermi energy $E_{\rm F}$ arises from the d band features A and B.
The shoulder on the left of this peak and the second peak arise from the p
band features C and D respectively. Sarma \etal \cite{SaShBa95} showed that
this good agreement between theory based on LSDA band calculations and the
observed XP spectrum extends to \chem{LaCoO_3}, \chem{LaFeO_3} and
\chem{LaMnO_3}. In the Fe and Mn compounds, both insulators, the calculated
valence band had to be shifted by 2.0 and 1.3 eV, respectively, to enlarge
the band-gap. The LDA invariably underestimates band-gaps in semi-conductors
and insulators (maybe LDA + small U should be considered, as discussed at the
end of section~\ref{sec:double-exchange-de}). In the ultraviolet
photoemission (UP) spectrum of \chem{LaNiO_3} shown in
figure~\ref{fig:fig7}(a) features A and B appear much more weakly than C and
D, in contrast to the XP spectrum, and this is consistent with the assigned d
and p angular momentum character \cite{BaChSa94}. The UP spectrum of
\chem{LaMnO_3}, both doped and undoped, is quite similar to that of
\chem{LaNiO_3} with clear p band features C and D \cite{ChMaSa93,SaBoMiNa95}.
However features A and B are not visible and seem only to contribute to a
tail in the spectrum which extends to $E_{\rm F}$ in metallic samples such as
\chem{La_{1-x}Sr_xMnO_3} \cite{SaBoMiNa95}. It seems even more difficult to
photo-excite d electrons near $E_{\rm F}$, with ultraviolet photon energies,
in doped \chem{LaMnO_3} than in \chem{LaNiO_3}. Furthermore only a weak
shoulder corresponding to features A and B is seen in the XP spectrum of
\chem{La_{0.7}Sr_{0.3}MnO_3}, unlike the large peak in \chem{LaNiO_3}
\cite{ChMaSa93}. However Park \etal \cite{PaChCh96}, in high resolution
spectra, see clear Fermi edges at 80 K in \chem{La_{0.67}Ca_{0.33}MnO_3} and
\chem{La_{0.7}Pb_{0.3}MnO_3}. A contributory factor to low density of states
at $E_{\rm F}$ may be the effect of electron-phonon coupling, which tends to
open a pseudo-gap in the one-electron spectrum as discussed in
sections~\ref{sec:many-body-cpa-hde} and~\ref{sec:pseudogaps}. Surface
preparation is clearly also very important since the probing depth in
photoemission may be as small as 5 {\AA}\ \cite{PaChCh98II}.

According to LSDA calculations, for example that shown in
figure~\ref{fig:fig8}, a doped manganite like \chem{La_{2/3}Ca_{1/3}MnO_3} is
almost `half-metallic'. This term applies to a ferromagnetic metal in which
the Fermi level lies in a gap in the DOS for states of one spin. In
figure~\ref{fig:fig8} the Fermi level lies just above a gap in the minority
spin DOS. According to this calculation states from just below $E_{\rm F}$ to
about 1.4 eV below it should be 100\% spin polarized. Spin polarization at
$E_{\rm F}$ can be measured directly by Andreev reflection and values of up
to 80\% have been found in \chem{La_{0.7}Sr_{0.3}MnO_3} at 4.2 K by Soulen
\etal \cite{SouByOsNaAm98} and Osofsky \etal \cite{OsNaSo99}. Large spin
polarization in \chem{La_{0.67}Sr_{0.33}MnO_3} is also deduced from
spin-polarized tunneling measurements \cite{Lu96,SuKrDu97}. Spin-polarized
photoemission can probe below $E_{\rm F}$ and Park \etal \cite{PaChCh98} find
100\% polarization in thin films of \chem{La_{0.7}Sr_{0.3}MnO_3} down to 0.6
eV below $E_{\rm F}$. This suggests a smaller gap in the minority spin DOS
than predicted theoretically. In early work on spin-polarized photoemission
from \chem{La_{0.7}Pb_{0.3}MnO_3} Alvarado \etal \cite{AlEiMu76} attributed
their failure to see more than 20\% polarization to a broad band of
unpolarized Pb 6s states extending almost up to $E_{\rm F}$.

\begin{figure}[htbp] 
  \centering
  \includegraphics[width=0.5\textwidth]{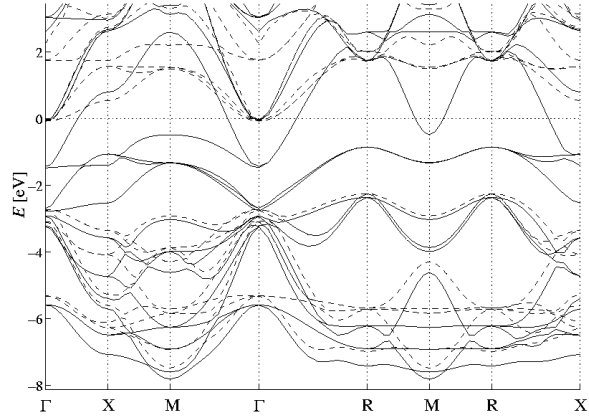}
  \caption{\label{fig:fig9}%
    Spin-polarized band structure of \chem{La_{0.7}Sr_{0.3}MnO_3}. The
    majority spin bands are shown as solid lines, and the minority as dashed
    lines \cite{Li99}.}
\end{figure}

\begin{figure}[htbp] 
  \centering
  \includegraphics[width=0.5\textwidth]{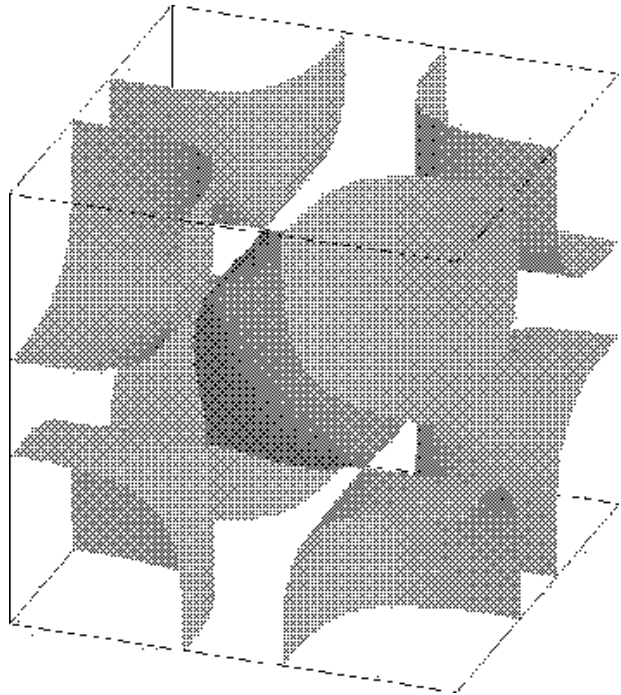}
  \caption{\label{fig:fig10}%
    Two sheets of the Fermi surface of \chem{La_{0.7}Sr_{0.3}MnO_3}. Hole
    cuboids at the $R$ points, coined `woolsacks'. Electron spheroid centred
    at the $\Gamma$ point \cite{Li99}.}
\end{figure}

A stringent test of a proposed electronic structure of a metal is whether it
gives the observed Fermi surface. In a disordered system like
\chem{La_{0.7}Sr_{0.3}MnO_3} the Fermi surface is not sharp as in a pure
metal. Even at $T=0$ there is no actual discontinuity in occupation number
between states inside and outside a perfectly-defined surface in
$\bi{k}$-space. Standard methods of determining the Fermi surface, like the
de Haas--van Alphen method, can generally not be applied, except for the case
of weak scattering in dilute alloys. However if scattering is not too strong
the transition between occupied and unoccupied states in $\bi{k}$-space is
sufficiently rapid to show up in the electron momentum distribution as a
fairly clear Fermi surface. The momentum distribution can be measured by
Compton scattering or positron annihilation. Livesay \etal \cite{Li99} have
used positron annihilation to investigate the Fermi surface of
\chem{La_{0.7}Sr_{0.3}MnO_3} and compare it with a LSDA band calculation.
They used a cubic perovskite structure and a virtual crystal approximation,
which makes the Fermi surface sharp, to find the band structure shown in
figure~\ref{fig:fig9}. The system is almost half-metallic and the two
majority spin Fermi surface sheets are shown in figure~\ref{fig:fig10}. These
comprise hole cuboids at the $R\ (\pm\pi/a,\pm\pi/a,\pm\pi/a)$ points that
touch an electron spheroid, centred at the $\Gamma\ (0,0,0)$ point, along the
$(111)$ directions. Similar Fermi surface sheets are described by Pickett and
Singh \cite{PiSi97}. The analysis of the positron annihilation data is not
simple, but Livesay \etal conclude that their results agree well with the
theory and establish the existence of the cuboid hole sheets. The virtual
crystal approach is reasonable since Pickett and Singh
\cite{PiSi96,SiPi98,PiSi97} have shown that scattering due to random A-site
occupation has little effect on the majority-spin bands around $E_{\rm F}$.
However the disorder localizes states at the bottom of the minority-spin band
near $E_{\rm F}$. Even if the bottom of the band falls just below $E_{\rm F}$
the system will still behave as half-metallic in transport \cite{PiSi97}.

The standard method of investigating band structure experimentally, away from
the Fermi level, is angle resolved photoemission (ARPES). McIlroy \etal
\cite{McWaZh96} have investigated \chem{La_xCa_{1-x}MnO_3} and
\chem{La_xBa_{1-x}MnO_3} with $x=0.35$ using this technique. As pointed out
above, the \chem{e_g} states close to the Fermi energy are hardly visible in
photoemission, so one is looking at the dense ensemble of bands between 2 and
8 eV below $E_{\rm F}$ (see figure~\ref{fig:fig9}). It is hardly to be
expected that any clear dispersion curves will emerge and McIlroy \etal
deduce that in the Ca-doped compound there are a lot of very flat bands. They
conclude that all the occupied states are extremely localized, but this seems
an improbable assessment of the broad p bands. Some dispersion is found in
the Ba compound, possibly associated with the steeply dropping bands in the
$\Gamma X$ direction at around 4 eV below $E_{\rm F}$. A discussion of ARPES
measurements by Dessau \etal \cite{De98} on a bilayer manganite is deferred
to section~\ref{sec:angle-resolv-phot}.

Clearly there is a lot of evidence that the electronic structure of the
manganites is well described by LSDA calculations and that states near the
Fermi level have predominantly Mn d character with \chem{e_g} symmetry. However
hybridization with O 2p states is by no means negligible. Sarma \etal
\cite{SaShBa95} find that at the valence-band top and conduction-band bottom
in \chem{LaMnO_3} the d component is 58\% and 73\% respectively. Simple
one-band or two-band models must be thought of as hopping between orbitals of
\chem{e_g} symmetry centred on Mn but extending onto neighbouring O atoms. The
simple nature of the bands crossing the Fermi level, and the simple Fermi
surface, suggest that a suitably parameterized two-band model, as used by
many authors, may be quite realistic. Sarma \etal \cite{SaShBa95} discuss the
reasons for the success of LSDA in the \chem{LaMO_3} compounds with M=Mn, Fe,
Co and Ni. In these perovskites the M--O bond is shorter than in the rocksalt
MO compounds, so that the M--O hopping parameter is larger. Also the effective
Coulomb interaction $U$ on the M site is smaller due to better screening by
the closer O ions and to larger effective d--d hopping.

We now briefly discuss some work which is often quoted in support of the
hypothesis that \chem{LaMnO_3} is an insulator with a charge-transfer (O p
$\rightarrow$ Mn d) gap. Instead of a band calculation for the crystal Saitoh
\etal \cite{SaBoMiNa95} performed configuration-interaction calculations for
an isolated \chem{MnO_6} cluster. The parameters of the model are the d--d
Coulomb interaction $U$, the p $\rightarrow$ d charge-transfer energy
$\Delta$ and p $\rightarrow$ d hopping parameter. They varied the parameters
to obtain the best fit to the observed Mn 2p core-level photoemission
spectrum, O 1s x-ray absorption spectrum and valence-band photoemission. They
conclude that for \chem{LaMnO_3} $U=7.8$ eV and $\Delta=4.5$ eV so that the
condition $U>\Delta$ for a charge-transfer insulator is satisfied. On the
other hand Chainani \etal \cite{ChMaSa93} in a similar, but slightly
different, analysis find $\Delta=5.0$ eV, $U=4.0$ eV which makes
\chem{LaMnO_3} more Mott-Hubbard-like than charge-transfer-like. Even for
spectroscopic properties the results of cluster calculations seem
inconclusive. Energy loss spectroscopy (EELS) on \chem{La_{1-x}Sr_xMnO_3}
in which electrons are excited from O 1s states to empty p-like states, has
been presented \cite{JuSoKr97} as confirmation of the hypothesis that
\chem{LaMnO_3} is a charge-transfer-type insulator. A peak at the Fermi level
grows with increasing $x$ and this is interpreted as proof that doping
produces holes in the oxygen p band. However it seems that the scenario
extensively discussed in this section, where the charge carriers have Mn
\chem{e_g} character, although significantly hybridized with O 2p states, is not
excluded.

In the next section we discuss the exchange mechanism responsible for
ferromagnetism in the doped manganites from the general viewpoint of
itinerant electron magnetism. It is clearly related to the nature of the
underlying electronic structure.


\section{Itinerant electron ferromagnetism and double exchange}
\label{sec:itin-electr-ferr}

\subsection{The double-exchange (DE) model}\label{sec:double-exchange-de}

All work on the electronic structure of the manganites is in agreement on one
point. The three majority spin \chem{t_{2g}} orbitals are all occupied and the
minority spin ones are unoccupied. This suggests that \chem{t_{2g}} electrons
play no part in transport and may be considered as localized. Consequently the
system may be considered as consisting of local spins $S=3/2$ on each Mn site
coupled to electrons in a predominantly \chem{e_g} conduction band by local
exchange interactions. According to the alternative view of the electronic
structure, discussed in section~\ref{sec:electronic-structure}, one \chem{e_g}
electron per Mn site is also localized and local spins $S=2$ are coupled to
holes in the oxygen p band. Although we mention this alternative view again
later we shall primarily adapt the former view for which strong evidence has
been given in section~\ref{sec:introduction}.

Metallic rare-earth materials can also be considered as
systems of local moments coupled to electrons in a conduction band by local
exchange interactions. We now discuss the distinction between these and the
manganites. The Hamiltonian for such a system is
\begin{equation}\label{eq:h_de}
  H = \sum_{ij\si} t_{ij}c_{i\si}^{\dag}c_{j\si}-J\sum_i
  \bi{S}_i\cdot\bsigma_i-h\sum_i\left(S_i^z+\si_i^z\right)\,,
\end{equation}
where $c^{\dag}_{i\si}$ creates an electron of spin $\si$ on lattice site
$i$, $\bi{S}_i$ is a local spin operator and
$\bsigma_i=\left(\si_i^x,\si_i^y,\si_i^z\right)$ is a conduction electron
spin operator defined by
\begin{equation}
  \si_i^+=\si_i^x+i\si_i^y=c_{i\up}^\dag
  c_{i\Nup}\,,\,\si_i^-=\si_i^x-i\si_i^y=c_{i\Nup}^\dag
  c_{i\up}\,,\,\si_i^z=\frac{1}{2} \left(n_{i\up}-n_{i\Nup}\right)
\end{equation}
with $n_{i\si}=c^\dag_{i\si}c_{i\si}$. The three terms of
equation~(\ref{eq:h_de}) describe hopping of the conduction electrons,
exchange coupling between local and itinerant spins and coupling to an
external magnetic field. If the local exchange coupling arises from
hybridization between the localized and itinerant electrons, as in anomalous
rare earth systems exhibiting heavy fermion behaviour, the exchange parameter
$J$ is negative. The Hamiltonian (\ref{eq:h_de}) is then often called the
Kondo lattice model in view of its connection with the Kondo impurity model
which has a local spin on one site only \cite{He93}. When Hund's rule
coupling is dominant $J>0$ and the system is sometimes called a ferromagnetic
Kondo lattice. This is misleading since for $J>0$ there is no connection with
the Kondo effect.

For $J>0$ it is useful to distinguish two distinct physical regimes,
depending on the magnitude of $J$ compared with the width $2W$ of the
conduction band. If $J\ll W$, as in a normal rare earth metal, the exchange
coupling can be treated as a perturbation which gives rise to the
Ruderman-Kittel-Kasuya-Yosida (RKKY) interaction between local moments. In
most rare earth metals this interaction, which oscillates in space, leads to
oscillatory or spiral configurations of the localized f electron moments.
The uniform ferromagnet \chem{Gd} is an exception. In this weak coupling
regime the Hamiltonian (\ref{eq:h_de}) is usually referred to as the $s-f$ or
$s-d$ model.

If $J\gg W$ the exchange coupling can no longer be treated as a perturbation.
A conduction electron can only hop onto a site with its spin parallel to the
local moment at that site. Furthermore if the number of conduction electrons
per atom $n\le1$ double occupation of a site is strongly suppressed. A single
electron at a site, with its spin parallel to the local spin $\bi{S}$, enjoys
an exchange energy $-JS/2$ which is lost if a second electron hops on.  The
system is therefore a {\em strongly correlated electron system}, just like
the Hubbard model in the regime of strong on-site Coulomb interaction $U$,
and for $n=1$ the system is a Mott insulator. In much of the theoretical work
on the present model the local spins are treated as classical vectors,
corresponding to $S\rightarrow\infty$. Since for $J\gg W$ the itinerant spin
must always be parallel to the local spin on each site, the effective hopping
integral for hopping between sites $i$ and $j$ becomes
$t_{ij}\cos\left(\theta_{ij}/2\right)$, where $\theta_{ij}$ is the angle
between the classical spins $\bi{S}_i$, $\bi{S}_j$. The cosine factor arises
from the scalar product of two spin $1/2$ eigenstates with different axes of
quantization. The resultant band narrowing in the paramagnetic state favours
ferromagnetism in order to lower the kinetic energy. This mechanism for
ferromagnetism was first introduced by Zener \cite{Ze51} and developed by
others \cite{AnHa55,Ge60,KuOh72}. Since it involves strong exchange coupling
on two adjacent atoms it is known as double-exchange. Consequently the
Hamiltonian (\ref{eq:h_de}) in the strong-coupling regime $J\gg W$ is called
the double-exchange (DE) model. We note that the p-hole picture is not
in the DE regime since the antiferromagnetic coupling between the Mn local
spins and the O p-holes is weak compared with the large oxygen
band-width. This review is largely concerned with the DE model,
with quantum and classical local spins, and with its extension to include
coupling of the electrons to local phonons. We call this extended model the
Holstein-DE model \cite{Gr01}. We now proceed to place the DE model in the
wider context of itinerant electron ferromagnetism.

\subsection{A wider view}\label{sec:wider-view}

The macroscopic exchange energy in a ferromagnet takes the form
\begin{equation}
  A \int\, \sum_i\left(\nabla n_i(\bi{r})\right)^2\rmd^3 r
\end{equation}
where $n_1$, $n_2$, $n_3$ are the direction cosines of the magnetization
direction at position $\bi{r}$. Here $A$ is the Bloch wall stiffness
constant. A long wavelength spin-wave is a macroscopic oscillation in which
the magnetization precesses around the equilibrium $z$ direction. The
transverse components of magnetization vary in space and time as
\begin{equation}
  M^+=M_x+\rmi M_y\propto \exp\left[\rmi(\bi{q}\cdot\bi{r}-\om t)\right]\,.
\end{equation}
The dispersion relation is
\begin{equation}
  \hbar\om=Dq^2
\end{equation}
and the spin-wave stiffness constant $D\propto A/M_s$ where $M_s$ is the
saturation moment. This macroscopic picture is valid for any ferromagnet, at
zero or finite $T$ ($<\Tc$) and with an ordered or disordered crystal
structure. In exciting a spin-wave quantum (magnon), for example by neutron
scattering, a single spin is flipped down. For ferromagnets with a
substantial moment per atom, of a Bohr magneton or two, long wavelength
spin-waves make an important, and often dominant, contribution to the
temperature dependence of the magnetization for $T\lesssim\Tc/2$, through the
well-known Bloch $T^{3/2}$ law. In manganites with $\Tc\gtrsim350$ K the
ratio $\delta=D/(k_{\rm B}\Tc a^2)$, where $a$ is the lattice constant, is
close to the value 0.286 for the spin $3/2$ nearest neighbour simple cubic
Heisenberg model. For manganites with lower $\Tc$, $\delta$ can be larger by
up to a factor 2 and this is attributed to strong electron-phonon coupling
\cite{HoEd01}. However in general $D$ provides a good measure of $\Tc$. This
is not the case, however, in very weak itinerant ferromagnets such as
\chem{ZrZn_2}, with small saturation moment, where longitudinal spin
fluctuations are a determining influence on $\Tc$.

A very useful feature of $D$ is that it is in effect a ground state
property. It is directly related to the Bloch wall stiffness constant A,
which can be determined from the ground state energy of the system in a
non-uniform transverse magnetic field. Thus in principle $D$ can be
calculated exactly in spin density functional theory, even though the band
structure used in the course of the calculations has no such rigorous
validity. In practice one obtains good values for $D$ in Fe and Ni, and their
alloys, on the basis of LSDA band calculations
\cite{EdMu85,MuCoEd85}. We are aware of only one calculation along these
lines for any manganite material. Solovyev and Terakura \cite{SoTe99} find
$D\approx300$ meV{\AA}$^2$ for \chem{La_{0.7}Ba_{0.3}MnO_3} which is
comparable to experimental values of somewhat less than 200 meV{\AA}$^2$ for
similar manganites with $x=0.3$. We discuss the general approach below since
it gives insight into the exchange mechanism in ferromagnetic metals and
makes clear the basis of the double-exchange model.

Edwards and Muniz \cite{EdMu85} showed how $D$ can be evaluated directly from
a multi-band tight-binding parameterization of a LSDA band calculation for
the ferromagnetic ground state. No input about the underlying
electron-electron interaction is required. However at a more microscopic
level one may consider the band structure to arise from a Hartree-Fock
approximation (HFA) to a Hamiltonian of the form
\begin{equation}
  H=\sum_{\bi{k}}\sum_{\mu\mu'\si}V_{\mu\mu'}(\bi{k})c^\dag_{\mu\bi{k}\si}
  c_{\mu'\bi{k}\si}+H_{\rm int}\equiv H_0 + H_{\rm int}\,,
\end{equation}
where $c^\dag_{\mu\bi{k}\si}$ creates an electron of spin $\si$ in a Bloch
state of wave-vector $\bi{k}$ formed from orbitals $\mu$. Here the first term
$H_0$ represents the electron `kinetic energy', actually the kinetic energy
plus a spin-independent local potential, and $H_{\rm int}$ is a local
(intraatomic) interaction term. Off-diagonal elements $V_{\mu\mu'}$ describe
hybridization between different orbitals. In general $H_{\rm int}$ contains
screened Coulomb interactions and exchange interactions of the Hund-rule type
between the various orbitals. It can also include an additional one-electron
term representing spin-independent diagonal disorder and may be generalized
to include interaction with local phonons, as we shall discuss later. The
Hamiltonian may be considered at two levels; either with effective
interactions designed to reproduce the LSDA band structure within the HFA, or
as a true many-body Hamiltonian in which correlation effects must be treated
explicitly, for example by dynamical mean field theory
\cite{GeKoKrRo96,pruschke_review}. In both approaches one may start with an
exact expression for $D$ \cite{EdFi71}:
\begin{equation}\label{eq:D_exact}
  D=\frac{\hbar^2}{N_\up-N_\Nup}\left(\lim_{q\rightarrow 0}(\hbar q)^{-1}
  \las\com{J^-_\bi{q}}{S^+_\bi{-q}}\ras-\lim_{\om\rightarrow 0}
  \chi_J(0,\om)\right)\,.
\end{equation}
Here $S^-_\bi{q}=\sum_{\bi{k}\mu} c^\dag_{\mu\bi{k+q}\Nup}c_{\mu\bi{k}\up}$
is a spin-flipping operator representing a transverse component of spin
density and $J^-_\bi{q}$ is the spin current operator defined by
\begin{equation}
  \hbar qJ^-_\bi{q}=\com{S^-_\bi{q}}{H}=\sum_\bi{k}\sum_{\mu\mu'}
  \left(V_{\mu\mu'}(\bi{k})-V_{\mu\mu'}(\bi{k+q})\right)c^\dag_{\mu\bi{k+q}\Nup}
  c_{\mu'\bi{k}\up}\,.
\end{equation}  
$N_\up$ and $N_\Nup$ are the total number of electrons of $\up$ and $\Nup$
spin and $\chi_J(\bi{q},\om)$, the spin current-spin current response function,
is given by the Fourier transform of a retarded Green function in the form
\begin{equation}
  \chi_J(\bi{q},\om)=\int\,\rmd t\llas
  J^-_\bi{q}(t),J^+_{-\bi{q}}\rras\rme^{-\rmi\om t}\,.
\end{equation}
The expression in brackets in equation~(\ref{eq:D_exact}) is proportional to
the Bloch wall stiffness constant $A$ ($\propto D(N_\up-N_\Nup)$) and the
second term is a generalized superexchange term which always makes a negative
contribution to $D$. The first term therefore yields an upper bound to $D$
and corresponds to the variational ansatz $S^-_\bi{q}\ket{F}$ for the state
with a spin-wave of wave-vector $\bi{q}$ excited, where $\ket{F}$ is the
ferromagnetic ground state. This zeroth approximation $D_0$ is often the
dominant term in itinerant electron ferromagnets with maximum spin alignment
such as the doped manganites. It is therefore worth examining in detail and
for a cubic crystal may be written in the form
\begin{equation}\label{eq:D0}
  D_0=\frac{1}{6(N_\up-N_\Nup)}\sum_\bi{k}\sum_{\mu\mu'\si}\nabla^2
  V_{\mu\mu'}(\bi{k})\las c^\dag_{\mu\bi{k}\si}c_{\mu'\bi{k}\si}\ras\,,
\end{equation}
where the expectation value is evaluated in the ferromagnetic ground state
and differentiation in $\nabla^2$ is with respect to $\bi{k}$. If a
particular orbital $\mu\si$ is completely full or completely empty in the
ground state it makes no contribution to $D_0$. This follows because under
these circumstances $\las c^\dag_{\mu\bi{k}\si}c_{\mu'\bi{k}\si}\ras$ and
$\las c^\dag_{\mu'\bi{k}\si}c_{\mu\bi{k}\si}\ras$ are both zero for
$\mu\neq\mu'$ and the integral of $\nabla^2V_{\mu\mu}(\bi{k})$ over the whole
zone may be written as a surface integral over the zone boundary which
clearly vanishes. Thus in the manganites the \chem{t_{2g}} orbitals do not
contribute to the first term in the Bloch wall stiffness constant $A$, as
long as their mixing into $\up$ spin states above $E_{\rm F}$ and into $\Nup$
spin states below $E_{\rm F}$ is negligible. Of course the \chem{t_{2g}}
electrons make the largest contribution to the factor $N_\up-N_\Nup$ in $D$,
and without their exchange coupling to the \chem{e_g} electrons there would
be no ferromagnetism. The considerations above provide a justification for
treating the \chem{t_{2g}} electrons as local spins, despite the substantial
width of the \chem{t_{2g}} bands, and for the model DE Hamiltonian of
equation~(\ref{eq:h_de}). We shall see that recent calculations of $D$ in
this model are special cases of the general formalism presented here.

Quijada \etal \cite{QuCeSi98} have considered a two-band generalization of
equation~(\ref{eq:h_de}) and calculate $D$ in an approximation equivalent to
the random phase approximation (RPA). Nearest-neighbour hopping integrals on
a simple cubic lattice are parameterized appropriately for \chem{e_g}
orbitals, assuming only one non-zero Slater-Koster parameter (dd$\si$). For
$J\rightarrow\infty$ the result is just $D_0$ given by
equation~(\ref{eq:D0}), with $N_\up-N_\Nup=(2S+n)N$ where $n=1-x$ is the
total number of electrons per atom in the conduction bands and $N$ is the
number of atoms in the crystal. For nearest-neighbour hopping
$\nabla^2V_{\mu\mu'}(\bi{k})=-a^2V_{\mu\mu'}(\bi{k})$, where $a$ is the
lattice constant and the on-site orbital energy taken as zero. Thus
\begin{equation}\label{eg:D0b}
  D_0=-\frac{K a^2}{6(2S+n)}\,,
\end{equation}
where $K=N^{-1}\las H_0\ras$ is the expectation value of the `kinetic energy'
per atom. The $K$ defined in reference \cite{QuCeSi98} is one-third of that
defined here and has the opposite sign. In general our $K\leq0$ and
equation~(\ref{eg:D0b}) emphasizes that exchange stiffness is driven by
one-electron (`kinetic') energy in the DE model.

It is important to notice that equations~(\ref{eq:D0}) and~(\ref{eg:D0b}) are
still valid if terms corresponding to spin-independent diagonal disorder and
local coupling to phonons are included in the Hamiltonian. This last point
has been exploited to investigate the effect of electron-phonon coupling on
$D$ \cite{HoEd01}, as discussed in section~\ref{sec:spinwaves}.
The result concerning disorder is important because it shows that the
double-exchange mechanism for
ferromagnetism does not depend on having mobile carriers. Localized states
contribute to the one-electron energy just as well as extended ones, as long
as the localization length is more than a few lattice constants. This
presumably explains why $\Tc$ varies continuously through the FM-FI phase
boundaries in figure~\ref{fig:fig2}. At low carrier density
($x\lesssim0.2$) the occupied states are below a mobility edge, making the
system an insulator, but double exchange still operates to give
ferromagnetism.

We have laid considerable stress on $D_0$ as the most important, typically
metallic, part of $D$. However in some systems the value of $D$ is subject to
enormous cancellation between $D_0$ and the second `superexchange' term. Thus
in a very weak itinerant ferromagnet such as \chem{ZrZn_2}, with very small
magnetization $M$, the bracket in equation~(\ref{eq:D_exact}) is proportional
to $M^2$, so that $D\propto M$. In a rare-earth metal like Gd, or in an
Anderson lattice model of a ferromagnetic heavy fermion system, it is a
delicate matter in this formalism to derive the correct RKKY-like result
\cite{EdRi92}. The situation would be similar in a calculation of $D$
based on the electronic structure picture of \chem{t_{2g}^3e_g^1} ($S=2$)
spins coupled via p holes. Zhao \cite{Zh00} criticizes the standard DE
picture with $S=3/2$ spins coupled by \chem{e_g} carriers. His criticism is
based on a two-band estimate of $D$ with finite $J$, which comes out very
small or negative. His model is exactly that of Quijada \etal \cite{QuCeSi98}
who estimate values of $D$ in good agreement with experiment. Zhao does not
estimate $D$ in the alternative p-hole picture which he favours. It would be
very surprising if the RKKY-like exchange of that picture were larger than
that of the itinerant electron DE picture. It would be interesting to carry
out proper calculations of $D$ in the multi-band tight-binding formalism
starting from a LSDA band structure (compatible with the DE picture) and from
LSDA+U or Hartee-Fock band structure (the p-hole case). It was pointed out
earlier that Solovyev and Terakura \cite{SoTe99} obtained a reasonable value
for $D$ for a $x=0.3$ manganite by means of the LSDA. They calculated the
energies of static spin spirals of different wave-vectors $\bi{q}$ and fitted
them to a Heisenberg expression for the magnon energy with exchange extending
over many neighbours (see \eg \cite{LiKa01}). For small $q$ this is
equivalent to calculating the Bloch wall stiffness constant $A$ which is
exactly related to $D$. However there is no reason to place much reliance on
the dispersion curve for larger $q$. Solovyev and Terakura had difficulty
reconciling their work with the observed doping dependence of $D$;
experimentally in LSMO \cite{EnHi97} $D$ increases with $x$ for
$0.175<x<0.3$, just as $\Tc$ does (see figure~\ref{fig:fig2}). It is clear,
from their analysis of a simplified two-band model, that their opposite
conclusion is due to the double-exchange being largest at $x=0$ since the
\chem{e_g} band is then half-full. At $x=0$ the system should be a Mott
insulator. Probably the best calculational method would be LDSA+U with a U
large enough to split the $\up$ spin \chem{e_g} band but not large enough to
reverse the ordering of p and d states. A picture like this is sketched in
figure~12 of reference \cite{SaBoMiNa95}. This does not change the general
picture of the electronic structure we presented in
section~\ref{sec:electronic-structure}.

\subsection{Spin-waves in the DE model}\label{sec:spin-waves-de}

The spin-wave stiffness constant $D$ in the one-band DE model
(equation~(\ref{eq:h_de})) is easily deduced from the general multi-band
formulation of section~\ref{sec:double-exchange-de}. The only role of the
local spins is to provide an exchange splitting between $\up$ and $\Nup$ spin
bands and to contribute $2SN$ to the local moment $N_\up-N_\Nup$. We can
therefore use equation~(\ref{eq:D_exact}) and need an approximation to
$\chi_J(0,\om)$. A reasonable approach is to make a local approximation in
which the electron self-energy $\Si$ is assumed to be a function of energy
only, with no dependence on wave-vector $\bi{k}$. This is formally exact in
infinite dimensions \cite{GeKoKrRo96} and is the basis of the coherent
potential approximation (CPA) and dynamical mean field theory (DMFT). In this
case the vertex correction in the spin current response function
$\chi_J(0,\om)$ makes no contribution and $\chi_J$ can be evaluated using a
simple bubble diagram which involves the product of two one-electron Green
functions. This result is well-known from the more familiar charge current
response function used to calculate electrical and optical conductivity
\cite{Ve69,pruschke_review}. Then equation~(\ref{eq:D_exact}) becomes
\cite{HiEd73,Fu73,EdHi76}
\begin{equation}\label{eq:D_DE}
  D=\frac{1}{3(N_\up-N_\Nup)}\left(\frac{1}{2}\sum_\bi{k}\las n_{\bi{k}\up}+
  n_{\bi{k}\Nup}\ras\nabla^2\ek-\frac{1}{\pi}\Im\int\,\rmd\en
  f(\en)\sum_\bi{k}G_{\bi{k}\up}(\en)G_{\bi{k}\Nup}(\en)|\nabla\ek|^2\right)\,,
\end{equation}
where $\ek=\sum_j t_{ij} \exp[\rmi\bi{k}\cdot(\bi{R}_i-\bi{R}_j)]$ is the
band energy, $\bi{R}_i$ being the position of site $i$, $f(\en)$ is the Fermi
function and the one-electron retarded Green function $G_{\bi{k}\si}$ for
spin $\si$ is given by
\begin{equation}\label{eq:def_G}
  G_{\bi{k}\si}(\en)=\left[\en-\ek-\Si_\si(\en)\right]^{-1}\,.
\end{equation}
The occupation number $\las n_{\bi{k}\si}\ras$ is determined by
\begin{equation}
  \las n_{\bi{k}\si}\ras=-\frac{1}{\pi}\int\,\Im G_{\bi{k}\si}(\en)
  f(\en)\rmd\en\,.
\end{equation}
On substituting this into equation~(\ref{eq:D_DE}), and applying Green's
theorem to the sum over $\bi{k}$ in the first term, we find
\begin{equation}\label{eq:Dexpression}
  D=\left[6\pi(N_\up-N_\Nup)\right]^{-1}\Im\int\,\rmd\en f(\en)\sum_\bi{k}
  \left[G_{\bi{k}\up}(\en)-G_{\bi{k}\Nup}(\en)\right]^2|\nabla\ek|^2\,.
\end{equation}
This compact expression was previously obtained \cite{Fu73,EdHi76} within the
framework of a CPA-RPA calculation for ferromagnetic alloys. However it is
clearly exact within any local approximation, for example DMFT, and can be
used in the presence of coupling to local phonons. It shows how the
cancellation between the two terms of equation~(\ref{eq:D_DE}) takes place as
$N_\up-N_\Nup\rightarrow0$. An equivalent result has been given by
Lichtenstein and Katsnelson \cite{LiKa01}. A convenient form of this
result is obtained by introducing the function $M(E)$ \cite{WaEdWo71} defined
by the integral
\begin{equation}\label{eq:def_M}
  M(E)=\left(\Omega/8\pi^3\right)\int\,|\nabla\ek|\rmd S
\end{equation}
over the surface $\ek=E$ in $\bi{k}$ space. Here $\Omega$ is the volume of
the unit cell. Then at $T=0$
\begin{equation}
  D=\left[6\pi(2S+n_\up-n_\Nup)\right]^{-1}\int_{-\infty}^{E_{\rm
  F}}\,\rmd\en\int\,\rmd E
  M(E)\Im\left[G_{E\up}(\en)-G_{E\Nup}(\en)\right]^2
\end{equation}
where $G_{E\si}(\en)$ is defined by equation~(\ref{eq:def_G}) with $\ek$
replaced by $E$ and $n_\si$ is the number of $\si$ spin electrons per atom in
the band in the ferromagnetic ground state. Normally, we consider $J$ large
enough such that $n_\Nup=0$, so that $n_\up=n=1-x$.

If we make the Hartree-Fock approximation (HFA) to $G_{\bi{k}\si}$ we have
$\Si_\up=-JS/2$, $\Si_\Nup=JS/2$ and it is easy to show that
equation~(\ref{eq:D_DE}) yields the RPA result of Wang \cite{Wa98} and
Furukawa \cite{FuPom} for $D$. The second term involves a factor
$\Delta^{-1}$ where $\Delta=JS$ is the exchange splitting in the conduction
band. If a Hubbard on-site Coulomb term $U\sum_i n_{i\up}n_{i\Nup}$ is added
to the Hamiltonian of equation~(\ref{eq:h_de}) the only effect on $D$,
assuming the spins are already aligned completely, is that $\Delta$ is
increased to $JS+Un$. If we then set $S=0$ we recover the well-known RPA
result for $D$ in the Hubbard model (see \eg \cite{IzKiKu63}). If
$J\rightarrow\infty$ in the DE model the RPA result for $D$ reduces to the
first term, which is called $D_0$ in the previous section. If we go beyond
the HFA for $G_{\bi{k}\Nup}$, in the state of complete spin alignment, we
find that even for $J\rightarrow\infty$ (but finite $S$) there is always some
$\Nup$ spin spectral weight near the Fermi level which may even cause the
totally spin-aligned state to be unstable, at least for $S=1/2$
\cite{BrEd98,Ok97,WuMu98}. Thus half-metallicity may be destroyed and, even
if not, the Fermi level lies at the top of the half-metallic gap.  This
effect was known long ago in the Hubbard model and is due to processes in the
self-energy $\Si_\Nup$ in which a $\Nup$ spin electron flips its spin and
appears as a $\up$ spin electron near $E_{\rm F}$ plus a magnon
\cite{Ed68,Ro69,HeEd73,EdHe73}. The $\up$ spin electron and the magnon may
bind to form low-lying $\Nup$ spin quasi-particle states. The magnon
dispersion curve necessarily runs into a continuum involving these states for
wave-vector $\bi{q}$ close to an $\up$ spin Fermi wave-vector \cite{EdHe73}.
Kaplan \etal \cite{KaMaSu01} have carried out some numerical studies of this
effect in one dimension but seem unaware of the earlier analytical work. When
the self-energy $\Si_\Nup$ discussed above is introduced into
$G_{\bi{k}\Nup}$ in equation~(\ref{eq:Dexpression}), the effect on $D$ is to
add a negative term to $D_0$ (the RPA value for $J\rightarrow\infty$) which
is attributed to magnon-electron scattering \cite{Ed67,HeEd73,EdHe73}. The
magnon excites $\up$ spin electron-hole pairs. This has been investigated
diagrammatically from the point of view of a $1/S$ expansion by Golosov
\cite{Go00} and the effect can reduce $D$ in the two-dimensional DE model by
a factor of 2. Golosov also discusses magnon damping due to
magnon-electron scattering; he finds that the inverse magnon lifetime is
proportional to the sixth power of the wave-vector $\bi{q}$, as expected for
a three-dimensional itinerant electron ferromagnet with complete spin
alignment in the ground state \cite{Th65}.

Another approach to this problem is a variational one similar to the ones of
references \cite{Ok97,WuMu98}. An obvious ansatz for the state with one
spin-wave excited from the ferromagnetic state of complete spin alignment is
\begin{equation}
  (S^-_\bi{q}+\sum_{\bi{kp}}A_{\bi{kp}}S^-_{\bi{q+p-k}}c^\dag_{\bi{k}\up}
  c_{\bi{p}\up}) \ket{F}
\end{equation}
where $S^-_\bi{q}$ is the total spin-flipping operator for local and
itinerant spins. It is difficult to choose the coefficients $A_\bi{kp}$
optimally, but a convenient choice is
\begin{equation}
  A_\bi{kp}=[N(n+2S)]^{-1}(\en_{\bi{p}}-\en_{\bi{p+q}})(\ek-\en_\bi{p}+
  \hbar\om^{\rm RPA}_\bi{q+p-k})^{-1}\,,
\end{equation}
where
\begin{equation}
  \hbar\om^{\rm RPA}_\bi{q}=[N(n+2S)]^{-1}\sum_\bi{l}\las n_{\bi{l}\up}
  \ras(\en_\bi{l+q}-\en_\bi{l})
\end{equation}
is the RPA spin-wave energy for $J\rightarrow\infty$. Then we find an upper
bound on $D$, for $J\rightarrow\infty$, in the form
\begin{align}\label{eq:long}\nonumber
  D &= D_0-\frac{1}{3[N(n+2S)]^2}\sum_\bi{kp}\frac{|\nabla\en_\bi{p}|^2}
  {\ek-\en_\bi{p}+\hbar\om^{\rm RPA}_\bi{p-k}}\\
  &-\frac{1}{3[N(n+2S)]^3}\sum_\bi{k}\sum_\bi{pp'}\frac{\left(\nabla\en_\bi{p}
  \cdot\nabla\en_\bi{p'}\right)\left(\en_\bi{p'+p-k}-\en_\bi{p'}-\en_\bi{p}+\ek
  \right)}{\left(\ek-\en_\bi{p}+\hbar\om^{\rm RPA}_\bi{p-k}\right)\left(
  \ek-\en_\bi{p'}+\hbar\om^{\rm RPA}_\bi{p'-k}\right)}\,.
\end{align}
Here the sums over $\bi{p}$, $\bi{p'}$ are restricted to occupied states in
the $\up$ spin band and the sum over $\bi{k}$ is restricted to unoccupied
ones. This result agrees with that of Golosov \cite{Go00} to order
$1/S^2$. The last term in equation~(\ref{eq:long}), involving
$\nabla\en_\bi{p}\cdot\nabla\en_\bi{p'}$ with $\bi{p}\neq\bi{p'}$ in general,
does not fit into the exact form of $D$ in the local approximation
(equation~(\ref{eq:D_DE}). This is not unexpected because the $\Nup$ spin
self-energy including electron-magnon scattering depends on wave-vector
$\bi{k}$, not only on energy. The last term in equation~(\ref{eq:long})
corresponds to a vertex correction first discussed by Edwards
\cite{Ed67}.

\section{Magnetism and transport in the DE model}\label{sec:magn-transp-de}

In the last section we saw that much is known about the low-lying excitations
of the DE model. Recent work on the properties of the model over the whole
temperature range often deals with the limit of classical local spins
($S\rightarrow\infty$). The most complete work along these lines is that of
Furukawa \cite{FuPom}, using dynamical mean field theory (DMFT) which is the
best local approximation. No analytical progress can be made within DMFT for
the quantum-spin DE model and numerical calculations have not yet
appeared. However there is no doubt that quantum spins introduce effects
which do not exist for classical spins. For example the low-lying $\Nup$ spin
spectral weight discussed in the last section has total weight $(1-n)/(2S+1)$
for large $J$ (see section~\ref{sec:many-body-cpa}) and therefore does not
appear for $S\rightarrow\infty$. In this section we shall therefore
concentrate on another local approximation, the many-body dynamical CPA
\cite{EdGrKu99,GrEd99} in which quantum spins can be treated analytically and
which agrees with DMFT in the classical limit. Like much current research on
the DE model (see section~\ref{sec:spin-waves-de}) the method has its roots in
work done on the Hubbard model in the sixties, in this case by Hubbard
himself \cite{Hu64}. Since such work is no longer common knowledge we start
with an introduction to CPA in the Hubbard model.

\subsection{CPA for the Hubbard model}\label{sec:cpa-hubbard-model}

To introduce the many-body CPA we consider the Hubbard model, which is a
simpler model for strongly correlated electrons than the DE model with
quantum spins. The Hamiltonian for this model is
\begin{equation}\label{eq:h_hubbard}
  H_{\rm H}=\sum_{ij\si}t_{ij}c^\dag_{i\si}c_{j\si}+U\sum_i n_{i\up}
  n_{i\Nup}
\end{equation}
and Hubbard \cite{Hu64} set out to find an approximation to the one-electron
retarded Green function $G_{\bi{k}\si}(\en)$ which is exact in the atomic
limit $t_{ij}=0$. He used the equation of motion method and the idea of
the alloy analogy described below. It turns out that Hubbard's approach,
without the minor ``resonance broadening correction'', is equivalent to the
CPA which was developed later \cite{ElKrLe74}. The CPA derivation of
Hubbard's result is much simpler than the original equation of motion
method. However we had to resort to an extension of the original method to
derive the many-body CPA for the DE model, with and without phonons, in the
general form needed to discuss magnetic properties. In this paper we restrict
the derivation to the paramagnetic state in zero magnetic field, although we
summarize some more general results.

The alloy analogy consists in considering the $\up$-spin electrons, say, to
move in the potential due to static $\Nup$-spin electrons, frozen in a random
configuration which must be averaged over. Thus a one-electron Hamiltonian
for $\up$-spin is obtained from (\ref{eq:h_hubbard}) by taking the last term
to be $U\sum_i n_{i\up} \las n_{i\Nup}\ras$ where $\las n_{i\Nup}\ras$ takes
the value $1$ with probability $n_{\Nup}$ and 0 with probability $1-n_\Nup$.
Here $n_\si$ is the number of $\si$-spin electrons per atom. It is important
to note that the alloy analogy is quite distinct from the Hartree-Fock
approximation in which $\las n_{i\si}\ras=n_\si$ for all $i$. In the alloy
analogy a $\si$-spin electron moves in a random potential given by $U$ on
$n_{\bar{\si}}$ sites and $0$ on $1-n_{\bar{\si}}$ sites. In the CPA the
random potential is replaced by a uniform, but energy-dependent, effective
potential $\Si_\si(\en)$ for an "effective medium". This effective potential,
in general complex, is called the coherent potential and is in fact the
electron self-energy. The procedure for determining $\Si_\si$ is to insist on
a zero average t-matrix for scattering by a central atom, with potential $U$
or $0$, set in the effective medium. Equivalently the average of the
site-diagonal element $G_\si(\en)$ of the Green function, for each type of
central atom, is put equal to the site-diagonal element of the Green function
for the effective medium. Thus
\begin{equation}\label{eq:cpa1}
  G_\si=n_{\bar{\si}}\frac{G_\si}{1-\left(U-\Si_\si\right)G_\si}+
  \left(1-n_{\bar{\si}}\right)\frac{G_\si}{1+\Si_\si G_\si}
\end{equation}
and
\begin{equation}\label{eq:cpa2}
  G_\si(\en)=\frac{1}{N}\sum_\bi{k}\frac{1}{\en-\en_\bi{k}-\Si_\si(\en)}=
  G_0\left(\en-\Si_\si(\en)\right)
\end{equation}
where the bare band Green function is given by
\begin{equation}\label{eq:cpa3}
  G_0(\en)=\int\,\rmd\en'\frac{N_0(\en')}{\en-\en'}\,.
\end{equation}
Here $\en_\bi{k}=\sum_{j}t_{ij}\exp[\rmi\bi{k}\cdot(\bi{R}_i-\bi{R}_j)]$ is
the band energy, where $\bi{R}_i$ is the position of site $i$, $N_0(\en)$ is
the corresponding density of states per atom and $N$ is the number of lattice
sites. Equations (\ref{eq:cpa1}) and (\ref{eq:cpa2}) are to be solved
self-consistently for $\Si_\si(\en)$ and hence for the local Green function
$G_\si(\en)$. Equation~(\ref{eq:cpa1}) may be written as
\begin{equation}\label{eq:cpa4}
  G_\si=\frac{n_{\bar{\si}}}{\Si_\si+G_\si^{-1}-U}+\frac{1-n_{\bar{\si}}}
  {\Si_\si+G_\si^{-1}}\,.
\end{equation}
This may be compared with the exact Green function for the atomic limit
($t_{ij}=0$) which is given by \cite{Hu64}
\begin{equation}\label{eq:cpa_AL}
  G_\si^{\rm AL}(\en)=\frac{n_{\bar{\si}}}{\en-U}+\frac{1-n_{\bar{\si}}}{\en}\,,
\end{equation}
where in this retarded Green function $\en$ has a small positive imaginary
part as usual. Hence
\begin{equation}\label{eq:cpa5}
  G_\si(\en)=G_\si^{\rm AL}\left(\Si_\si+G_\si^{-1}\right)\,.
\end{equation}
Clearly this CPA equation is exact in the atomic limit, when
$N_0(\en)=\delta(\en)$ and it follows from equation~(\ref{eq:cpa2}) that
$\Si_\si+G_\si^{-1}=\en$. Solution of the CPA equation becomes simple if the
density of states $N_0(\en)$ is taken to be of the elliptic form
\begin{equation}\label{eq:ell_dos}
  N_0(\en)=\frac{2}{\pi W^2}\left(W^2-\en^2\right)^{1/2}
\end{equation}
where $2W$ is the bandwidth. Then from equation~(\ref{eq:cpa3})
\begin{equation}\label{eq:G_0}
  G_0(\en)=\frac{2}{W^2}\left[\en-\left(\en^2-W^2\right)^{1/2}\right]\,.
\end{equation}
Introducing this expression for $G_\si$ in equation~(\ref{eq:cpa2}), and
solving for $\en-\Si_\si(\en)$, we find
\begin{equation}\label{eq:cpa6}
  \Si_\si+G_\si^{-1}=\en-W^2 G_\si/4\,.
\end{equation}
Hence equations (\ref{eq:cpa_AL}) and (\ref{eq:cpa5}) give an algebraic
equation for $G_\si$.

Solving this type of equation for $G_\si$, Hubbard \cite{Hu64} calculated the
density of states $N_\si(\en)=-\pi^{-1}\Im G_\si(\en)$, considering
particularly the paramagnetic state $n_\up=n_\Nup=n/2$ and concentrating on
the half-filled band case $n=1$. He showed that for $U/W$ greater than a
critical value, equal to $1$ in the present approximation, a gap opens in
$N(\en)$ at the Fermi level so that the system becomes an insulator as
envisaged by Mott. For $U\gg W$ the density of states consists of two peaks
centred on $\en=0$ and $\en=U$, these being broadened versions of the
$\delta$-functions at these energies in the atomic limit. Furthermore, for
general band-filling $n$, the spectral weights in the two peaks are the same
as in the atomic limit.

The CPA for the Hubbard model has some serious defects. There are no
self-consistent solutions with magnetic order. Furthermore in the
paramagnetic metallic state, for $n<1$ or for $n=1$ with $U/W$ less than the
critical value, the system is never a Fermi liquid. There is never a sharp
Fermi surface at $T=0$ with a Migdal discontinuity in the Bloch state
occupation number, as pointed out by Edwards and Hewson \cite{EdHe68}. This
is due to the absence of states with infinite lifetime at the Fermi level,
since within the alloy analogy all states are scattered by disorder. A
modification of the CPA to remedy this defect, retaining the analytic
simplicity of the method, had some limited success
\cite{EdHe90,LuEd95,PoHeNo98}. However the most satisfactory approach is DMFT
which involves numerical solution of an associated self-consistent impurity
problem \cite{GeKoKrRo96,BuHePr98}. DMFT may be regarded as the best local
approximation, in which the self-energy is a function of energy only, and is
exact in infinite dimension.

The many-body CPA is considerably more satisfactory for the DE model than it
is for the Hubbard model, as discussed in the next section. There is one
limit, the case of classical spins ($S=\infty$), in which the CPA is
identical to DMFT. This is because classical spins are static and an alloy
analogy of frozen disordered spins is completely justified. DMFT for the DE
model has only been implemented fully for classical spins \cite{Fu94,Fu96}
and the many-body CPA discussed in the next section provides an approximate
analytic extension of DMFT to quantum spins. The system orders
ferromagnetically below a Curie temperature $\Tc$, as it should, and the
disordered spin state above $\Tc$ should be well described. However, the
accuracy of the ground state at $T=0$ for finite $S$ is unclear. The
saturated state with all itinerant and local spins completely aligned, which
is the ground state for $S=\infty$ (we are always considering large $J$ in
the DE model), is never a self-consistent CPA solution for finite $S$
\cite{EdGrKu99}. This is due to low-lying $\Nup$ spin states of the type
discussed in section~\ref{sec:spin-waves-de}. In the treatment described
there involving magnon emission the local approximation corresponds to
neglecting the magnon energy, in which case $\Nup$ spin quasi-particle states
always appear below the Fermi level and the saturated state is unstable. In
their approximate treatments of DMFT Meyer \etal \cite{MeSaNo01} also find
$\Nup$ spin states below the Fermi level for any finite $S$. Possibly this is
a general result of the local approximation.  The true parameter range of
stability of the saturated ground state is unknown. It has been shown
rigorously that for $J=\infty$ it is unstable for $S=1/2$ and $0.12<n<0.45$,
with a simple cubic nearest-neighbour tight-binding band \cite{BrEd98}. Wurth
and M\"uller-Hartmann \cite{WuMu98} find no instability with respect to a
single spin-flip for any $n$ when $J=\infty$ and $S=3/2$. If the true ground
state is not saturated it seems unlikely to be a uniform (spatially
homogeneous) ferromagnet, with partially ordered local and itinerant spins,
as in the uniform CPA ground state for finite $S$. Such a state would
probably not be a Fermi liquid, just as in CPA, unless the electrons making
up the spin $S$ became partially delocalised with spectral weight at the
Fermi level. Numerical DMFT results for quantum spins are urgently required
to throw light on the matter.

\subsection{The many-body CPA for the DE model}\label{sec:many-body-cpa}

Edwards \etal \cite{EdGrKu99} developed the many-body CPA for the DE
model using an extension of Hubbard's equation~of motion method.
Hubbard's ``scattering correction'' becomes more complicated owing to the
form of the interaction term in the DE model whereby electrons can flip their
spin via exchange of angular momentum with the local spins. This dynamical
effect couples the equations for $G_\up$ and $G_\Nup$ and was first treated
by Kubo \cite{Ku74} in a one-electron dynamical CPA. The main feature of the
many-body CPA is that we recover Kubo's one-electron CPA as $n\rightarrow 0$
and the correct atomic limit for general band-filling $n$ as
$t_{ij}\rightarrow 0$. A second paper \cite{GrEd99} showed the
equivalence to DMFT in the limit $S\rightarrow\infty$, $J\rightarrow\infty$.
The full equation of motion derivation of the many-body CPA is required
to obtain general results in the presence of a magnetic field and/or magnetic
order \cite{GrEd99}. However it turns out that in the zero-field paramagnetic
state we can deduce the CPA equation from the atomic limit Green function
$G_\si^{\rm AL}$ and equation~(\ref{eq:cpa5}), just as in the Hubbard model.
We shall therefore not repeat the full derivation in this paper although we
shall discuss results on magnetic properties in
section \ref{sec:magnetism-de-model}.

The atomic limit Green function, $G_\up^{\rm AL}$, say, is easily obtained by
the equation of motion method using the Hamiltonian (\ref{eq:h_de}) with
$t_{ij}=0$ \cite{EdGrKu99}. The result for zero field ($h=0$) is
\begin{equation}\label{eq:de_AL_general}
  \begin{split}
    G_\up^{\rm AL}(\en)=\frac{1}{2S+1}\left[\frac{\la\left(S+S^z\right)
    n_\Nup-S^-\si^+\ra}{\en+J(S+1)/2}+\frac{\la\left(S-S^z\right)
    \left(1-n_\Nup\right)-S^-\si^+\ra}{\en-J(S+1)/2}\right.\\
    \left.{}+\frac{\la\left(S+1+S^z\right)\left(1-n_\Nup\right)+S^-\si^+\ra}
    {\en+JS/2}+\frac{\la\left(S+1-S^z\right)n_\Nup+S^-\si^+\ra}{\en-JS/2}
  \right]
  \end{split}  
\end{equation}
and for $h\ne 0$ one merely has to replace $\en$ by $\en+h/2$. The angle
brackets $\las\dots\ras$ represent thermal averages and all operators within
them correspond to the same site $i$, this suffix thus being omitted. This
expression, with four poles, is considerably more complicated than the
two-pole Hubbard model expression of equation~(\ref{eq:cpa_AL}). The poles at
$\en=\pm JS/2$, $\pm J(S+1)/2$ correspond to energies to add or remove an
electron from the atom, that is to transitions between singly-occupied states
and either unoccupied or doubly-occupied states. The singly-occupied states
have total spin $S+\frac{1}{2}$ or $S-\frac{1}{2}$ with energies $-JS/2$ and
$J(S+1)/2$ respectively; the unoccupied and doubly-occupied states have zero
energy. In a state of complete spin alignment, with all local and itinerant
spins $\up$, $G^{\rm AL}_\up$ has a single pole at $\en=-JS/2$ and $G^{\rm
  AL}_\Nup$ has three poles at $\en=\pm JS/2$ and $J(S+1)/2$. The weight in
the low-lying $\Nup$ spin level at $-JS/2$ is $(1-n)/(2S+1)$.

\begin{figure}[htbp]
  \centering
  \subfigure[]{\label{fig:DOS_DE_model_a}
  \begin{minipage}[b]{0.45\textwidth}
    \centering \includegraphics[width=\textwidth]{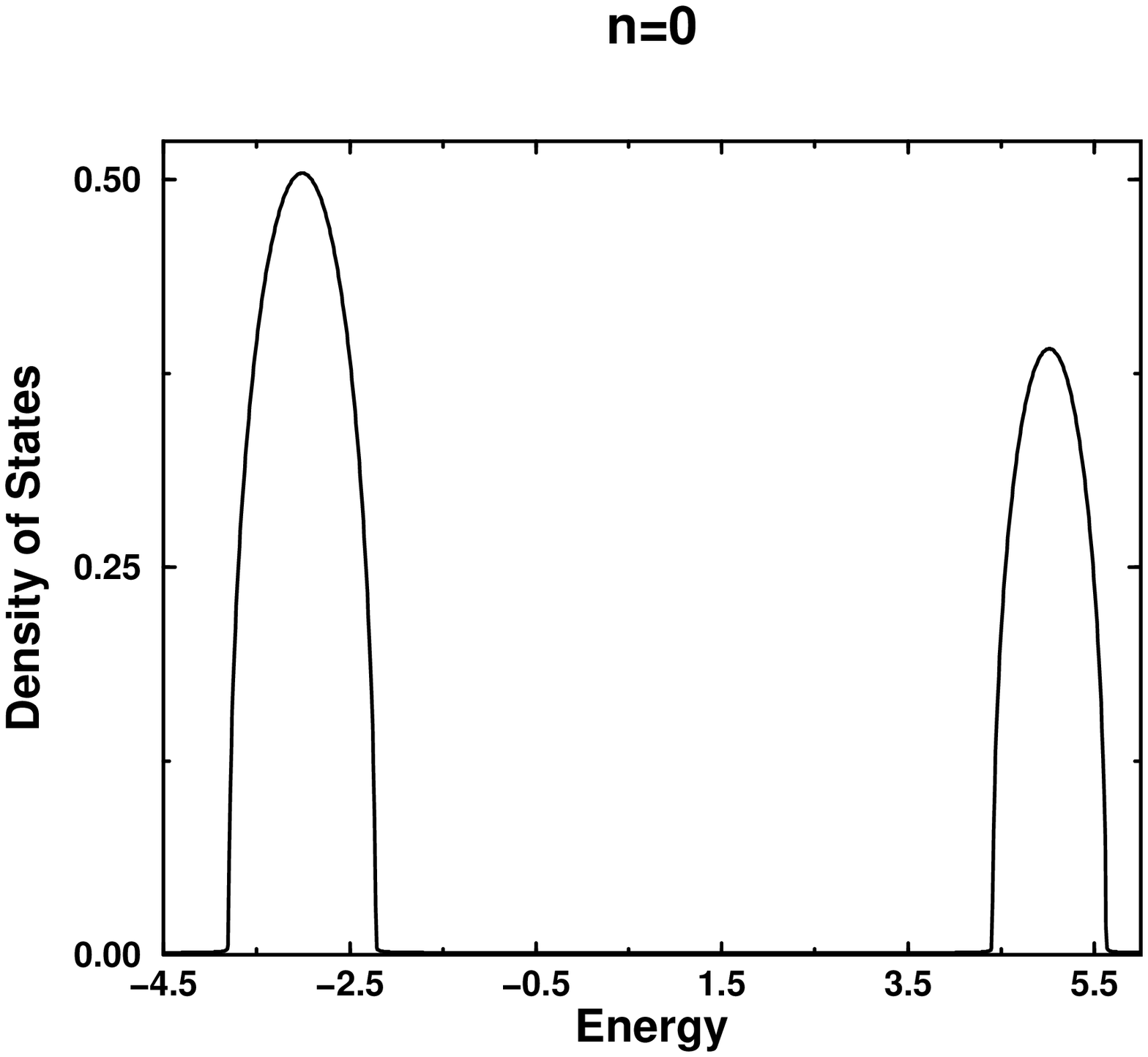}
  \end{minipage}}%
\subfigure[]{\label{fig:DOS_DE_model_b}
  \begin{minipage}[b]{0.45\textwidth}
    \centering \includegraphics[width=\textwidth]{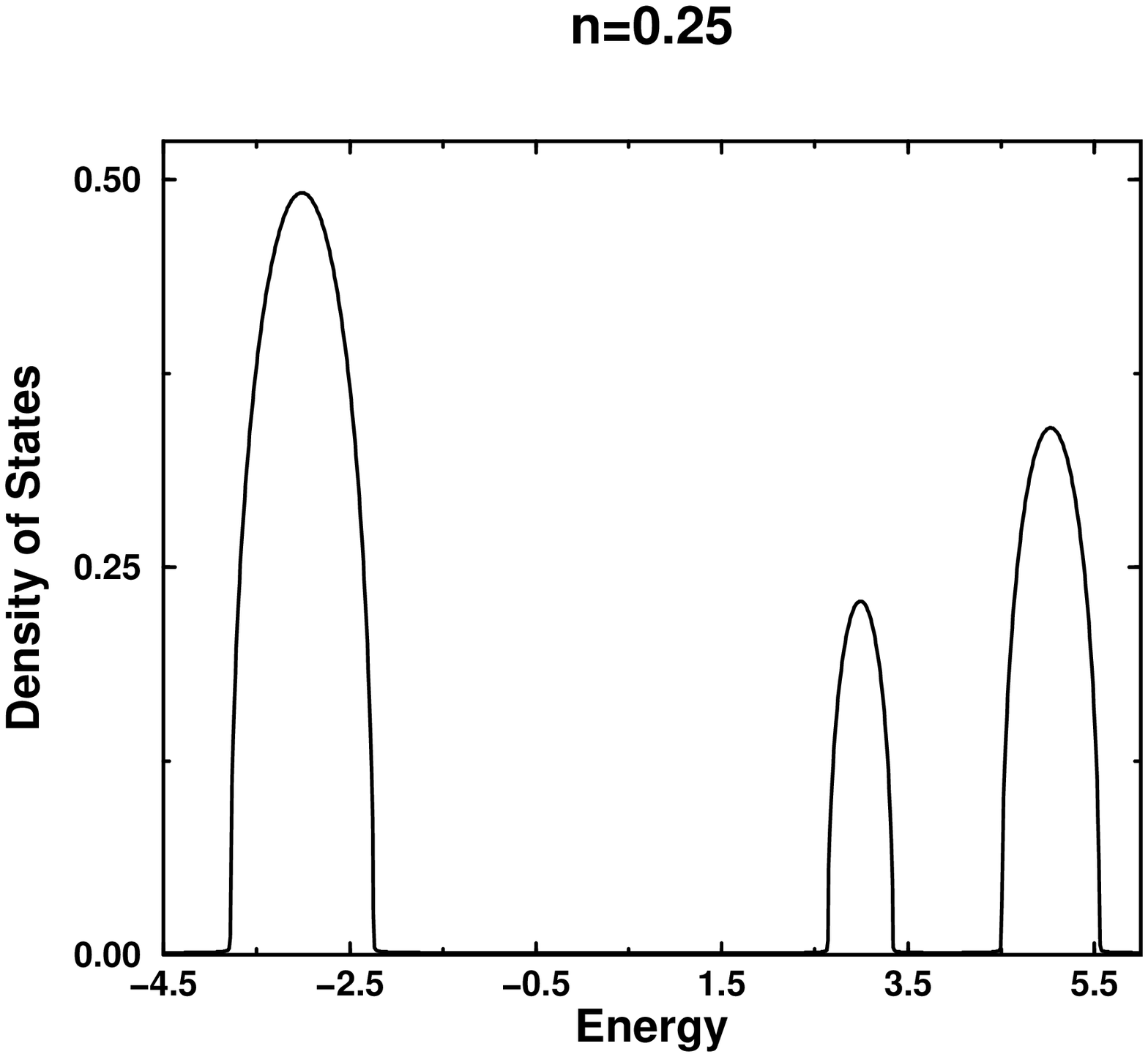}
  \end{minipage}}
\subfigure[]{\label{fig:DOS_DE_model_c}
  \begin{minipage}[b]{0.45\textwidth}
    \centering \includegraphics[width=\textwidth]{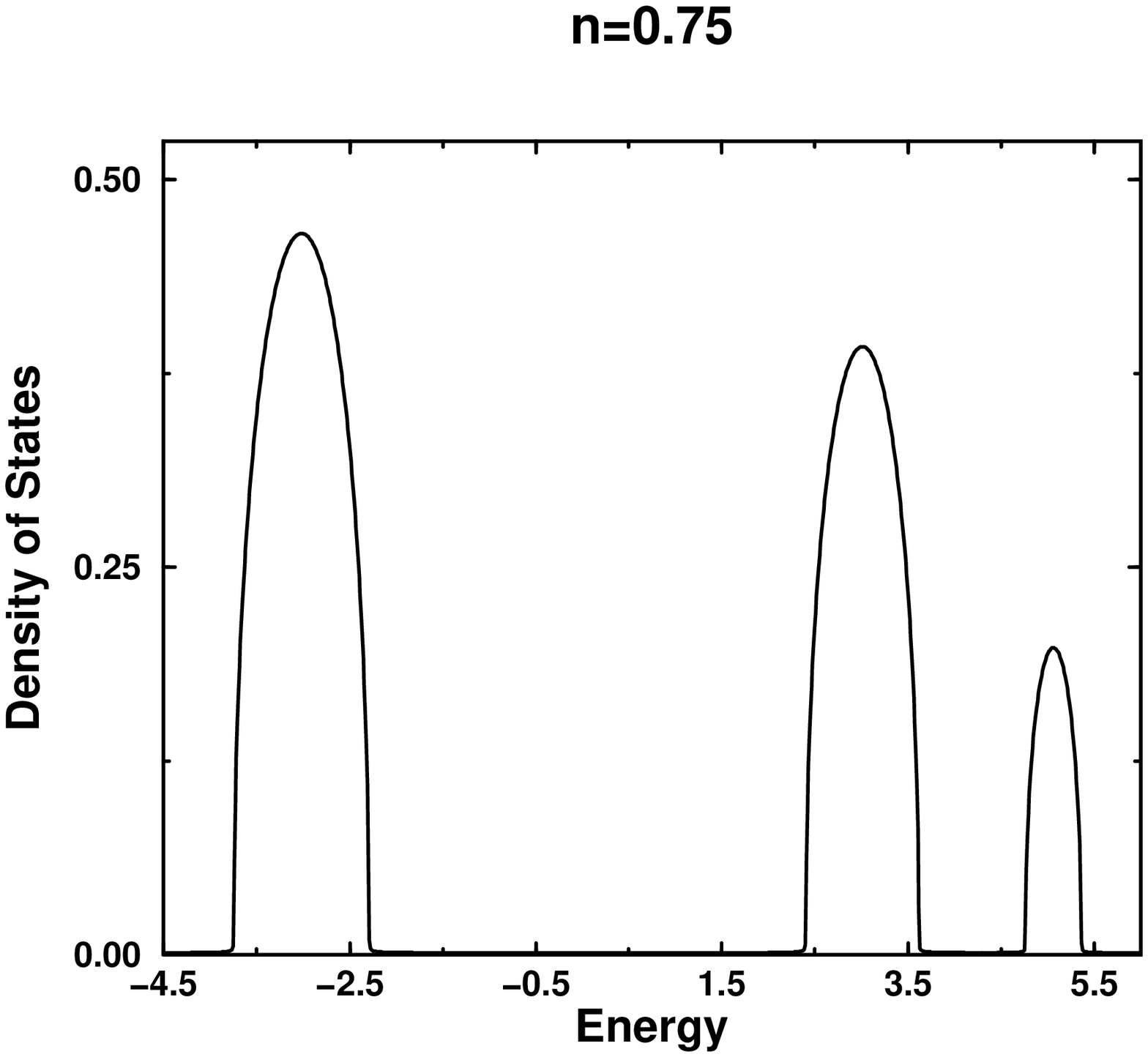}
  \end{minipage}}%
\subfigure[]{\label{fig:DOS_DE_model_d}
  \begin{minipage}[b]{0.45\textwidth}
    \centering \includegraphics[width=\textwidth]{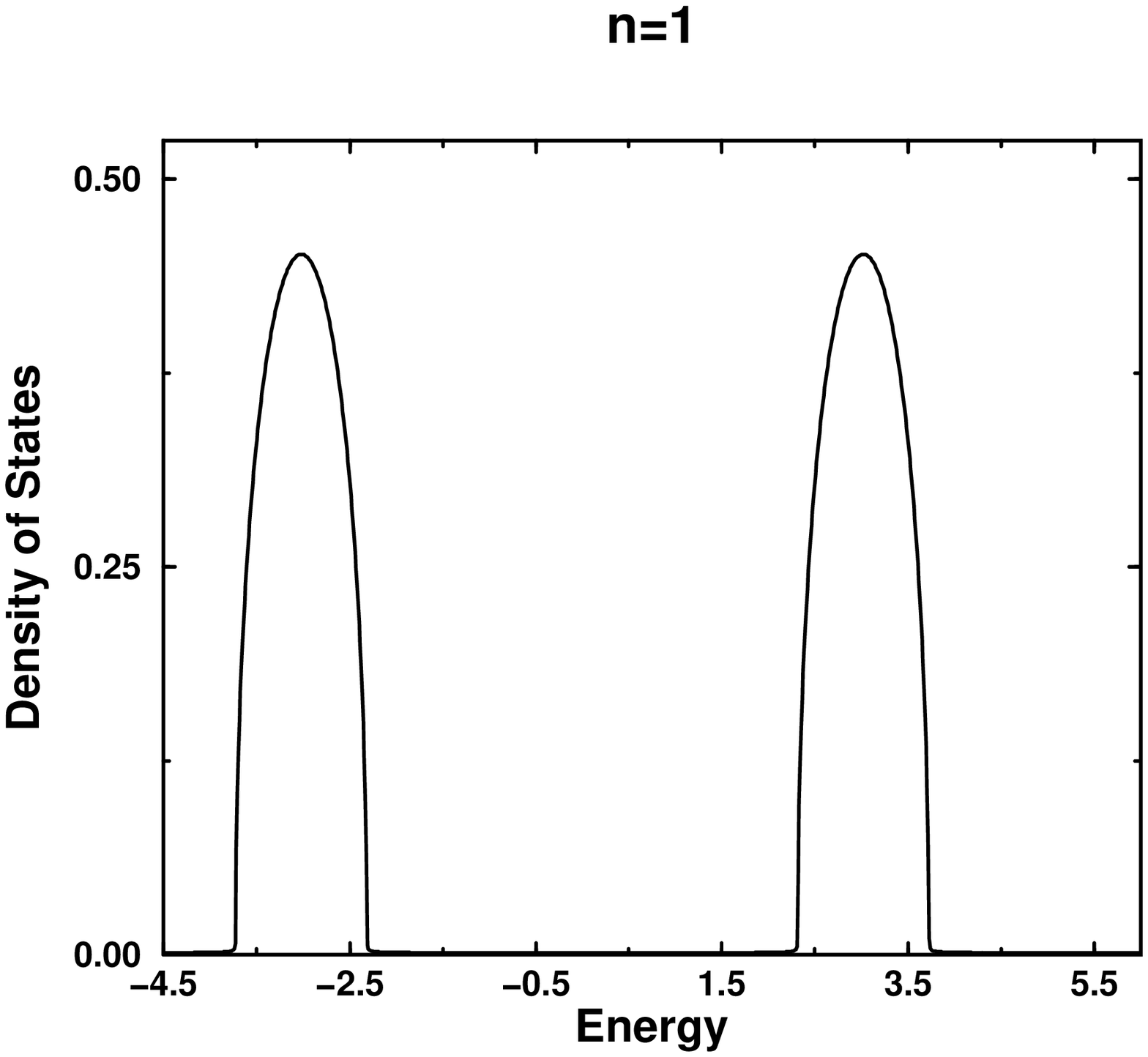}
  \end{minipage}}
  \caption{\label{fig:DOS_DE_model}%
    The density of states in the paramagnetic state of the double-exchange
    model for $S=3/2$, $J=4W$, and $n=0,\,0.25,\,0.75$ and 1. The Fermi level
    is always within the lowest band until this is just filled for $n=1$.
    Energy units of $W$ are used. (from reference \cite{EdGrKu99})}
\end{figure}

In the zero-field paramagnetic case it turns out that the CPA equation for
$G(\en)$ with the redundant suffix $\si$ omitted, is given by
equation~(\ref{eq:cpa5}) as in the Hubbard model. Thus, taking the band to
have the elliptic form (\ref{eq:ell_dos}), the CPA equation for $G$ is
\begin{equation}\label{eq:de_CPA}
  G(\en)=G^{\rm AL}\left(\en-W^2 G/4\right)
\end{equation}
with $G^{\rm AL}$ given by
\begin{equation}\label{eq:de_AL}
  \begin{split}
    G_\up^{\rm AL}(\en)=\frac{1}{2S+1}\left[\frac{nS/2-\la\bi{S}\cdot\bsigma\ra}
    {\en+J(S+1)/2}+\frac{S(1-n/2)-\la\bi{S}\cdot\bsigma\ra}{\en-J(S+1)/2}\right.\\
    \left.{}+\frac{(S+1)(1-n/2)+\la\bi{S}\cdot\bsigma\ra}{\en+JS/2}
    +\frac{n(S+1)/2+\la\bi{S}\cdot\bsigma\ra}{\en-JS/2}\right]\,.
  \end{split}
\end{equation}
The spin symmetry of the paramagnetic state has been used to simplify the
expectations in the previous form of $G^{\rm AL}$,
equation~(\ref{eq:de_AL_general}).  It is easy to show that
$\las\bi{S}\cdot\bsigma\ras\rightarrow nS/2$ as $J\rightarrow\infty$ and
$\las\bi{S}\cdot\bsigma\ras$ will be very near this limit as long as $JS\gtrsim
2W$. We make this approximation in calculating $G(\en)$, and hence the
density of states, $N(\en)=-\pi^{-1}\Im G(\en)$ from
equations~(\ref{eq:de_CPA}) and (\ref{eq:de_AL}). The results are shown in
figure~\ref{fig:DOS_DE_model} for $S=3/2$ and $J=4W$ for various $n$.
Clearly, from equation~(\ref{eq:de_AL}), the approximation to
$\las\bi{S}\cdot\bsigma\ras$ has the effect of removing the weak band centred
on $\en=-J(S+1)/2$ but it does not affect the total weight or the
distribution of weight between the two lower and two upper bands. It may be
seen that as $n$ increases from $0$ the band near $\en=J(S+1)/2$ is reduced
in weight and a new band appears near $JS/2$, until at $n=1$ no weight
remains in the band near $J(S+1)/2$. The weight in the band near $-JS/2$ is
$(S+1-n/2)/(2S+1)$ per spin so if $JS$ is sufficiently large to separate the
bands ($JS\gtrsim 2W$) this band will just be filled at $n=1$ producing a Mott
insulator as expected. This redistribution of weight between bands as they
fill with electrons is characteristic of the many-body CPA and
was missing from Kubo's one-electron CPA \cite{Ku74} which was restricted to $n=0$.

In the strong-coupling limit $J\rightarrow\infty$, which is taken with a
shift of energy origin $\en\rightarrow\en-JS/2$, equation~(\ref{eq:de_AL})
simplifies to
\begin{equation}\label{eq:de_AL_Jinf}
  G^{\rm AL}(\en)=\en^{-1}(S+1-n/2)/(2S+1)\,.
\end{equation}
Equation~(\ref{eq:de_CPA}) then becomes a quadratic equation for $G$ with
solution
\begin{equation}\label{eq:de_gf}
  G(\en)=\alpha^2\frac{2}{D^2}\left[\en-\sqrt{\en^2-D^2}\right]
\end{equation}
where $\alpha^2=(S+1-n/2)/(2S+1)$ and $D=\alpha W$. By comparing with
equations~(\ref{eq:ell_dos}) and (\ref{eq:G_0}) we see that the density of
states is a single elliptical band of weight $\alpha^2$ and bandwidth
$2\alpha W$. As $S\rightarrow\infty$ the band-narrowing factor
$\alpha\rightarrow 1/\sqrt{2}=0.707$, which is close to the classical result
of $2/3$, obtained by averaging $\cos\left(\theta/2\right)$ over the solid
angle.

In the classical spin limit $S\rightarrow\infty$ we rescale $J$, replacing it
by $J/S$, and the CPA equation becomes
\begin{equation}\label{eq:de_Sinf}
  G=\frac{1}{2}\left[\frac{1}{\Si+G^{-1}+J/2}+\frac{1}{\Si+G^{-1}-J/2}\right]\,.
\end{equation}
Here we have used the general equation~(\ref{eq:cpa5}), valid for arbitrary
band-shape, rather than equation~(\ref{eq:de_CPA}).
Equation~(\ref{eq:de_Sinf}) is precisely the equation obtained by Furukawa
\cite{FuPom} within DMFT.

\subsection{Resistivity in the paramagnetic state of the DE model}
\label{sec:resist-param-state}

The Kubo formula for the conductivity $\si$ involves the two-particle
current-current response function. However in the local approximation of CPA
or DMFT there is no vertex correction \cite{Ve69,pruschke_review} and $\si$
may be expressed in terms of the one-particle spectral function
\begin{equation}\label{eq:spec_function}
  A_\bi{k}(\en)=-{\pi}^{-1}\Im G_\bi{k}(\en)=-{\pi}^{-1}\Im
  \left[\en-\en_\bi{k}-\Si(\en)\right]^{-1}\,.
\end{equation}
In the paramagnetic state $G$ is $T$-independent if we assume
$\las\bi{S}\cdot\bsigma\ras=nS/2$, and $\si$ depends on temperature only
weakly through the Fermi function. If we neglect this thermal smearing around
the Fermi energy $\mu$ we may calculate at $T=0$ but consider the results to
apply to the actual paramagnetic state at $T>\Tc$. We find \cite{EdGrKu99}
\begin{equation}\label{eq:si_1}
  \si=\frac{2\pi \rme^2}{3Na^3\hbar}\sum_\bi{k} \bi{v}_{k}^2
  \left[A_\bi{k}(\mu)\right]^2
\end{equation}
where $\bi{v}_\bi{k}=\nabla\en_\bi{k}$ is the electron velocity and $a^3$ is
the volume of the unit cell. Just as in section~\ref{sec:spin-waves-de} we can
introduce the function $M(E)$, defined by equation~(\ref{eq:def_M}), and write
\begin{equation}\label{eq:4.19}
\si=\frac{2\pi e^2}{3 a^3\hbar}\int\,\rmd E M(E) |A_E(\mu)|^2\,,
\end{equation}
where $A_E(\mu)$ is defined by the right-hand expression in
equation~(\ref{eq:spec_function}) with $\en=\mu$, $\ek=E$. For a simple cubic
tight-binding band $\ek=-2t\sum_\beta \cos k_\beta a$, with $\beta$ summed
over $x$, $y$, $z$, $\nabla^2\ek=-a^2\ek$. Then it is straight-forward to
show that
\begin{equation}\label{eq:4.20}
  M(E)=-a^2\int_{-\infty}^{E}E N_{\rm c}(E)\rmd E
\end{equation}
where $N_{\rm c}(E)$ is the density of states for the simple cubic band. If
$N_{\rm c}(E)$ is replaced by a suitably-scaled Gaussian $N_{\rm
  g}(E)=(3/\pi)^{1/2}W^{-1}\exp\left[-3(E/W)^2\right]$, corresponding to a
hypercubic lattice in infinite dimensions, $M(E)$ becomes $M_{\rm
  g}=(a^2W^2/6)N_{\rm g}(E)$. However if $N_{\rm c}(E)$ is replaced by the
elliptic density of states $N_0(E)$ it becomes
$M_0(E)=[a^2(W^2-E^2)/3]N_0(E)$. We refer to previous treatments of the
$M(E)$ factor in $\si$ when we discuss the optical conductivity in
section~\ref{sec:optical-conductivity}.

\begin{figure}[htbp]
  \centering \includegraphics[width=0.65\textwidth]{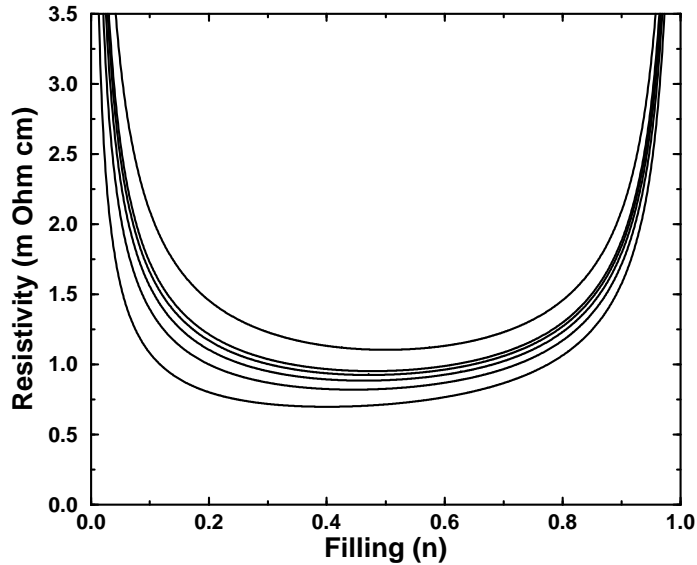}
    \caption{\label{fig:resis_elliptic}%
      The zero field paramagnetic state resistivity $\rho=\si^{-1}$ versus
      band-filling $n$ for the double-exchange model from reference
      \cite{EdGrKu99}. Here $J=\infty$, $a=5$ {\AA}, and
      $S=1/2,\,1,\,3/2,\,2,\,5/2$ and $\infty$, $\rho$
      increasing with $S$. The elliptical density of states and
      formula (\ref{eq:4.19}) with $M(E)\equiv M_0(E)$ are used (from reference
      \cite{EdGrKu99}).}
\end{figure}

\begin{figure}[htbp]
  \centering \includegraphics[width=0.65\textwidth]{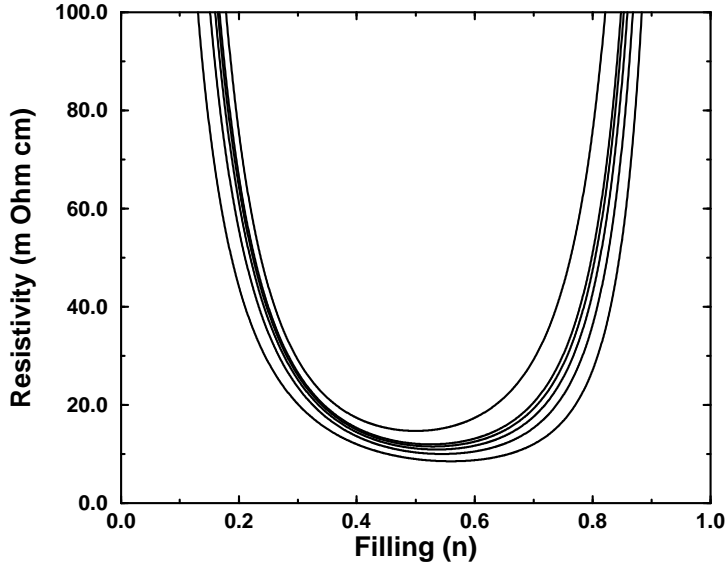}
    \caption{\label{fig:resis_lorentz}%
      As in figure~\ref{fig:resis_elliptic} but using a Lorentzian density of
      states and formula (\ref{eq:4.19}) with
      $M(E)=(W^2a^2/6\pi)(E^2+W^2)^{-1}$. (from reference \cite{EdGrKu99})}
\end{figure}

Since it is convenient to use $N_0(\en)$ to calculate the Green function and
self-energy, as in the previous section, it is reasonable to evaluate $\si$
using equation~(\ref{eq:4.19}) with $M(E)=M_0(E)$. Edwards \etal
\cite{EdGrKu99} took the strong-coupling limit $J\rightarrow\infty$ for
simplicity and the results for the resistivity $\rho=\si^{-1}$ are plotted
against band-filling $n$ for various $S$ in figure~\ref{fig:resis_elliptic}.
The lattice constant was taken as $a=5$ {\AA} which is comparable with the
Mn--Mn spacing in manganites. For $J=\infty$ the band-width $W$ is the only
energy-scale and, since the integral in equation~(\ref{eq:4.19}) is
dimensionless when a factor $a^2$ is taken out, $\rho$ does not depend on
$W$. In fact in the DE regime $JS\gtrsim 2W$ the resistivity is almost
independent of both $J$ and $W$. It is seen in
figure~\ref{fig:resis_elliptic} that $\rho$ diverges correctly at $n=0$,
owing to the absence of carriers, and at $n=1$ where the system becomes a
Mott insulator. Well away from these insulating limits $\rho$ does not depend
strongly on $S$, so that quantum spin effects are not very important.
Furthermore $\rho\approx1$ \res over a wide range of band-filling, which is
much smaller than observed in some manganites above $\Tc$, for example in
LCMO (see figure~\ref{fig:fig5}). This agrees with the conclusion of Millis
\etal \cite{MiLiSh95} that the DE model, with electrons scattered purely by
disordered local spins, cannot describe the physics of the manganites
completely. Early work by Furukawa \cite{Fu94} seemed to point to another
conclusion, although the DMFT is equivalent to the CPA approach. Edwards
\etal \cite{EdGrKu99} showed that the confusion arose from Furukawa's use of
a convenient, but rather unreasonable, Lorentzian density of states. This
makes the calculation of $A_E(\mu)$ very simple and Furukawa effectively took
$M(E)$ in equation~(\ref{eq:4.19}) to be Lorentzian also. In later work
Furukawa \cite{FuPom} used the elliptic band. Results of such calculations
for the limit $J\rightarrow\infty$ are shown in
figure~\ref{fig:resis_lorentz} and it is remarkable that $\rho$ is at least
an order of magnitude larger than one finds in
figure~\ref{fig:resis_elliptic} for the more reasonable elliptic band.

\subsection{Magnetism in the DE model}\label{sec:magnetism-de-model}

As discussed at the beginning of section~\ref{sec:many-body-cpa}, the full
equation of motion approach to many-body CPA is required to determine
magnetic properties such as spin susceptibility $\chi$ and Curie temperature
$\Tc$. In reference \cite{EdGrKu99} this involved a hierarchy of Green
functions satisfying $4S+1$ coupled algebraic equations for local spin $S$;
only the $S=1/2$ case was briefly discussed. In reference \cite{GrEd99}
a major simplification was achieved by introducing generating Green functions
which generate all the required Green functions by differentiation with
respect to a parameter. The coupled equations are then replaced by a single
first-order linear differential equation, the parameter being the independent
variable, whose analytic solution yields the required CPA equations for the
Green functions. The classical limit $S=\infty$ can then be taken and for
$J=\infty$ the equations coincide with those of DMFT, which are only
obtainable in the classical limit. The many-body CPA is therefore an analytic
approximation to DMFT for arbitrary quantum spin $S$ which becomes exact for
$S=\infty$. The many-body CPA also coincides with Kubo's \cite{Ku74}
one-electron CPA in the limit $n\rightarrow 0$ where that is valid.

\begin{figure}[htbp]
  \centering
  \subfigure[]{\label{fig:Tc_n_elliptic}
  \begin{minipage}[b]{0.45\textwidth}
    \centering \includegraphics[width=\textwidth]{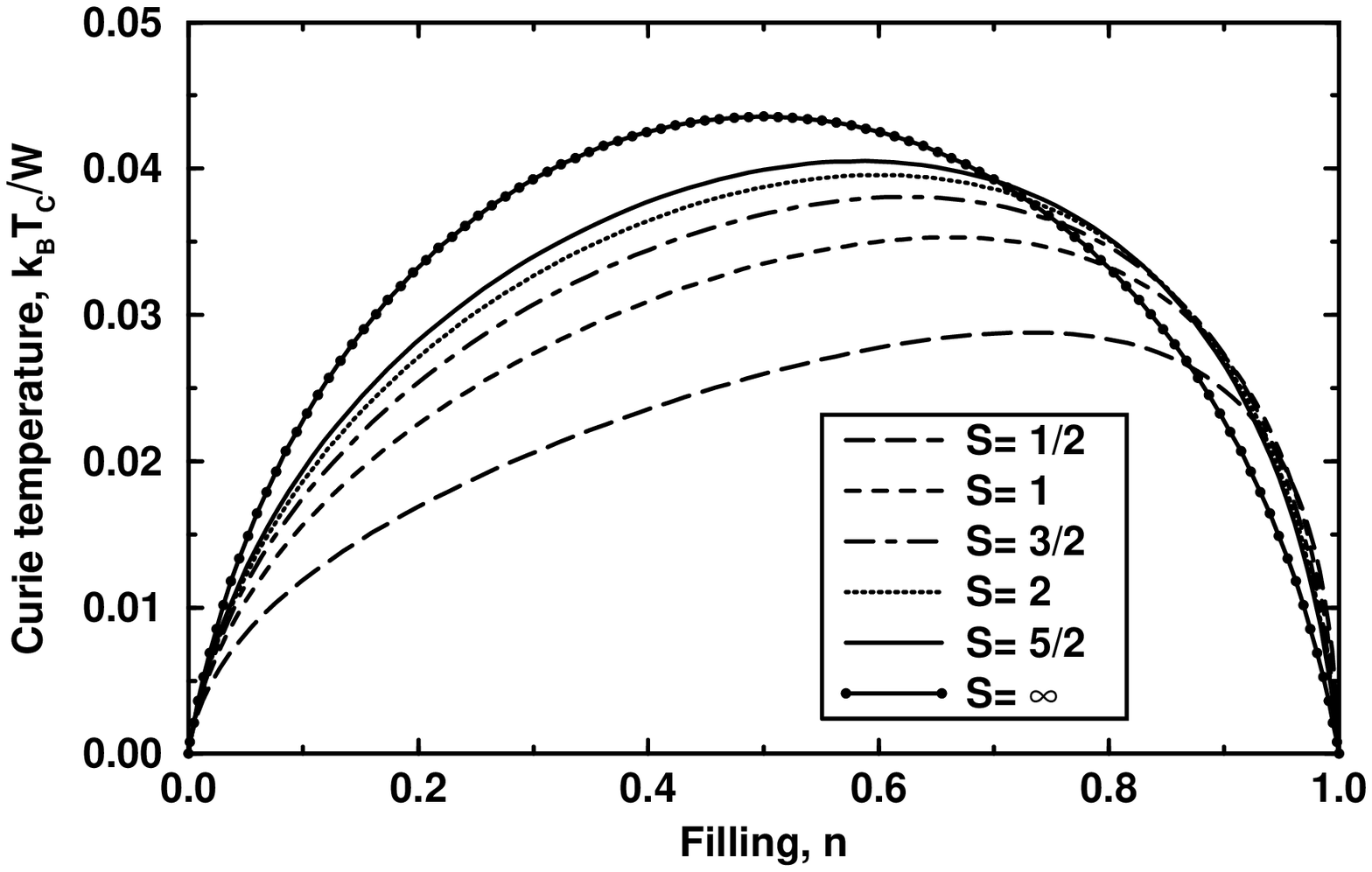}
  \end{minipage}}%
\subfigure[]{\label{fig:Tc_n_ell_cubic}
  \begin{minipage}[b]{0.45\textwidth}
    \centering \includegraphics[width=\textwidth]{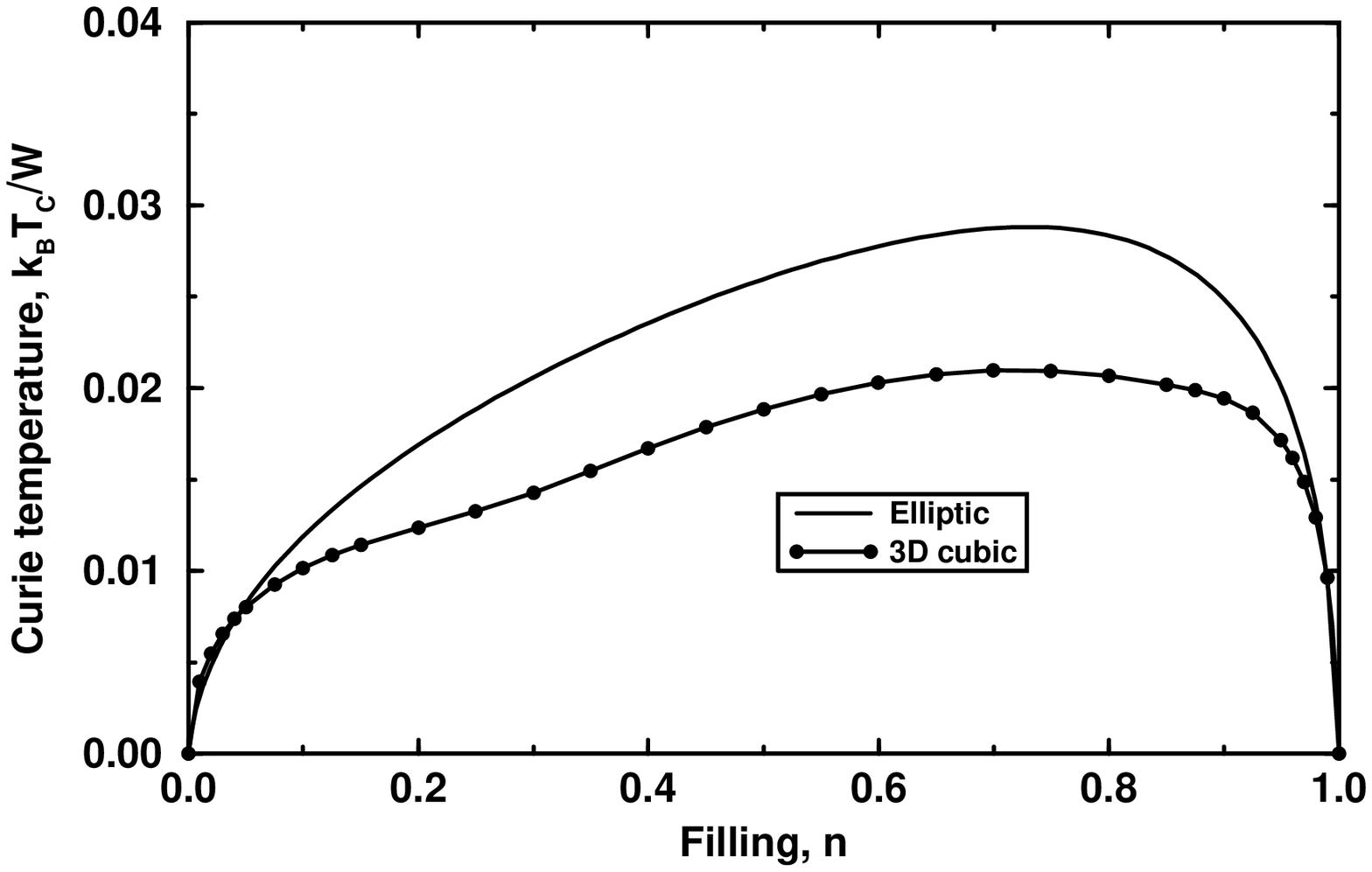}
  \end{minipage}}
  \caption{\label{fig:Tc_n}%
    The Curie temperature $k_{\rm B}\Tc/W$ of the double-exchange model
    versus band-filling $n$ for various $S$, calculated with $J=\infty$ using
    the elliptic band (a); the effect on $\Tc$ for $S=1/2$ of changing the
    elliptic band to the density of states for a simple cubic tight-binding
    band with nearest neighbour hopping (b). (from reference \cite{GrEd99})}
\end{figure}

To determine the magnetic properties one problem remains; the CPA and DMFT
equations contain one set of correlation functions $\las(S^z)^m\ras$ which
cannot be obtained directly from the Green functions. There is an indirect
procedure for determining these correlation functions within CPA but it
proves to be unsatisfactory, never yielding ferromagnetic solutions.
However, for $S=\infty$, DMFT provides a way to calculate the probability
distribution function $P(S^z)$, and hence $\las(S^z)^m\ras$, and Green and
Edwards \cite{GrEd99} used an empirical extension of this for finite $S$.
This extension guarantees that the spin susceptibility exhibits the correct
Curie laws for band occupations $n=0$ and $n=1$. Thus for $n=0$ we have a
Curie law over the whole temperature range, corresponding to $N$ independent
spins $S$. For $n=1$, with $J=\infty$, we have independent spins
$S+\frac{1}{2}$. For $0<n<1$ we find a finite Curie temperature $\Tc$ and
some results \cite{GrEd99} are shown in figure~\ref{fig:Tc_n}. In
figure~\ref{fig:Tc_n_elliptic} $\Tc$ is plotted as a function of $n$ for
various $S$ with $J=\infty$, using the elliptic band.  Clearly for finite $S$
ferromagnetism is more stable for $n>0.5$ than for $n<0.5$, in agreement with
the findings of Brunton and Edwards \cite{BrEd98}. For $S=\infty$ the result
agrees closely with that of Furukawa \cite{FuPom}. In
figure~\ref{fig:Tc_n_ell_cubic} we see the effect on $\Tc$ for $S=1/2$ of
changing the bare band-shape from elliptic to simple cubic tight-binding. A
dip in $\Tc$ occurs around $n=0.3$ which is the region where the ground state
of the simple cubic DE model with $S=1/2$ is rigorously not one of complete
spin alignment \cite{BrEd98}.


\section{The metal-insulator transition}\label{sec:metal-insul-trans}

In section~\ref{sec:introduction} we pointed out the wide range of behaviour
in manganites. LSMO with $x\approx0.3$ has a metal-poor metal transition at
$\Tc$, whereas in LCMO the resistivity $\rho(T)$ decreases with rising
temperature above $\Tc$ (see figure~\ref{fig:fig5}). Moreover the resistivity
near $\Tc$ is an order of magnitude larger in LCMO than in LSMO. This
behaviour of LCMO, and many other manganites, is characterized as a
metal-insulator (MI) transition. Furukawa has pointed out that the
contrasting behaviour of LSMO is quite well described by the DE model
. Furukawa's $\rho(T)$ curve (figure~9 of reference \cite{FuPom}) is similar
to that of figure~\ref{fig:rho_T_g0.1} in this paper, which corresponds to
weak electron-phonon coupling \ie essentially to the pure DE model. A central
problem in the theory of the manganites is to explain the origin of the MI
transition in systems like LCMO and to explain why it does not occur in LSMO.
Clearly the pure DE model is insufficient, as first emphasized by Millis
\etal \cite{MiLiSh95}. However a common element of many theories of the MI
transition is the narrowing of the band in the paramagnetic state above $\Tc$
due to the DE mechanism. This narrowing of the band encourages any tendency
to localize the electrons and produce insulating behaviour. However theories
differ greatly as to what actually drives the localization. Millis \etal
\cite{MiLiSh95,MiMuSh96II} and R\"oder \etal \cite{RoZaBi96} propose that the
driving force is strong electron-phonon coupling associated with the dynamic
Jahn-Teller effect. On the other hand Varma \cite{Va96} and M\"uller-Hartmann
and Dagotto \cite{MuDa96} proposed that localization could occur due to a
combination of magnetic disorder and non-magnetic A-site disorder. Nagaev's
theory \cite{Na96,Naga96} is also based on A-site disorder while Furukawa
\cite{FuPom} argues in favour of phase separation models of LCMO with, for
example, ferromagnetic and charge-ordered regions. We begin with a discussion
of models based on disorder.

\subsection{The role of disorder}\label{sec:role-disorder}

Sheng \etal \cite{ShXiShTi97} pursued a line similar to Varma's
\cite{Va96}. They studied localization due to both non-magnetic randomness
and the off-diagonal disorder associated with the classical spin DE model. As
mentioned in section~\ref{sec:double-exchange-de}, the effective hopping
integral is $t_{ij}\cos(\theta_{ij}/2)$ where the angle $\theta_{ij}$ between
neighbouring spins varies randomly in the paramagnetic state. Sheng \etal
showed using scaling theory that this off-diagonal disorder alone can only
localize a small fraction of electron states close to the band edges but fails
to cause localization of the electron states at the Fermi level for the
ferromagnetic regime with $x=0.2$--0.5. Li \etal \cite{LiZaBiSo97} came to a
similar conclusion using a transfer-matrix method. This is an important
result because local approximations such as DMFT and CPA are not able to
detect localization. Both groups further showed that to obtain a MI transition
between the ferromagnetic and paramagnetic state, by means of the mobility
edge moving through the Fermi level on introducing the off-diagonal disorder
of the paramagnetic state, one needs large diagonal disorder with random site
energies in the range $(-W,W)$. Here $2W$ is the bandwidth in the
ferromagnetic state and $W\sim1$ eV typically. The random energies at Mn
sites in \chem{La_{1-x}Ca_xMnO_3}, say, arise from varying local environments
of \chem{La^{3+}} and \chem{Ca^{2+}} ions. Neglecting metallic screening, the
change in potential at a Mn site due to  changing the charge on one of the
neighbouring A sites by $e$ is about 0.4 eV, assuming a local dielectric
constant of 10. Pickett and Singh \cite{PiSi97} deduce from a band calculation
on ordered \chem{La_{2/3}Ca_{1/3}MnO_3} that in the disordered case Mn site
energies have a Gaussian-like distribution with full width at half maximum
equal to 0.6 eV. This is much less than $2W$ so the condition for a MI
transition is not fulfilled. Smolyaninova \etal \cite{SmXiZhRa00} searched
for scaling behaviour of their resistivity measurements on
\chem{La_{0.67}Ca_{0.33}MnO_3} and \chem{Nd_{0.7}Sr_{0.3}MnO_3} thin films;
they concluded their results were incompatible with an Anderson localization
transition.

Nagaev's \cite{Na96,Naga96} theory is rather different. He assumes the charge
carriers are holes in the O p band so that DE does not apply. His idea
relies on the fact that a charge disturbance in a metallic ferromagnet
produces a magnetic response extending over a magnetic correlation length
$\xi$. In a state of complete spin alignment at $T=0$, $\xi$ is essentially
the lattice spacing but $\xi$ diverges at $\Tc$. Nagaev therefore proposes
that A-site disorder leads to extensive static variation in the exchange
potential which could localize carriers, particularly for $T\approx\Tc$. An
applied magnetic field reduces $\xi$ and will lower the resistance as
required for the CMR effect. Unfortunately, no quantitative work seems to
have been done on the model which makes it difficult to asses.

Diagonal disorder in the DE model with classical or Ising local spins has
also been treated using CPA \cite{AuKo00} or DMFT \cite{ZhDoWa98,LeFr01}.
Within these local approximations localization does not occur and the only
possibility of a MI transition is for a gap to open in the density of states
at the Fermi level. A random site potential is considered, within a one-band
model, which takes the value $\Delta/2$ on a fraction $x$ of the sites, and
$-\Delta/2$ on the others. The idea is that for a suitable $\Delta$ a gap
will open as the band narrows in the paramagnetic state. These models are
faced with several problems. A more Gaussian-like distribution of site
energies, as appropriate on Mn sites due to A-site disorder, would not
produce a gap and, even with the binary distribution, one requires
$\Delta\approx W$, the half band-width, which is unrealistically large.
Another difficulty is the position of the gap relative to the Fermi level.
The band which should be split by disorder is the lowest one in the density
of states shown in figure~\ref{fig:DOS_DE_model}. In the case of classical or
Ising local spins the total weight in this band is one state per site
(including both spins) and, when split by disorder, the weight below the gap
is $1-x$. The gap is therefore positioned at the Fermi level as required for
$n=1-x$. However for quantum spins $S$ it is clear from
section~\ref{sec:many-body-cpa} that the corresponding weight is
$(1-x)(2S+2-n)/(2S+1)$; thus the electron density $n$ would have to be
assigned arbitrarily a strange dependence on $x$, $n=5(1-x)/(5-x)$ for
$S=3/2$, to place the Fermi level in the gap.

The conclusion of this section is that disorder alone cannot account for the
MI transition in manganites with $x\approx0.3$. The effect of A-site disorder
is sufficiently weak for a virtual crystal approximation to be a reasonable
starting-point for considering the effects of electron-phonon coupling. Of
course if this coupling is so strong that narrow polaron bands appear then
disorder will become important \cite{LiZaBiSo97}. We discuss this in
section~\ref{sec:polarons-bipolarons}.

\subsection{Electron-phonon coupling}\label{sec:electr-phon-coupl}

For sufficiently strong coupling to local phonon modes an electron produces a
strong distortion of the lattice in its immediate neighbourhood. This
distortion moves with the electron and the whole structure is called a small
polaron. At low temperatures the small polaron occupies a narrow band of
coherent Bloch-like states which narrows even further as the temperature
rises. At high temperatures, typically larger than that corresponding to
half the phonon energy, the polaron moves diffusively with an activation
energy (see \eg \cite{Ma90}). A condition for small polarons to exist is
$E_{\rm p}>W z^{-1/2}$ \cite{Eagl66}, where $E_{\rm p}$ is the polaron
binding energy, $W$ is the half-width of the bare electron band and $z$ is
the coordination number of the lattice. The idea proposed by Millis \etal
\cite{MiLiSh95} is that as the band narrows on passing into the paramagnetic
state, by the DE effect, this criterion just becomes satisfied. Thus in the
low-temperature ferromagnetic state, where the criterion is not satisfied,
the system is metallic and electron-phonon interaction merely produces a
small enhancement of the quasi-particle mass; the quasi-particles in this
Fermi-liquid are sometimes called large polarons. As $\Tc$ is approached
small polarons are formed and their motion is governed by an activation
energy related to $E_{\rm p}$. This picture of the MI transition was
developed in different ways by Millis \etal \cite{MiMuSh96II} and R\"oder
\etal \cite{RoZaBi96}. Millis \etal used DMFT and treated the phonons
classically, whereas R\"oder \etal used the quantum-mechanical approach of
small-polaron theory. This approach is discussed further in
section~\ref{sec:polarons-bipolarons}.

The electrons may couple to different local phonon modes, which may have the
$Q_1$, $Q_2$, $Q_3$ symmetries shown in figure~\ref{fig:fig1}. The breathing
mode $Q_1$ is the simplest but the $Q_2$ JT mode is usually regarded as the
most important and was emphasized by Millis \etal \cite{MiLiSh95}. To
describe the JT coupling properly one should consider a two-band model with
the correct \chem{e_g} symmetry. This was done by Millis \etal
\cite{MiMuSh96II} and Zang \etal \cite{ZaBiRo96}. R\"oder \etal
\cite{RoZaBi96} and Zang \etal \cite{ZaBiRo96} treat DE in a simple
mean-field way, by means of a band-narrowing, but avoid double occupation of
sites by effectively using spinless electrons. Millis \etal \cite{MiMuSh96II}
treat the Hund's rule coupling $J$ but in a two-band model large $J$ no
longer produces a Mott insulator for $n=1$, as it does in the one-band model.
It is therefore necessary to introduce on-site Coulomb interaction
\cite{BeZe99,HeVo00}. Millis \etal \cite{MiMuSh96II} do not do this so that
for $n=1$, which should correspond to the undoped insulator, they have a
metal. Also, for weak electron-phonon interaction, they find $\Tc$ is largest
for this value of $n$. This contrasts strongly with figure~\ref{fig:Tc_n}
where, in the one-band DE model, $\Tc=0$ at $n=0$ and 1 and has a maximum in
between. Held and Vollhardt \cite{HeVo00}, using DMFT, obtain very similar
results to those of figure~\ref{fig:Tc_n} for $S=\infty$ in a two-band model
with strong on-site Coulomb interaction included. However the values for
$\Tc$ are about twice those in figure~\ref{fig:Tc_n}. Nevertheless, if
on-site Coulomb interaction is neglected for simplicity, it is sensible to
use a one-band model to describe the manganites. In fact Green \cite{Gr01}
developed the theory of such a model (the Holstein-DE model) in which
electrons couple to quantum spins and quantum phonons. He used the many-body
CPA method \cite{EdGrKu99,GrEd99} which was discussed in
section~\ref{sec:magn-transp-de}. In the limit of classical spins and phonons
this method is equivalent to the DMFT of Millis \etal \cite{MiMuSh96II}.
Green's work therefore bridges the gap between this classical phonon
treatment and the polaron theories. We therefore describe it in some detail
in the next section. Comparison between theory and experiment indicates that
electron-phonon coupling in the manganites is just in the intermediate regime
where small-polarons are on the verge of forming. At present the many-body
CPA is the only theoretical method which can deal satisfactory with this
cross-over regime and in section~\ref{sec:theory-experiment} we show how it
can be used to tie together experimental data using many different
techniques.


\section{The many-body CPA for the Holstein-DE model}
\label{sec:many-body-cpa-hde}

This section is based on Green's \cite{Gr01} recent study of the Holstein-DE
model in which the electrons of the DE model couple to local phonons as in
the Holstein treatment of small polarons \cite{Ho59a,Ho59b}. The Hamiltonian
is
\begin{equation}\label{eq:h_hde}
  \begin{split}
    H = &\sum_{ij\si} t_{ij}c_{i\si}^{\dagger}c_{j\si}-J\sum_i
    \bi{S}_i\cdot\bsigma_i-h\sum_i\left(S_i^z+\si_i^z\right)\\
    & -g\sum_i n_i\left(b_i^{\dagger}+b_i\right)+\om\sum_i
    b_i^{\dagger}b_i\,.
  \end{split}
\end{equation}
The first three terms constitute the DE Hamiltonian of
equation~(\ref{eq:h_de}) while the first, fourth and fifth terms form the
Holstein model. Einstein phonons on site $i$, with energy $\om$ and creation
operator $b_i^\dag$, couple to the electron occupation number $n_i=\sum_\si
n_{i\si}$ with coupling strength $g$. The electron-phonon coupling is of the
form $-g'\sum_i n_i x_i$, where $x_i$ is the displacement of a shell of atoms
surrounding site $i$, and in application to the manganites it may be regarded
as an effective Jahn-Teller coupling. Previous studies of this model have
either concentrated on coherent polaron bands, like R\"oder \etal
\cite{RoZaBi96}, or have treated the phonons classically \cite{MiMuSh96II} so
that there are no polaron bands at all. The many-body CPA approach is able to
encompass both aspects and to describe the crossover from quantum polarons to
the classical picture as temperature and/or model parameters are varied. The
relationship to previous theoretical work and to experimental studies of the
manganites is discussed fully in section~\ref{sec:theory-experiment}. However
we mention briefly below some related work on the pure Holstein model,
without coupling to local spins.

Sumi considered the Holstein model with one electron in the band, first
treating the phonons classically \cite{Su72} and later quantum mechanically
\cite{Su74}.  The classical case, with frozen displacements $x_i$,
corresponds to a multicomponent alloy for which CPA is the best local
approximation. In his dynamical CPA treatment of quantum phonons, Sumi
\cite{Su74} treated the one-site dynamics correctly and his work is
completely equivalent to the more recent DMFT treatment of Ciuchi \etal
\cite{CiPaFrFe97}. As a general rule dynamical CPA and DMFT are the same for
one-electron problems. DMFT is the correct extension of CPA to the many-body
problem of finite electron density but for the Holstein model, as for the DE
model, it cannot be carried through analytically in the quantum case.
Numerical work \cite{FrJaSc93} applying DMFT to the Holstein model
has been aimed mostly at understanding superconducting transition
temperatures and charge-density-wave instabilities rather than the polaron
physics with which we are mainly concerned. An unfortunate feature of the
Holstein model for spin $1/2$ electrons is that in a quantum treatment the
true ground state for strong electron-phonon coupling consists of unphysical
singlet bipolarons with two electrons bound on the same site. This problem
does not occur in the one-band Holstein-DE model since strong coupling $J$ to
local spins prevents double occupation of sites, as pointed out earlier. It
is also bypassed if the phonons are treated classically, as in the work of
Millis \etal \cite{MiMuShI} on the Holstein model. The Holstein model is more
complicated than the DE model and it turns out that Green's many-body CPA no
longer reduces to the correct one-electron dynamical CPA/DMFT
\cite{Su74,CiPaFrFe97} as band-filling $n\rightarrow 0$.  Although correct in
the atomic limit $t_{ij}=0$, the theory is clearly cruder for the Holstein
and Holstein-DE model than for the pure DE model.

We start by deriving the Green function for the Holstein-DE model in the
atomic limit. The Hamiltonian $H_{\rm AL}$ in this limit is given by
equation~(\ref{eq:h_hde}) with the first term omitted and with site indices
and summation suppressed.  We remove the electron-phonon coupling by the
standard canonical transformation \cite{Ma90} $\tilde{H}=\rme^s H_{\rm
  AL}\rme^{-s}$ where $s=-(g/\om)n(b^\dag-b)$. Under this transformation
$b\rightarrow b+(g/\om)n$ and the Hamiltonian separates into a fermionic and
bosonic component:
\begin{equation}\label{eq:hde_1}
  \tilde{H}=H_{\rm f}+H_{\rm b}
\end{equation}
\begin{equation}\label{eq:hde_2}
  H_{\rm f}=-J\bi{S}\cdot\bsigma-h\left(S^z+\si^z\right)-
  \left(g^2/\om\right)n^2\,,\,H_{\rm b}=\om b^\dag b\,.
\end{equation}
The transformation corresponds to a displacement of the equilibrium position
of the phonon harmonic oscillator in the presence of an electron and the
downward energy shift $g^2/\om$ is a polaron binding energy which we write as
$\lambda\om$, where $\lambda=g^2/\om^2$. If two electrons occupy the site
($n=2$), which will not occur for large $J$, the energy shift becomes
$4g^2/\om^2$ corresponding to an on-site bipolaron. Writing out explicitly
the thermal average in the definition of the one-particle retarded Green
function we have
\begin{eqnarray}\label{eq:hde_3}
  G_\si^{\rm AL}(t)&=&-\rmi\theta(t)\la\left[c_\si(t),c_\si^\dag\right]_+
  \ra\nonumber\\
  &=&-\rmi\theta(t)\frac{\Tr\left\{\rme^{-\beta H_{\rm AL}}\left[c_\si(t),
  c_\si^\dag\right]_+\right\}}{\Tr\left\{\rme^{-\beta H_{\rm AL}}\right\}}
\end{eqnarray}
and the canonical transformation introduced above can be carried out within
the traces, using the property of cyclic invariance. Thus $H_{\rm
  AL}\rightarrow\tilde{H}$, $c_\si^\dag\rightarrow X^\dag c_\si^\dag$ and
$c_\si(t)$ becomes
\begin{equation}\label{eq:hde_4}
  \rme^{\rmi\tilde{H}t}Xc_\si \rme^{-\rmi\tilde{H}t}
\end{equation}
where $X=\exp\left[g(b^\dag-b)/\om\right]$. Using equation~(\ref{eq:hde_1}),
we can write the traces in equation~(\ref{eq:hde_3}) as products of fermionic
and bosonic traces. Hence we find
\begin{equation}\label{eq:hde_5}
  G_\si^{\rm AL}(t)=-\rmi\theta(t)\left\{\la c_\si(t)c_\si^\dag\ra_{\rm f}F(t)+
    \la c_\si^\dag c_\si(t)\ra_{\rm f} F^*(t)\right\}
\end {equation}
where $F(t)=\las X(t)X^\dag\ras_{\rm b}$ and the thermal averages
$\las\ \ras_{\rm f}$, $\las\ \ras_{\rm b}$ correspond to the systems with
Hamiltonians $H_{\rm f}$ and $H_{\rm b}$ respectively. It may be shown \cite{Ma90}
that
\begin{equation}\label{eq:hde_6}
  F(t)=\rme^{-\lambda(2b+1)}\exp\left\{2\lambda\left[b(b+1)\right]^{1/2}\cos
    \left[\om(t+i\beta/2)\right]\right\}
\end{equation}
where $b=b(\om)=\left(\rme^{\beta\om}-1\right)^{-1}$ is the
Bose function with $\beta=\left(k_{\rm B}T\right)^{-1}$. The last factor is
of the form $\exp\left(z\cos\phi\right)$
which generates the
modified Bessel functions ${\rm I}_r(z)$:
\begin{equation}\label{eq:hde_7}
  \exp\left(z\cos\phi\right)=\sum_{r=-\infty}^{\infty}
  {\rm I}_r(z)\rme^{\rmi r\phi}\,.
\end{equation}
\begin{figure}[htbp]
  \centering \includegraphics[width=0.65\textwidth]{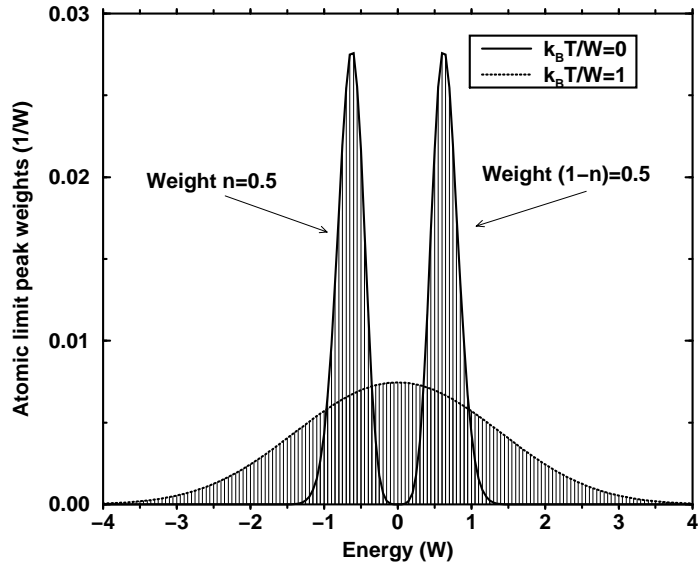}
    \caption{\label{fig:hde_AL}%
      One-electron spectra of the Holstein-DE model in the atomic limit at
      zero and very high temperature. They consist of delta-functions, with
      energy spacing $\om$, whose strength is indicated by the envelope
      curves. The plots are for the paramagnetic state with $S=J=\infty$,
      $h=0$, $n=0.5$, $\om/W=0.05$ and $g/W=0.18$, where $W$ is a unit of
      energy later to be identified with the half-width of the electron band
      in the full Hamiltonian \cite{Gr01}.}
\end{figure}
To evaluate the fermionic averages we
consider for simplicity the limit $J\rightarrow\infty$ in zero field
($h=0$). Then the last term in $H_{\rm f}$ may
be written $-(g^2/\om)n$, since $n=0$ or 1 only, and this may be
absorbed into the chemical
potential which is finally determined to give the correct number of electrons
$n$ per atom. Thus $H_{\rm f}$ is just the DE Hamiltonian in the atomic limit
and the sum of
the two fermionic averages corresponds to the function $G^{\rm AL}(t)$ whose
Fourier
transform is given by equation~(\ref{eq:de_AL_Jinf}). It is easy to see that the
first and second
thermal averages in equation~(\ref{eq:hde_5}) take constant values
$(1-n)(S+1)/(2S+1)$ and
$n/2$ respectively. Hence, from
equations~(\ref{eq:hde_5})-(\ref{eq:hde_7}), we obtain the
Fourier transform of $G^{\rm AL}$, with $J\rightarrow\infty$ and $h=0$, in
the form 
\begin{equation}\label{eq:G_hde_AL}
  G^{\rm AL}(\en)=\sum_{r=-\infty}^{\infty}\frac{{\rm I}_r\{2\lambda
      \left[b(\om)(b(\om)+1)\right]^{1/2}\}}
  {(2S+1)\exp\{\lambda\left[2b(\om)+1\right]\}}
  \frac{(2S+1)\frac{n}{2}\rme^{r\beta\om/2}+(S+1)(1-n)\rme^{-r
      \beta\om/2}}{\en+r\om}\,.
\end{equation}
The density of states $-\pi^{-1}\Im\,G^{\rm AL}(\en)$ is shown in
figure~\ref{fig:hde_AL} for
the classical spin limit $S\rightarrow\infty$ at
quarter-filling $n=0.5$. It consists of delta-function peaks separated in
energy by $\om$
and the envelope curves show the weight distribution at low and high
temperature. $W$ is an energy unit which, when we go beyond the atomic limit,
will be the half-width of the itinerant electron band, as usual. The values
adopted for the parameters $\om/W$ and $g/W$ relate to the manganites, as
discussed in section~\ref{sec:theory-experiment}. The symmetry of the spectrum about
zero energy is due to the
choice of filling $n=0.5$; in general at $T=0$ the lower and upper `bands'
have weights $n$
and $1-n$ respectively. By counting weights it may be seen that for any $n$ the
chemical potential lies in the peak at $\en=0$, which has very small weight
$\rme^{-\lambda}/2$ per
spin. The shape of the envelope function at $T=0$, with two maxima and very small
values at the centre of the pseudogap between them, may be understood
physically as follows. The delta-function at $\en=r\om$ ($r\ge 0$)
corresponds to an excitation
from the ground state, with no electron and the undisplaced oscillator in its
ground state, to a state with one electron and the displaced oscillator in
its $r^{\rm th}$ excited state. The strength of the delta-function depends on
the square
of the overlap integral between the displaced and undisplaced oscillator wave
functions. Clearly this is very small for $r=0$ and goes through a maximum with
increasing $r$ as the normalized displaced wave function spreads out. At
$T=0$ it is
easily seen from equation~(\ref{eq:G_hde_AL}), using ${\rm I}_r(z)\sim
(z/2)^{|r|}/|r|!$ for small
$z$, that the weight of the
delta-functions at $\en=\pm r\om$ is proportional to $\lambda^r/r!$ . Hence
the maxima in the envelope curve occur at $\en\approx\pm\lambda\om$, which
is the polaron binding energy.

We now turn to the
Holstein-DE model with finite band-width. As for the DE model it is 
necessary to use the full equation~of motion method to derive the many-body 
CPA in the presence of a magnetic field and/or magnetic order \cite{Gr01}. In the 
present case it is very difficult to determine self-consistently all the 
expectation values which appear. Green therefore approximated them by their 
values in the atomic limit. It then turns out that in the zero field 
paramagnetic state, for $J=\infty$ and with the elliptic band, the CPA Green
function $G$ again satisfies equation~(\ref{eq:de_CPA}), with $G^{\rm AL}$ now given
by equation~(\ref{eq:G_hde_AL}).

\begin{figure}[htbp]
  \centering \includegraphics[width=0.65\textwidth]{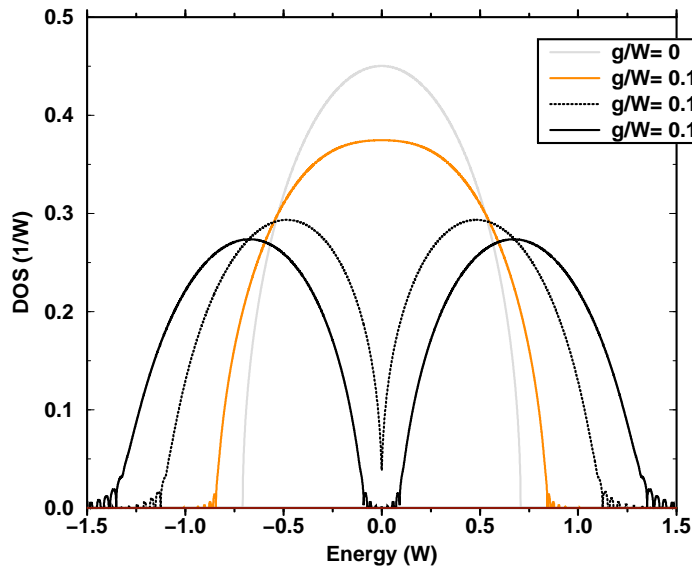}
    \caption{\label{fig:DOS_g}%
      The one-electron density of states (DOS) for the Holstein-DE model with
      half-bandwidth $W$, for the hypothetical paramagnetic state at $T=0$,
      with various strengths of electron-phonon coupling $g/W$.  Other
      parameters as in figure~\ref{fig:hde_AL} \cite{Gr01}.}
\end{figure}

\begin{figure}
  \centering \includegraphics[width=0.65\textwidth]{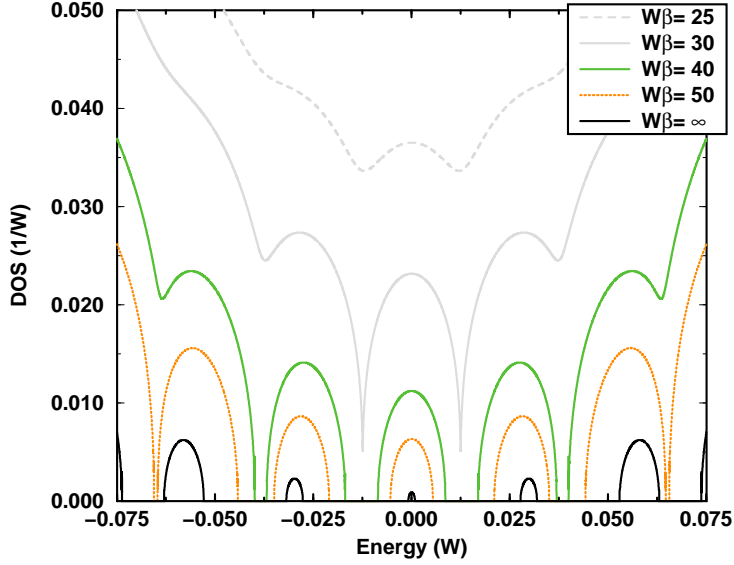}
    \caption{\label{fig:DOS_polarons}%
      Evolution with temperature $\beta=\left(k_{\rm B}T\right)^{-1}$ of the
      polaron subbands in the pseudogap around the chemical potential $\mu=0$
      for $g/W=0.18$. These subbands at $T=0$ can just be seen in
      figure~\ref{fig:DOS_g} \cite{Gr01}. All parameters as in figure~\ref{fig:hde_AL}.}
\end{figure}

The densities of states calculated for $T=0$ using equations~(\ref{eq:G_hde_AL})
and (\ref{eq:de_CPA}) with $S=\infty$, $n=0.5$, $\om/W=0.05$ and various
values of $g/W$ are
shown in figure~\ref{fig:DOS_g}. Apart from lacking the perfect symmetry about
the chemical
potential $\mu=0$ the results are qualitatively similar for other values of
$n$ not too close to 0 or 1. For $g=0$ we recover the elliptic band with
half-width $W/\sqrt{2}$ as for the DE model with $J=\infty$, $S=\infty$. As
$g$ increases the density of states broadens and small
subbands are split off from the band edges. As $g$ increases further a pseudogap
develops near the chemical potential. At a critical value $g=g_{\rm c}$ a gap
appears
which contains a small polaron band around the chemical potential.
Increasing $g$ further causes more bands to be formed in the gap, with weights
similar to those of the relevant atomic limit. It should be pointed out that
the paramagnetic state considered here at $T=0$ is not the actual ground state,
which is ferromagnetic. We discuss the magnetic state later. The effect of
increasing temperature on the density of states in the gap region is shown in
figure~\ref{fig:DOS_polarons} for $g=0.18W>g_{\rm c}$. With increasing $T$ the
polaron bands grow rapidly and eventually merge to fill the gap.

It is important to compare these results with the standard small
polaron theory developed by Holstein \cite{Ho59a,Ho59b}. Holstein distinguished
between `diagonal transitions', in which the number of phonons is unchanged
as the electron moves from site to site, and `nondiagonal transitions' in
which phonon occupation numbers change. The former give rise to a coherent
Bloch-like polaron band of half-width $W\rme^{-\lambda(2b+1)}$ which decreases
with increasing
temperature. The nondiagonal transitions are inelastic processes which
destroy phase coherence and the polaron moves by diffusive hopping. The
hopping probability increases with temperature so that polaron motion crosses
over from coherent Bloch-like at $T=0$ to diffusive hopping as $k_{\rm B}T$
approaches the
phonon energy $\om$. The paramagnetic state of the Holstein-DE model
differs from
this standard picture in one important respect. There are no well-defined
Bloch states, owing to strong scattering by the disordered local spins, so no
coherent polaron band will form. This is fortunate because the CPA treatment
of electron-phonon scattering will never lead to coherent states of infinite
lifetime at the Fermi surface at $T=0$. However in the presence of strong spin
disorder it should be satisfactory. We interpret the central band around the
chemical potential in figure~\ref{fig:DOS_polarons} as an incoherent polaron
band whose increasing
width as the temperature rises is due to life-time broadening of the atomic
level. The life-time decreases as the hopping probability increases with
rising temperature.

To substantiate this picture we study the central
polaron band in the limit of very strong electron-phonon coupling. In this
limit it can be shown that we need retain only the $r=0$ term in
equation~(\ref{eq:G_hde_AL}) and it is
then easy to solve equation~(\ref{eq:de_CPA}) for $G$. The result is of the same
form as equation~(\ref{eq:de_gf}) but with
\begin{equation}\label{eq:hde_8}
  D^2=\frac{1}{2}W^2\rme^{-\lambda(2b+1)}{\rm
  I}_0\left(2\lambda\left[b(b+1)\right]^{1/2}\right)
\end{equation}
and $\alpha^2=D^2/W^2$ in the case $S=\infty$.
The central band is thus elliptical with half-width $D$ and
weight $D^2/W^2$. It is now easy to calculate the conductivity $\si$ from
equation~(\ref{eq:si_1}) and, using $D^2\ll W^2$, we find
\begin{equation}\label{eq:hde_9}
  \si=\frac{\pi \rme^2}{6\hbar a}\frac{D^2}{W^2}\approx\frac{\pi \rme^2}{12\hbar a}
  \left(\frac{\beta\om}{4\pi\lambda}\right)^{1/2}\rme^{-\beta\lambda\om/4}\,.
\end{equation}
The last step follows by using the asymptotic forms for
strong coupling and high temperature ${\rm I}_0(z)\sim\left(2\pi
  z\right)^{-1/2}\exp{z}$ and $b\sim\left(\beta\om\right)^{-1}$. This form of
$\si$ is similar to that for small polaron hopping conduction in the
adiabatic limit ($W\gg\om$) \cite{EmHo69}
but with activation energy $\lambda\om/4$ equal to one quarter, instead of
one half, of the polaron binding energy. Nevertheless this establishes the link
between the many-body CPA and standard small polaron theory in the strong coupling
limit. However the results shown in figure~\ref{fig:DOS_polarons}, with
parameters relevant to
typical manganites, are far from this limit. They correspond to intermediate
coupling and in the actual paramagnetic state above the Curie temperature the
polaron bands are largely washed out. In this regime, with increasing
temperature, there is a crossover from polaronic behaviour to a situation
where the phonons behave classically, the case considered by Millis \etal
\cite{MiMuSh96II}. For electron-phonon coupling greater than a critical value these
authors find a gap in the density of states which gradually fills with
increasing temperature. However in their classical treatment there are no
polaron bands in the gap so that the link with standard polaron physics is
not established.

Apart from the symmetry about $\en=0$ the above results for $n=0.5$ are not
untypical of the general case. For general $n$ the main lower and upper bands,
separated by a gap for $g>g_{\rm c}$, have approximate weights $n$ and $1-n$
respectively. The chemical potential at $T=0$ is always confined to the
polaron band arising from the $r=0$ term of equation~(\ref{eq:G_hde_AL}), and
moves from the bottom at $n=0$ to the top at $n=1$, so that we correctly have
an insulator in these limits.

\begin{figure}
  \centering \includegraphics[width=0.65\textwidth]{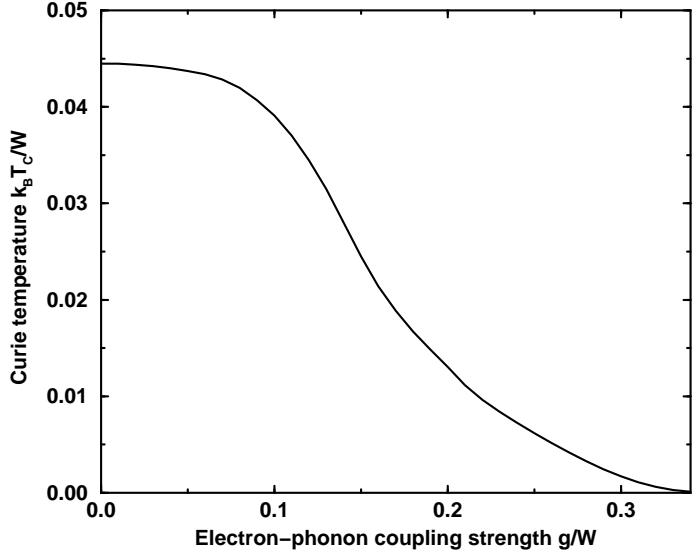}
    \caption{\label{fig:Tc_coupling}%
      Suppression of the Curie temperature of the Holstein-DE model with
      increasing electron-phonon coupling $g/W$. The plot is for
      $S=J=\infty$, $h=0$, $n=0.5$ and $\om/W=0.05$ \cite{Gr01}.}
\end{figure}

\begin{figure}
  \centering \includegraphics[width=0.65\textwidth]{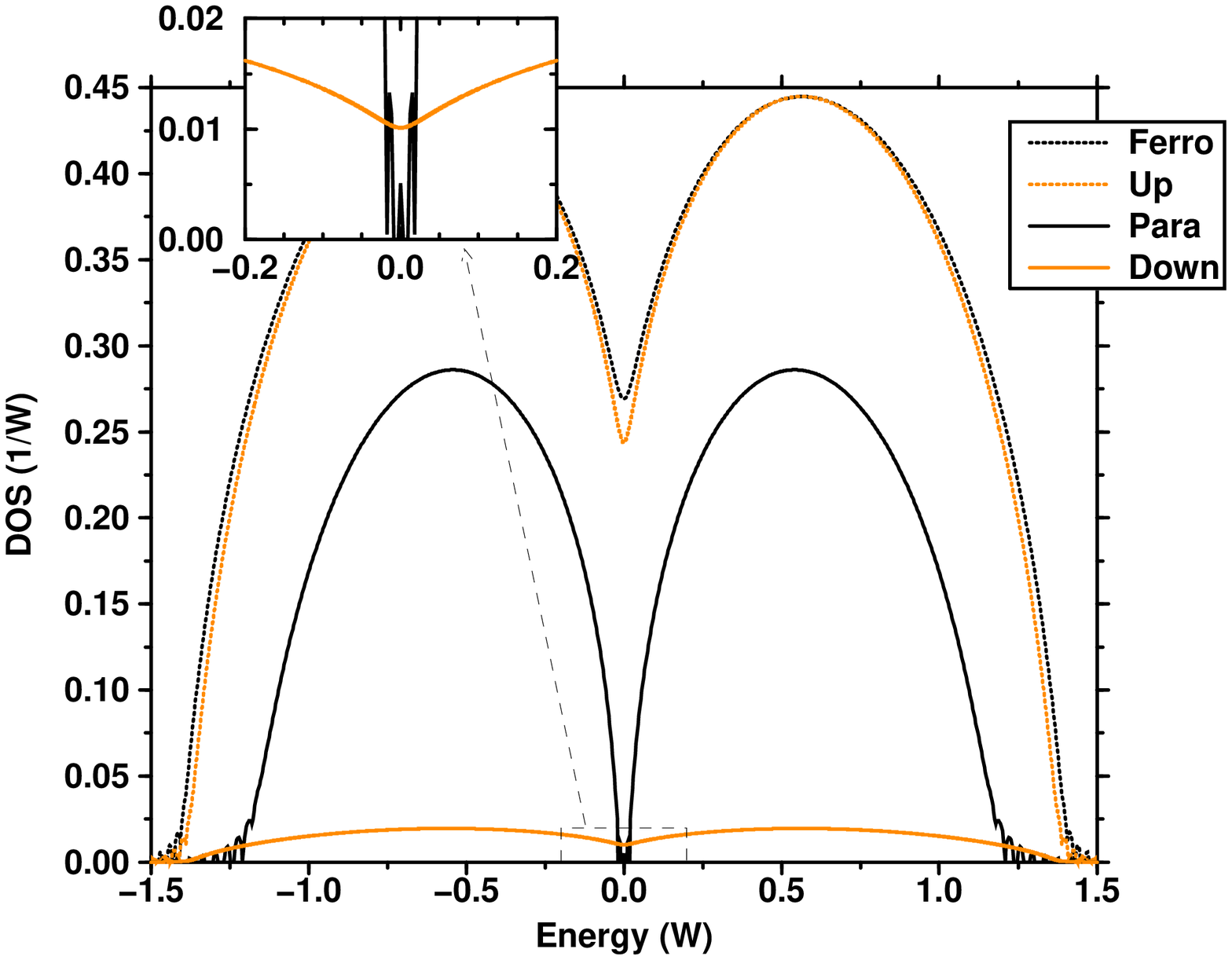}
    \caption{\label{fig:DOSs_PM_FM}%
      The up- and down-spin density of states of the Holstein-DE model with
      $g/W=0.16$ for $k_{\rm B}T=0.005W\ll k_{\rm B}\Tc$ where $\las
      S^z\ras=0.915$. Also shown are the DOS for the saturated ferromagnetic
      state at $T=0$ and for the hypothetical paramagnetic state at $T=0$.
      All plots are for $S=J=\infty$, $h=0$, $n=0.5$, $\om/W=0.05$ and
      $g/W=0.16$ \cite{Gr01}.}
\end{figure}

\begin{figure}
  \centering \includegraphics[width=0.65\textwidth]{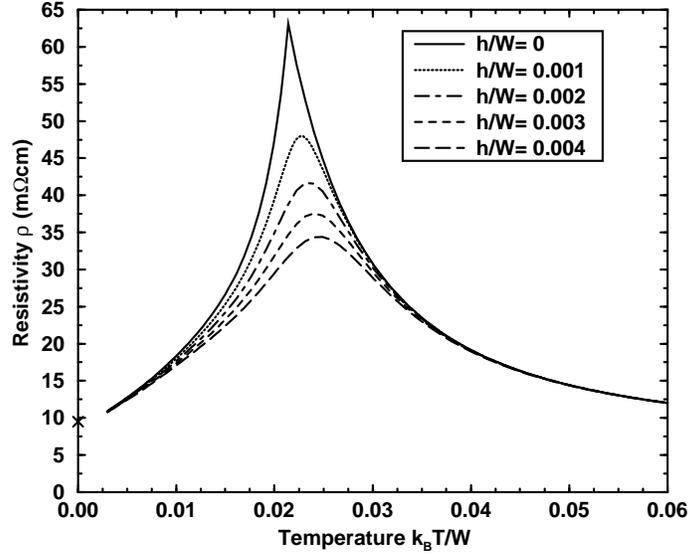}
    \caption{\label{fig:rho_T_g0.16}%
      Resistivity $\rho$ versus temperature for the Holstein-DE model with
      $S=J=\infty$, $n=0.5$, $\om/W=0.05$, intermediate coupling $g/W=0.16$
      and various applied fields $h$. The lattice constant is taken as $a=5$
      {\AA}, slightly larger than the Mn--Mn spacing in the manganites \cite{Gr01}.}
\end{figure}

\begin{figure}
  \centering \includegraphics[width=0.65\textwidth]{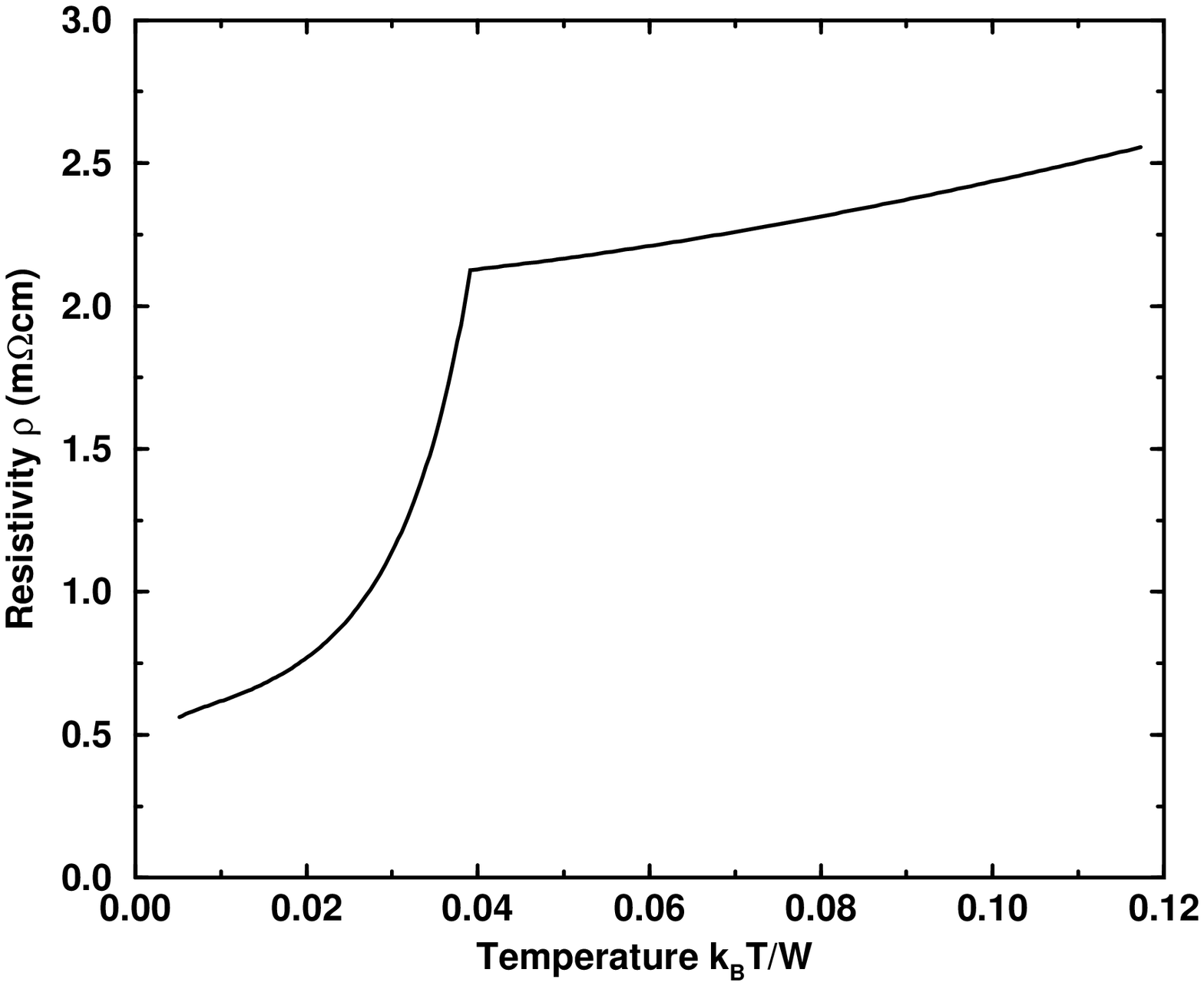}
    \caption{\label{fig:rho_T_g0.1}%
      The same plot as figure~\ref{fig:rho_T_g0.16} but for weak
      electron-phonon coupling $g/W=0.10$.}
\end{figure}

To calculate the Curie temperature $\Tc$ we need the full CPA theory
combined with an exact result of DMFT for $S=\infty$ \cite{Gr01}. Results on $\Tc$
for the same parameters as before are plotted as functions of electron-phonon
coupling $g$
in figure~\ref{fig:Tc_coupling}. The suppression of $\Tc$ with increasing $g$
was first noted by R\"oder \etal
\cite{RoZaBi96} and the CPA results are quite similar to those of Millis
\etal \cite{MiMuSh96II}. In CPA there is no reliable means of calculating
the probability distribution
function $P(S^z)$, so to go below $\Tc$ Green \cite{Gr01} used the mean-field approximation for the
ferromagnetic Heisenberg model with classical spins and nearest neighbour
exchange. The exchange constant is determined by $\Tc$.
We plot the up- and
down-spin density of states for $T=0.005W/k_{\rm B}\ll\Tc$ and $g=0.16W>g_{\rm
  c}$ in figure~\ref{fig:DOSs_PM_FM}, also showing curves for the
saturated ferromagnetic state and paramagnetic state at $T=0$ for comparison. The
value $g=0.16W$ is closer to $g_{\rm c}$ than the value of $0.18W$ used in
figures~\ref{fig:DOS_g} and \ref{fig:DOS_polarons} and we
discuss these results in relation to the manganites in section~\ref{sec:theory-experiment}. In
figure~\ref{fig:rho_T_g0.16} the resistivity $\rho$ is plotted as a function
of temperature, for the same parameter set, with different applied fields
$h$. The resistivity peaks sharply
at $\Tc$, and for comparison we show results for weaker electron-phonon
coupling $g/W=0.10$ in figure~\ref{fig:rho_T_g0.1}. The curve in
figure~\ref{fig:rho_T_g0.1} is almost indistinguishable from that of
figure~7 in reference~\cite{Gr01} for $g/W=0.01$. This
is not surprising since we see from figures~\ref{fig:DOS_g} and
\ref{fig:Tc_coupling} of this paper that the density of states and $\Tc$
change very little between $g/W=0$ and $g/W=0.1$. These results are all
discussed further in section~\ref{sec:theory-experiment}.


\section{Polarons and bipolarons}\label{sec:polarons-bipolarons}

Much experimental data on the manganites in both the paramagnetic and
ferromagnetic state is interpreted in terms of the standard Holstein
small-polaron theory \cite{Ho59a,Ho59b,Ma90}. However there are usually
difficulties in finding parameters which are both plausible and capable of
explaining data from more than one type of measurement. This is not
surprising in view of the conclusions of the last section, namely that many
manganites are only just at the threshold of small-polaron formation even in
the paramagnetic state. This means that one is still far from the strong
coupling limit where standard theory applies. Let us review some of the
problems in applying the standard theory, starting with the low temperature
state.

The width of the coherent polaron band at $T=0$ is reduced from the bare
band-width by a factor $\rme^{-E_{\rm p}/\om}$, where $E_{\rm p}$ is the
polaron binding energy ($g^2/\om$ in the notation of the last section) and
$\om$ is the phonon energy. For the typical values $E_{\rm p}=0.5$ eV,
$\om=0.05$ eV this factor is less than $10^{-4}$. The polaron band is so
narrow that any disorder in the system will produce Anderson localization and
a metallic state is impossible. To avoid this problem, by reducing the
narrowing factor to $10^{-1}$ say, one requires $E_{\rm p}/\om\approx2$ so
that $E_{\rm p}\approx0.1$ eV. But then the condition for small-polaron
formation $E_{\rm p}>Wz^{-\frac{1}{2}}$, where $W$ is half the bare band-width
and $z$ is the number of nearest neighbours \cite{Eagl66}, is far from being
met. Even a band-narrowing factor of 0.1 is too much to be compatible with
heat capacity measurements \cite{Coey99}; the term linear in $T$ at low
temperatures indicates a quasi-particle density of states not much larger
than that given by band calculations.

To examine the high temperature state one must consider the adiabatic limit
(hopping integral $>$ phonon energy). Worledge \etal \cite{WoMiGe98} used the
result of Emin and Holstein \cite{EmHo69} for the conductivity in this limit,
$\si=(A/T)\exp(-E_{\rm a}/k_{\rm B}T)$, to obtain excellent fits to their
data on \chem{La_{1-x}Ca_xMnO_3} films for $T=300$--1200 K over the whole
range of $x$. However for $x=0.3$ the activation energy $E_{\rm
  a}\approx0.08$ eV which is the same as in \chem{Nd_{0.7}Sr_{0.3}MnO_3}
films \cite{ZhKacond-mat}; according to the theory this should be about $E_{\rm
  p}/2$ which makes $E_{\rm p}$ too small for consistency with small-polaron
formation. Also according to small-polaron theory \cite{Ma90} the optical
conductivity peaks at phonon energy $2E_{\rm p}\approx4E_{\rm a}$; almost
universally this peak in the paramagnetic state of pseudo-cubic manganites is
observed at about 1 eV \cite{QuCeSi98} which is therefore inconsistent with
$E_{\rm a}\approx0.08$ eV. This problem does not occur in the many-body CPA
calculations where $E_{\rm a}$ and the optical conductivity peak are both
given correctly for one set of parameters \cite{HoEd01} (see
section~\ref{sec:optical-conductivity}).

Alexandrov and Bratkovsky (AB) \cite{AlBrat99,AlBr99,AlexBr99} have recently
proposed a new polaronic theory of the manganites, including CMR. It is based
on an extended Holstein model \cite{AlKo99} in an attempt to avoid some of
the difficulties discussed above. AB assume the O p-hole model of the
manganite electronic structure so that DE does not come into play. The
antiferromagnetic interaction between the local Mn spins $S=2$ and the
carriers is then assumed weak enough to treat within mean field theory. It is
supposed that small polarons are the carriers in the ferromagnetic state, but
near $\Tc$ they mostly combine to form immobile singlet, and possibly
triplet, bipolarons. This leads to the rapid rise in resistivity near $\Tc$;
it falls slowly above $\Tc$ as the bipolarons dissociate thermally. It is
necessary to assume that it is not favourable for polarons in the
ferromagnetic state to combine as triplet bipolarons. AB propose that this is
because the immobility of the bipolarons suppresses their exchange
interaction with the local spins. The two holes in the bipolarons are
supposed to reside on nearest neighbour oxygen sites. An applied magnetic
field tends to split up the immobile bipolarons and reduce the resistivity;
this is the CMR effect.

To avoid the problem of very narrow polaron bands at $T=0$ Alexandrov and
Kornilovich \cite{AlKo99} introduced an extended Holstein model (EHM) in
which the interaction between an electron on site $i$ and a local phonon at
site $j$ is long range, corresponding to the effect of unscreened Coulomb
interactions. This may also be regarded as an extension of the Fr\"ohlich
dielectric continuum model \cite{Fr54} to a discrete lattice. The model has
also been studied by Fehske \etal \cite{FeLoWe99}. They stress that the EHM
polaron is a large polaron in the whole electron-phonon coupling range. That
is the lattice distortion is spread over large distances even if the coupling
is strong, a regime where a small polaron is formed in the Holstein
model. Alexandrov and Kornilovich describe a `small' Fr\"ohlich polaron, with
a large size lattice distortion but an electronic wave function of small
radius. It is unclear how these two length scales can be distinguished
\cite{FeLoWe99} unless the electron is tightly bound to a defect. The main
point is that the EHM polaron band is much less narrow than the small
Holstein polaron band with the same binding energy. This is because the
lattice distortion undergoes smaller relative changes as the electron moves
from site to site. AB \cite{AlexBr99} also find that the peak energy in
optical conductivity is $2\gamma E_{\rm p}$, with $\gamma\approx0.2$--0.4,
and $E_{\rm a}=\gamma E_{\rm p}/2$. The low observed activation energy
discussed above can therefore be accommodated but, in a theory involving
single polarons only, the inconsistency with the energy of the peak in optical
conductivity above $\Tc$ would persist. However, according to AB, this peak
above $\Tc$ is associated with splitting up a bipolaron, which requires a
larger energy than $2\gamma E_{\rm p}$. The theory is therefore consistent
with the observed shift of optical conductivity towards higher energy on
going from the ferromagnetic to the paramagnetic state. This is discussed
again in section~\ref{sec:optical-conductivity}.

AB's theory presents certain problems. First, the model based on O p-hole
conduction is open to considerable doubt, as discussed in
section~\ref{sec:electronic-structure}. Secondly, it is unclear why, in a
$x=0.3$ metallic manganite with residual resistivity less than 0.1 \res, the
long-range Fr\"ohlich interaction is not screened. Chakraverty \etal
\cite{ChRaFe98,ChRaFe99} argue that it will be reduced to a local Holstein
interaction. Also the small enhancement of the quasi-particle density of
states at the Fermi level, as deduced from the specific heat, is hardly
compatible with a polaronic band at $T=0$. Zhao \etal \cite{ZhSmPrKe00} argue
for small-polaron metallic conduction in the ferromagnetic state of LCMO on
the basis of fitting their measured resistivity $\rho(T)$ below 100 K to the
theoretical predictions. From other data they infer mass enhancements of 9
and 35 in \chem{La_{0.7}Ca_{0.3}MnO_3} and \chem{Nd_{0.7}Sr_{0.3}MnO_3}
which, although small for small polarons, seem too large to be compatible
with specific heat measurements.
Another consideration is that AB's `small' Fr\"ohlich polaron is
really a large polaron. If a bipolaron can form it will be an extended object
in which the two carriers are only confined within the large region of
lattice distortion. Nagaev argues strongly against AB's theory and we refer
again to the public correspondence \cite{Comment49,AlexBr99} between him and
AB in section~\ref{sec:transp-prop-curie}.

R\"oder \etal \cite{RoZaBi96} adopted a different method of avoiding the
extreme polaronic band-narrowing effect. They used a variational Lang-Firsov
transformation which contains parameters $\gamma$ and $\Delta$ whose role is
to quench part of the dynamical polaron effect as a static distortion. Thus
the operator $s$, introduced just before equation~(\ref{eq:hde_1}), becomes
$-(g/\om)(\gamma n+\Delta)(b^\dag-b)$ and the transformation is applied to
every site in the lattice. The band-narrowing factor is then modified from
$\rme^{-E_{\rm p}/\om}$ to $\rme^{-\gamma^2 E_{\rm p}/\om}$ and
$0<\gamma<1$. The main result is that the $\Tc$ of the pure DE model is
reduced by this factor due to electron-phonon coupling; the resistivity was
not calculated. The most recent work along similar lines is that of Perroni
\etal \cite{PeFiCaIa01,CaFiIa01}. The DE effect is treated by a simple
band-narrowing but these authors go much further than R\"oder \etal In
particular they calculate a one-electron Green function which is quite
similar to Green's \cite{Gr01} and, like his, yields the correct atomic limit
(equation~(\ref{eq:G_hde_AL})) for the classical spin case
considered. However instead of a smooth evolution of a pseudogap with
increasing temperature, they find phase separation above $0.65\Tc$ with
ferromagnetic regions of low electron density (large polaron behaviour) and
paramagnetic insulating regions of high electron density (localized small
polarons). However the effect of the nanoscale structure of such regions,
enforced by Coulomb interaction, is not considered.


\section{Theory and experiment}\label{sec:theory-experiment}

The experimental evidence and theoretical arguments reviewed in previous
sections point to the Holstein-DE model as a plausible model of the
manganites \chem{A_{1-x}A'_xMnO_3}, at least for $x\approx0.33$. For this
degree of doping with the divalent element \chem{A'} one is normally in the
middle of the ferromagnetic regime and inhomogeneity due to charge and
orbital ordering is least likely. In order to confront theory with experiment
effectively it is essential to keep the number of adjustable parameters to a
minimum. Ideally one would like to understand the difference between
materials, as reflected in the results of many different types of
experiments, by changes in one crucial parameter. The many-body CPA approach
to the Holstein-DE model gives us the opportunity to attempt this and from
section~\ref{sec:many-body-cpa-hde} it is clear that the crucial parameter is
the electron-phonon coupling $g/W$. In this section we take fixed reasonable
values of the parameters $\om$ and $W$. Band calculations \cite{PiSi96,SaShBa95}
suggest $W=1$ eV as an appropriate half-bandwidth for the \chem{e_g} band.
Then it is reasonable to take $\om/W=0.05$ to correspond to the observed
transverse optic phonons with $\om\approx40$--70 meV which couple strongly to
the electrons in LCMO \cite{KiGuChPa96}. It is also reasonable to take the DE
limit $J\rightarrow\infty$ \cite{Sa96}. Furthermore we see from
figures~\ref{fig:Tc_n_elliptic} and~\ref{fig:resis_elliptic} that for the DE model
neither $\Tc$ nor the resistivity $\rho$ vary enormously with $S$ so that
$S=\infty$ is a convenient approximation to the $S=3/2$ Mn spins. These are
the parameters which have been used in
figures~\ref{fig:DOS_g}--~\ref{fig:rho_T_g0.1}. The band-filling is also
restricted to $n=1-x=0.5$ rather than $n=0.6$--0.7. This is very convenient
because the chemical potential remains fixed for all $T$ by electron-hole
symmetry in the case $S=\infty$. This will change the critical value of $g/W$
for small polaron formation somewhat but the correct general picture should
emerge. Since we consider a homogeneous state we are not concerned with the
existence of charge ordering for $x=0.5$. We now compare the theory with
different types of experiment.

\subsection{Transport properties and Curie temperature}
\label{sec:transp-prop-curie}

Perhaps the most striking feature of the manganites is the very different
behaviour observed in apparently similar materials such as LSMO and LCMO. For
LSMO, with $x\approx 0.33$, $\Tc\approx 370$ K whereas for LCMO, with a
similar $x$, $\Tc\approx 240\,{\rm K}$. The difference in behaviour of the
resistivity $\rho$ above $\Tc$ is much more striking (see
figures~\ref{fig:fig4} and~\ref{fig:fig5}). For LSMO $\rho\approx 4\,{\rm
  m\Omega\,cm}$ and increases slowly with temperature as in a poor metal
\cite{UrMoAr95}. The $\rho(T)$ curve is very similar to that of
figure~\ref{fig:rho_T_g0.1} for $g/W=0.1$ except for a much larger
resistivity at low temperature in the calculations.  Since this feature
persists even for $g/W=0.01$ (figure~7 in reference~\cite{Gr01}) it
presumably arises from overestimated spin disorder scattering at low
temperatures due to use of the classical spin Heisenberg model to
determine $P(S^z)$. In LCMO the resistivity rises to a maximum at $\Tc$ of
about 40 \res and then falls with increasing temperature above $\Tc$. In
contrast to LSMO there is thus a transition from metallic to insulating
behaviour. Also the resistivity peak is strongly reduced and shifted to
higher temperature with increasing applied magnetic field
(figure~\ref{fig:fig5}). This is the CMR effect. This type of behaviour is
seen in figure~\ref{fig:rho_T_g0.16} for $g/W=0.16$. The decrease of $\rho$
with increasing temperature above $\Tc$ is associated with the increasing
density of states at the Fermi level, as seen in
figure~\ref{fig:DOS_polarons}. The main differences
between theory and experiment are a more rapid observed drop in $\rho$ with
decreasing temperature below $\Tc$ and a more sensitive observed CMR effect.
$h/W=0.004$ corresponds to a field of about 20 T for $W=1$ eV and the
corresponding reduction in $\rho$ in figure~\ref{fig:rho_T_g0.16} is achieved
with a field of about 5 T experimentally. Millis \etal \cite{MiMuSh96II}
noted a similar problem in their work using classical phonons. Both of the
discrepancies mentioned might be remedied by introducing a dependence of $g$
on $\rho$, corresponding to more efficient screening of the electron-phonon
interaction with increasing density of states. The huge reduction in resistivity
peak on reducing $g/W$ from 0.16 to 0.10 shows the extreme sensitivity
of $\rho$ to changes in $g$. The main point to notice is that we can
understand the enormous difference between LCMO and LSMO within the
Holstein-DE model by
assuming the electron-phonon coupling changes from $g/W=0.16$ in LCMO to
$g/W=0.10$, or slightly greater, in LSMO. The observed ratio of the Curie
temperatures, slightly less than 2, is then in accord with
figure~\ref{fig:Tc_coupling}. As discussed in section
\ref{sec:many-body-cpa-hde} the critical coupling $g_{\rm c}$ for the
formation of a polaron band is $g_{\rm c}/W\approx 0.15$, with phonon
energy $\om/W=0.05$, and to obtain the right order of magnitude for $\rho$
above $\Tc$ in LCMO, $g/W$ is pinned down closely to 0.16. A larger value for
$g$ leads to too high a resistivity and too low a Curie temperature. It is
interesting that neither $\rho(T)$ nor $\Tc$ change when $g/W$ is varied
between 0.1 and 0. This means that the resistivity of LSMO can be described
quite well by the pure DE model, as stressed by Furukawa \cite{FuPom},
but electron-phonon coupling is not negligible and shows up in the optical
conductivity, for example, which we discuss in section~\ref{sec:optical-conductivity}.
However, from the results of Millis \etal \cite{MiMuSh96II} for classical
phonons, one can understand why a coupling small enough to give a LSMO-like
$\rho(T)$ curve does not lead to a change in slope of the rms oxygen
displacements, as a function of temperature, at $\Tc$. No such change is
found in LSMO \cite{MaEn96}, in contrast to the case of LCMO \cite{DaZhMo96}.
It is more difficult to understand the observation \cite{LoEgBr97} of static
local Jahn-Teller distortions in LSMO at room temperature, apparently
associated with localized carriers in the presence of metallic conduction.

From figure~\ref{fig:DOS_g} we see that for $g/W=0.1$, appropriate to LSMO,
there is no sign of a pseudogap in the density of states. An actual gap in
the hypothetical paramagnetic state at $T=0$ appears at $g=g_{\rm c}$ with
$g_{\rm c}/W$ between 0.15 and 0.16. From figure~\ref{fig:DOSs_PM_FM} we see
that for $g/W=0.16$, appropriate to LCMO, a few polaron subbands have
appeared in the gap. These are seen much more clearly in
figure~\ref{fig:DOS_polarons} for $g/W=0.18$ when there is a larger gap.
However the subband structure is washed out completely for $\beta W=25$,
corresponding to $T=464$ K for $W=1$ eV, and this effect will occur at a much
lower temperature for $g/W=0.16$. Thus in the actual paramagnetic state of
LCMO above $\Tc$ we do not expect the quantum nature of phonons to manifest
itself, so we are essentially in the classical regime of Millis \etal
\cite{MiMuSh96II}. The same is true in the saturated ferromagnetic state at
$T=0$ where only a pseudogap appears in figure~\ref{fig:DOSs_PM_FM}. As the
temperature rises towards $\Tc$ a minority spin band grows, also with a
pseudogap, while the majority spin band loses weight. The width of the
ferromagnetic bands decreases with increasing temperature, corresponding to
the DE effect, but the narrowing in the paramagnetic state is not so marked
as in the pure DE model. Thus double exchange is not so effective in the
presence of strong electron-phonon coupling, which is consistent with the
reduction in $\Tc$ shown in figure~\ref{fig:Tc_coupling}. We discuss this
again in section~\ref{sec:spinwaves} in connection with spin-waves.

Clearly the picture of the manganites which emerges here is close in spirit
to that of Millis \etal \cite{MiMuSh96II}, although the relationship to
polaron physics is not so clear in their classical approximation. We have
described earlier how some other authors adopt completely different
viewpoints. In particular Nagaev \cite{Naga99} argues against any polaronic
effects, while Alexandrov and Bratkovsky \cite{AlBr99,AlexBr99} assume strong
electron-phonon coupling with small polarons even in the ferromagnetic state
and with immobile bipolarons forming near $\Tc$. The public correspondence
\cite{Comment49,AlexBr99} between Nagaev and Alexandrov and Bratkovsky (AB)
centres on estimating the magnitude of the polaron binding energy $E_{\rm p}$
and the criterion for small polaron formation \cite{AlBrat99}. Since the
picture of LCMO presented here lies between their extreme views it is
interesting to compare our estimates with theirs. For LCMO we find $E_{\rm
  p}=g^2/\om\approx0.5$ eV for $W=1$ eV whereas, for manganites in general,
Nagaev estimates $E_{\rm p}\approx0.1$--0.3 eV and AB estimate $E_{\rm
  p}\approx1$ eV. Our condition for small-polaron formation in a paramagnetic
state at $T=0$ is $g>g_{\rm c}\approx0.15W$ which corresponds to $E_{\rm
  p}>0.45W$. Nagaev adopts the criterion $E_{\rm p}>W$, remembering that $W$
is the half-bandwidth in our notation, while AB \cite{AlBrat99} propose
$E_{\rm p}>2W(8z)^{-1/2}=0.29W$ with number of nearest neighbours $z$ taken
as 6. AB's condition is less stringent than Eagle's \cite{Eagl66} condition
for `nearly small polarons' $E_{\rm p}>Wz^{-1/2}=0.41W$ which is close to
ours. Both as regards this criterion and the value of $E_{\rm p}$ for LCMO,
the results of the Holstein-DE model are intermediate between Nagaev's and
AB's, as expected. For LSMO, on the other hand, our estimate of $E_{\rm p}$
is 0.2 eV. In this case we agree with Nagaev and Furukawa \cite{FuPom} that
electron-phonon coupling is not so important. AB's work is reviewed in
section~\ref{sec:polarons-bipolarons}.

\subsection{Isotope and pressure effects}\label{sec:isot-press-effects}

There is an excellent review of oxygen isotope effects in manganites by Zhao
\etal \cite{ZhaoPom}. The most striking effects are a decrease in $\Tc$ and
an increase in resistivity $\rho$, particularly near $\Tc$, when
\chem{^{16}O} is replaced by \chem{^{18}O}. However other important effects
are an induced metal-insulator transition, a shift of the charge-ordering
transition and effects on the thermal-expansion coefficient and electron
paramagnetic resonance (EPR) measurements. Although Zhao \etal attempt to
understand these effects using the concepts of small-polarons and bipolarons,
they point out clearly the difficulties in such a theory. We believe that
most of these difficulties can be resolved within the many-body CPA approach
to the Holstein-DE model, which frees us from the strong coupling regime of
standard small-polaron theory.

The oxygen isotope exponent is defined as $\alpha_0=-(\Delta\Tc/\Tc)/(\Delta
M/M)$, where $\Tc$ and $M$ are the Curie temperature and oxygen mass for the
\chem{^{16}O} sample. $\alpha_0$ ranges from 0.07 with $\Tc=367$ K in
\chem{La_{0.67}Sr_{0.33}MnO_3} to 0.85 in \chem{La_{0.8}Ca_{0.2}MnO_{3+y}}
($\Tc=206$ K) and up to 4 in \chem{(La_{0.25}Nd_{0.75})_{0.7}Ca_{0.3}MnO_3}
($\Tc\approx110$ K). Zhao \etal \cite{ZhaoPom} point out that all available
data (for $x\approx0.2$--0.35) can be fitted by an equation
\begin{equation}\label{eq:alpha_0}
  \alpha_0=21.9 \exp(-0.016\Tc)\,.
\end{equation}
According to the DE model and small-polaron theory
$\Tc\propto\exp(-\gamma^2E_{\rm p}/\om)$, which is the usual band-narrowing
factor (see section~\ref{sec:polarons-bipolarons}) with $0\le\gamma^2\le1$
\cite{RoZaBi96}. When the mass $M$ changes, $\om$ varies as $M^{-1/2}$ but
the polaron binding energy $E_{\rm p}$ is independent of $M$ since it
corresponds to a static response of neighbouring oxygen atoms to an
\chem{e_g} electron on a Mn site. Hence $\Tc$ decreases rapidly with
increasing $M$ and it is easy to show that $\Tc\propto\exp(-2\alpha_0)$
\cite{ZhaoPom}. Zhao \etal \cite{ZhaoPom} point out that this is incompatible
with the experimental relation~(\ref{eq:alpha_0}). They go on to make the
very important observation that the effect of pressure on $\Tc$, again over a
wide range of manganites, is summarized accurately by the equation
\begin{equation}\label{eq:Tc-P}
  \rmd \ln\Tc/\rmd P = 4.4 \exp(-0.016\Tc)\,.
\end{equation}
Combining equations~(\ref{eq:alpha_0}) and~(\ref{eq:Tc-P}) one has
$\alpha_0=5(\rmd\ln\Tc/\rmd P)$. Zhao \etal deduce from this simple
proportionality between the isotope exponent and the pressure-effect
coefficient that the major effect of pressure is to increase the phonon
frequency. We come to a completely different conclusion from the following
analysis using the many-body CPA approach to the Holstein-DE model.

\begin{figure}
  \centering \includegraphics[width=0.65\textwidth]{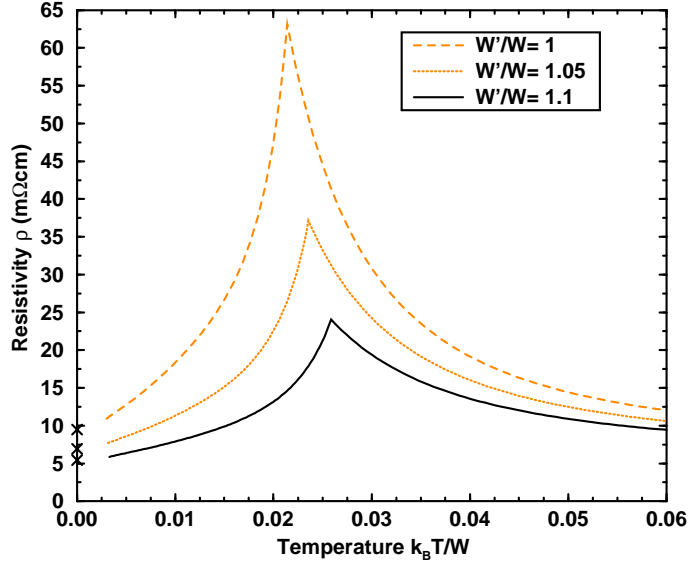}
    \caption{\label{fig:pressure}%
      The effect of pressure on the resistivity and Curie temperature in LCMO
      by increasing the half-bandwidth $W$ to $W'=1.05W$ and $1.1W$ \cite{Gr01}.}
\end{figure}

In discussing the pressure effect, Green \cite{Gr01} observed that the
increase in $\Tc$, and decrease in $\rho$, stems mainly from a decrease in
the effective coupling constant $g^2/(\om W)=E_{\rm p}/W$. As pointed out in
section~\ref{sec:introduction} (see figure~\ref{fig:fig3}), the effect of
pressure is equivalent to an increase in $\las r_{\rm A}\ras$ and broadens
the \chem{e_g} band, that is $W$ increases. Green \cite{Gr01} therefore
modelled the strong suppression of the resistivity peak and the increase in
Curie temperature in LCMO by increasing the bandwidth, and assuming other
terms in the Hamiltonian are constant. Calculated results are shown in
figure~\ref{fig:pressure}. A simple estimate, using the known compressibility
and dependence of $W$ on lattice constant, shows that the theoretical
pressure for a given effect is about four times larger than that required
experimentally.  This is the same factor that was found in the case of the
magnetic field required for a given CMR effect, so both discrepancies could
possibly be removed with the same dependence of $g$ on $\rho$, due to
screening, postulated in section~\ref{sec:transp-prop-curie}. We return to
this point later.
In fact near $\Tc$ the CMR effect (figure~\ref{fig:rho_T_g0.16}) and the
pressure effect (figure~\ref{fig:pressure}) are both driven by a change of
bandwidth. In the first case the magnetic field produces a substantial
magnetization for $T\approx\Tc$ and this increases $W$ by the DE effect.

\begin{figure}
  \centering \includegraphics[width=0.65\textwidth]{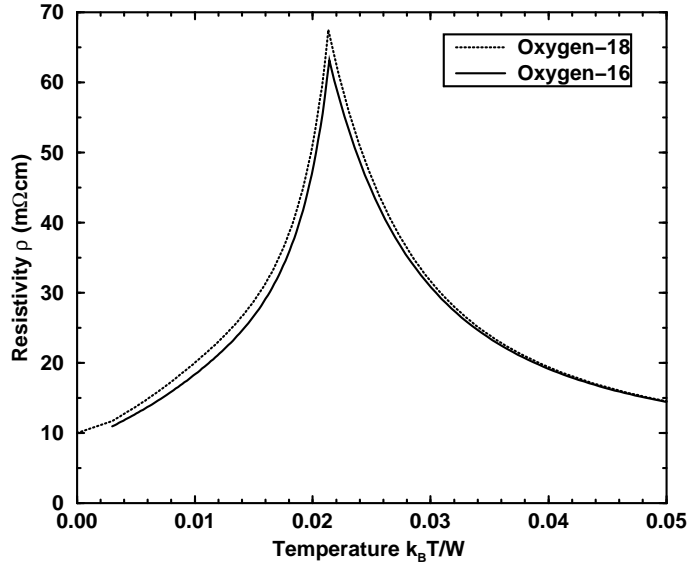}
    \caption{\label{fig:isotope_LCMO}%
      The absence of an isotope effect in LCMO due to a simple mass
      scaling $\left(\propto M^{-1/2}\right)$ of the phonon frequency.}
\end{figure}

To investigate the isotope effect in our model, assuming only an effect via
$\om$, some care is needed. It was pointed out that in
equation~(\ref{eq:h_hde}) the electron-phonon coupling term corresponds to a
term $-g'\sum_i n_i x_i$ where $x_i$ is to be associated with oxygen
displacement around a \chem{Mn} atom. Here $g'$ should be independent of the
oxygen mass $M$. In the second-quantized form of equation~(\ref{eq:h_hde})
one finds $g=g'\left(2M\om\right)^{-1/2}$ so that the polaron binding energy
$E_{\rm p}=g^2/\om={g'}^2/\left(2M\om^2\right)$. Since for an oscillator $\om\propto
M^{-1/2}$ the polaron binding energy $g^2/\om$ is independent of $M$ as
expected.  However $g$ varies as $M^{-1/4}$. We therefore recalculated the
resistivity $\rho(T)$ with the same parameters as used in
figure~\ref{fig:rho_T_g0.16} for \chem{^{16}O} but with $g$ and $\om$ scaled
appropriately for \chem{^{18}O}. The results are compared in
figure~\ref{fig:isotope_LCMO}. The almost complete absence of an isotope
effect in both $\Tc$ and $\rho$ confirms Green's observation that $\rho(T)$
depends almost exclusively on the parameter $E_{\rm p}/W$. Since $E_{\rm p}$
should not depend on $M$ we have to conclude that the isotope effect is due
to a decrease of bandwidth $2W$ when $M$ is increased.

Thus we make the hypothesis that the isotope effect occurs via the \chem{e_g}
bandwidth $2W$ rather than the phonon energy $\om$, whose undoubted change
has no effect. We must stress that $2W$ is the bare band-structure
bandwidth. Clearly a sort of scaling theory, linking CMR, pressure effect and
isotope effect, can be constructed but this will be done elsewhere. The
change in the $\rho(T)$ curve is typically quite similar for
\chem{^{18}O\rightarrow}\chem{^{16}O} back-substitution to an applied field of
5--10 T or a pressure of 5--10 kbar. Babushkina \etal \cite{BaBeOzGo98} have
already suggested that a metal-insulator transition induced by oxygen isotope
substitution in \chem{(La_{1-y}Pr_y)_{0.7}Ca_{0.3}MnO_3} may proceed by a
change of bandwidth. They point out that the mean-square displacement of an
ion from its nominal position depends on the ionic mass $M$, even at $T=0$, due
to zero-point vibrations. Such displacements could affect the Mn--O--Mn bond
angle on which the bandwidth critically depends. Similar effects were
observed earlier in \chem{(La_{0.5}Nd_{0.5})_{0.67}Ca_{0.33}MnO_3}
\cite{ZhKeHo97}. Babushkina \etal also see large shifts of the $\rho(T)$
curves with applied magnetic field, as one would expect from the above
discussion. Clearly similar ideas can be applied to shifts in charge-ordering
transitions due to isotopic substitution and magnetic field. The isotope
effect in the jump of the thermal expansion coefficient at $\Tc$ is easily
understandable since it is related by thermodynamics to the pressure effect
\cite{ZhaoPom}.

We now have to address the question of whether replacing \chem{^{16}O} by
\chem{^{18}O} can produce the required decrease of $W$ to account for the
effect on $\rho(T)$. The change in $\rho(T)$ is typically like that
calculated for a 5--10\% decrease in $W$ (cf. figure~\ref{fig:pressure}). If
we allow for the factor of 4 proposed earlier, due to improved screening of
the electron-phonon coupling as the system becomes more metallic this is
reduced to 1--2\%. According to Zhao \etal \cite{ZhHuKe97} the volumes of the
unit cells of \chem{^{16}O} and \chem{^{18}O} samples of
\chem{La_{0.67}Ca_{0.33}MnO_3} at room temperature are the same within the
accuracy of the x-ray determination. Singh and Pickett \cite{SiPi98} have
investigated the change in band structure of a \chem{La_{2/3}Ba_{1/3}MnO_3}
virtual crystal on going from a cubic perovskite structure to a Pnma
structure consistent with the neutron refinement of Dai \etal \cite{DaZhMo96}
for \chem{La_{0.65}Ca_{0.35}MnO_3}, the two structures having the same unit
cell volume. Large band-narrowings occur owing to bending of the Mn--O--Mn
bonds; in particular a gap of a few tenths of an eV opens between the
majority spin \chem{e_g} and \chem{t_{2g}} bands. The distortion from the
cubic structure has no static JT components, consisting essentially of
rotations of O octahedra. It is not inconceivable that isotopic substitution
could modify these rotations sufficiently to give a 1--2\% reduction in the
\chem{e_g} bandwidth. Further theoretical and experimental investigation of
this possibility is highly desirable. Finally we come back to the enhancement
factor of 4 postulated above. It means that a 1\% increase in $W$ corresponds
to a 4\% decrease in $E_{\rm p}/W$, the large relative change in $E_{\rm p}$
being assumed to arise from increased screening with increased
metallization. This is not unreasonable since it is known from optical
conductivity measurements, discussed in
section~\ref{sec:optical-conductivity}, that the polaron binding energy
$E_{\rm p}$ decreases from 1.2 eV in \chem{Nd_{0.7}Sr_{0.3}MnO_3} to about
0.6 eV in LSMO. The change in \chem{e_g} bandwidth between these two materials is
certainly much less than a factor 2 so that in general a much larger relative
change in $E_{\rm p}/W$ than in $W$ is a reasonable hypothesis.

In summary our picture of the isotope effect due to
\chem{^{16}O\rightarrow}\chem{^{18}O} substitution is as follows. The change
in phonon frequency plays no part. The effect is driven by a 1--2\% reduction
in the \chem{e_g} bandwidth $2W$ associated with small changes in the
rotations of O octahedra probably arising from modified vibrational amplitude
of the O ions due to their mass change. This primary effect is strongly
enhanced by an increased polaron binding energy $E_{\rm p}$ arising from
reduced metallic screening. The scaling parameter $E_{\rm p}/W$ links CMR,
pressure-effect and the various isotope effects.

We should mention that Nagaev \cite{Naga99,Naga98} has taken a completely
different line on the isotope effect in which electron-phonon coupling has no
role. He considered several possibilities, all based on the isotope
dependence of the number of excess or deficient oxygen atoms in thermodynamic
equilibrium. Since he quoted experimental evidence for this \cite{FrIsCh98}
his arguments were quite persuasive. However Zhao \etal \cite{ZhCoKeMu00}
have criticized the experimental work of Franck \etal \cite{FrIsCh98} and
they refute Nagaev's model in which the isotope effect is not intrinsic.

\subsection{Pseudogaps}\label{sec:pseudogaps}

According to theory we expect pseudogaps at the Fermi level to be observable
in the density of states of LCMO both below and above $\Tc$. These should
appear in experiments such as scanning tunnelling spectroscopy, photoemission
and optical conductivity measurements. No pseudo-gap is expected in LSMO
below $\Tc$ and the effect should be much less than in LCMO above $\Tc$.
The pseudogap is a feature of the atomic limit, typified by the envelope
function in figure~\ref{fig:hde_AL} for $T=0$ with maxima determined by the
polaron binding energy $g^2/\om$, and is completely washed out when $g^2/(\om
W)<0.2$ (see figure~\ref{fig:DOS_g}).  Early results of scanning tunnelling
spectroscopy on LCMO \cite{WeYeVa97} with $x=0.3$ seem unlikely to relate to
the bulk. In the ferromagnetic state at 77 K there is a huge gap of about 1
eV. It is not clear why the authors interpret this as evidence for
half-metallic ferromagnetism. A gap of this size associated with
small-polaron formation in the bulk would imply an unrealistically large
electron-phonon coupling, certainly incompatible with metallic conduction and
a Curie temperature of reasonable magnitude for LCMO. More recently Biswas
\etal \cite{BiElRaBh99} reported a scanning tunnelling spectroscopy study of
several manganites. The results are very much in accord with the theory.
There is no gap in the low temperature ferromagnetic state but a small gap
(pseudogap) appears for $T\approx\Tc$ in the low (high) $\Tc$ materials. As
$T$ increases above $\Tc$ the pseudogap or gap gets filled in as we would
expect (see figure~\ref{fig:DOS_polarons}). The decrease in resistivity with
increasing $T$ is due to the gradual filling of the pseudogap.

\subsubsection{Angle-resolved photoemission spectroscopy (ARPES)}
\label{sec:angle-resolv-phot}

\begin{figure}[htbp]
  \centering \includegraphics[width=0.65\textwidth]{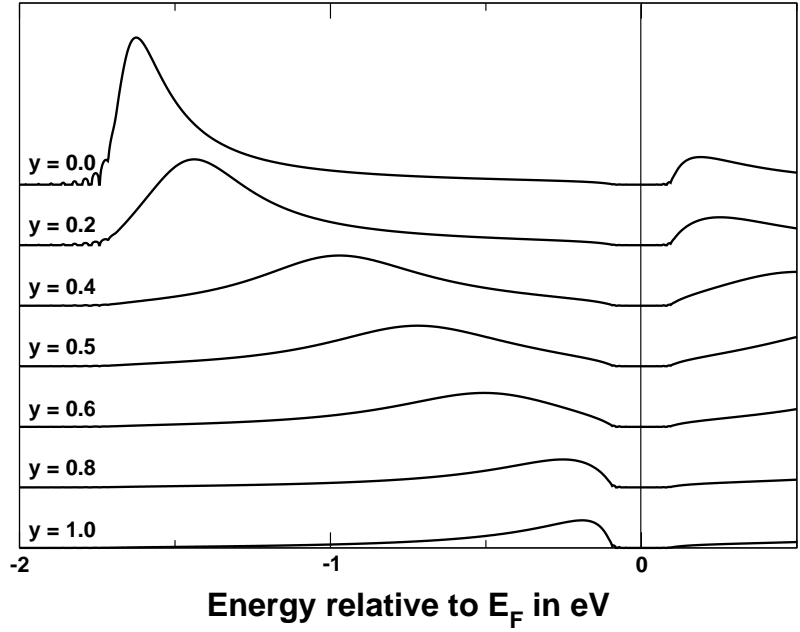}
\caption{\label{fig:spec_func_th}%
  The spectral function $A_{\bi{k}}(\en)$ in the ferromagnetic state at $T=0$
  for the Holstein-DE model with $J=S=\infty$, $n=0.5$ and strong
  electron-phonon coupling $g/W=0.20$, and $\bi{k}=\pi(1,y)$ \cite{HoEd01}.}
\end{figure}

\begin{figure}[htbp]
  \centering \includegraphics[width=0.65\textwidth]{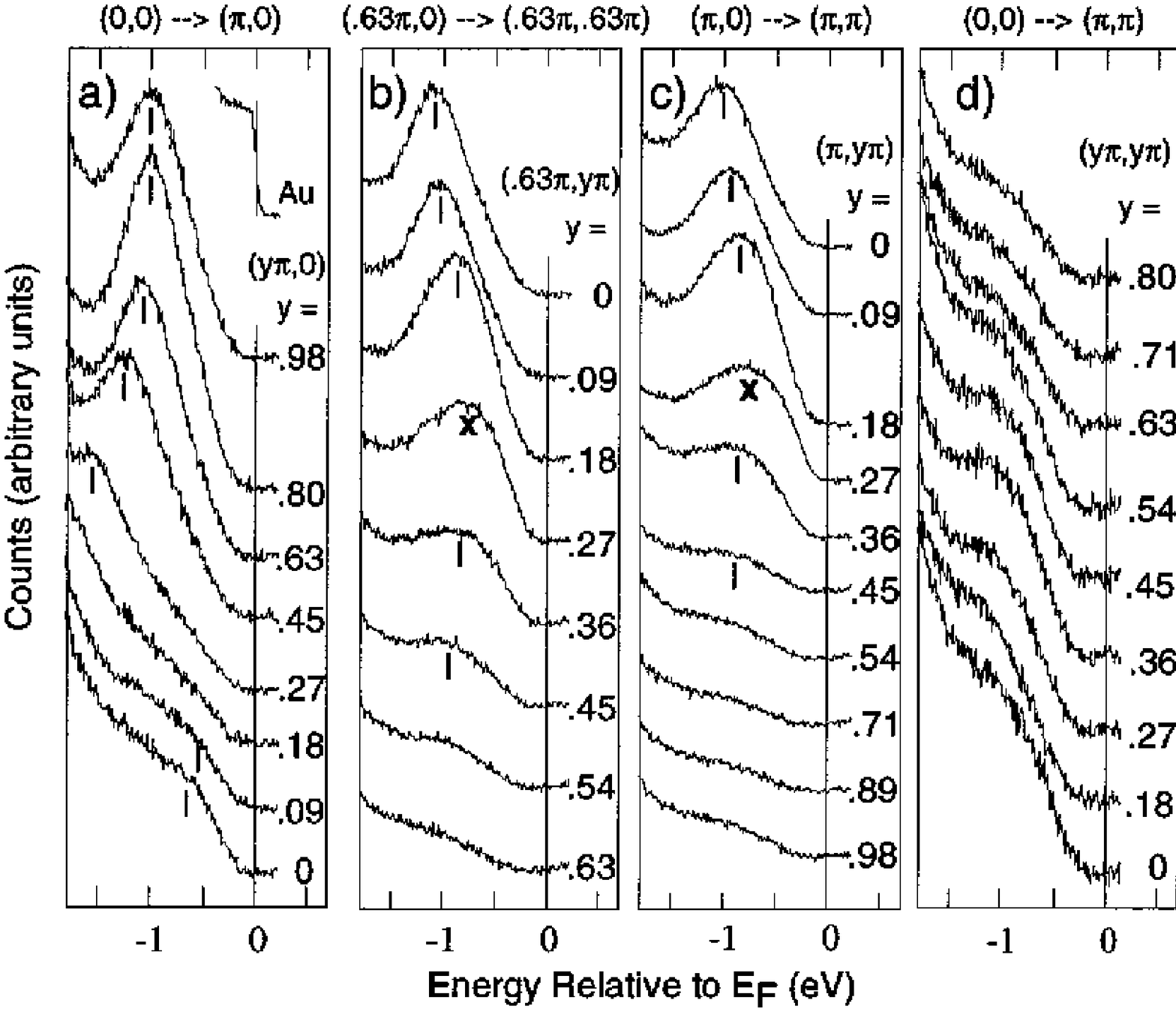}
\caption{\label{fig:spec_func_exp}%
  ARPES spectra of \chem{La_{1.2}Sr_{1.8}Mn_{2}O_{7}} ($\Tc=126\,{\rm K}$) in
  the ferromagnetic state at $T=10\,{\rm K}$, reproduced from Dessau \etal
  \cite{De98}.}
\end{figure}

In an extremely interesting paper on ARPES for the bilayer manganite
\chem{La_{1.2}Sr_{1.8}Mn_2O_7}, nominally with $n=0.6$, Dessau \etal
\cite{De98} interpret their results very much in the spirit of the Holstein
model. The low $\Tc=126$ K in this bilayer manganite is partly due to
quasi-two dimensional fluctuations, but the large resistivity $\rho\simeq3$
\res at low temperatures indicates that small-polaron bands might exist even
in the ferromagnetic state. Consequently the electron-phonon coupling should
be stronger than in cubic manganites like LCMO and to model the system by the
Holstein-DE model Hohenadler and Edwards \cite{HoEd01} chose $g/W=0.2$. The
one-electron spectral function $A_\bi{k}(\en)$ is given by
equation~(\ref{eq:spec_function}) and the band is taken to be of the form
$\ek=-W\cos\pi y$ for $\bi{k}=\pi(1,y)$, $0\le y\le1$ with $W=1$ eV as usual.
This roughly models a band, calculated by Dessau \etal \cite{De98}, which
crosses the Fermi level $E_{\rm F}$ at $\bi{k}=\pi(1,\frac{1}{2})$. The
calculated results \cite{HoEd01} for $A_\bi{k}$ are shown in
figure~\ref{fig:spec_func_th}. Well away from the Fermi level, a well-defined
peak exists which broadens as $\bi{k}$ approaches the Fermi momentum at
$y=0.5$. For larger $y$ the weight below the Fermi level is strongly reduced.
The peaks never approach the Fermi level closely which is an important
feature of the observed spectra \cite{De98} reproduced in
figure~\ref{fig:spec_func_exp}. The theoretical curves in
figure~\ref{fig:spec_func_th} resemble quite closely the data of
figure~\ref{fig:spec_func_exp}(c). There is a pseudogap in the calculated
spectra extending about 0.1 eV on each side of the Fermi level. In fact this
pseudogap contains polaron bands like those shown in
figure~\ref{fig:DOS_polarons}. However, their amplitude is too small to show
up in figure~\ref{fig:spec_func_th} and in the experimental data.
Nevertheless, it is the central polaron band around the Fermi level which is
presumably responsible for the low but finite conductivity of the system.
This comparison between theory and experiment supports the conclusion of
Dessau \etal \cite{De98} that, in the manganites with a layered structure,
strong electron-phonon coupling (with the appearance of a pseudogap) is
already important below $\Tc$. This contrasts with the usual pseudocubic
manganites where the pseudogap only appears above $\Tc$. It should be
mentioned that Moreo \etal \cite{MoYu99} interpret the observed pseudogap not
as an intrinsic property but in terms of phase separation.

\subsubsection{Optical conductivity}\label{sec:optical-conductivity}

\begin{figure}[htbp]
  \centering \subfigure[ ]{
    \label{fig:oc:a}
    \includegraphics[width=0.45\textwidth]{oc_0.16_bw.eps}} \subfigure[ ]{
    \label{fig:oc:b}
    \includegraphics[width=0.45\textwidth]{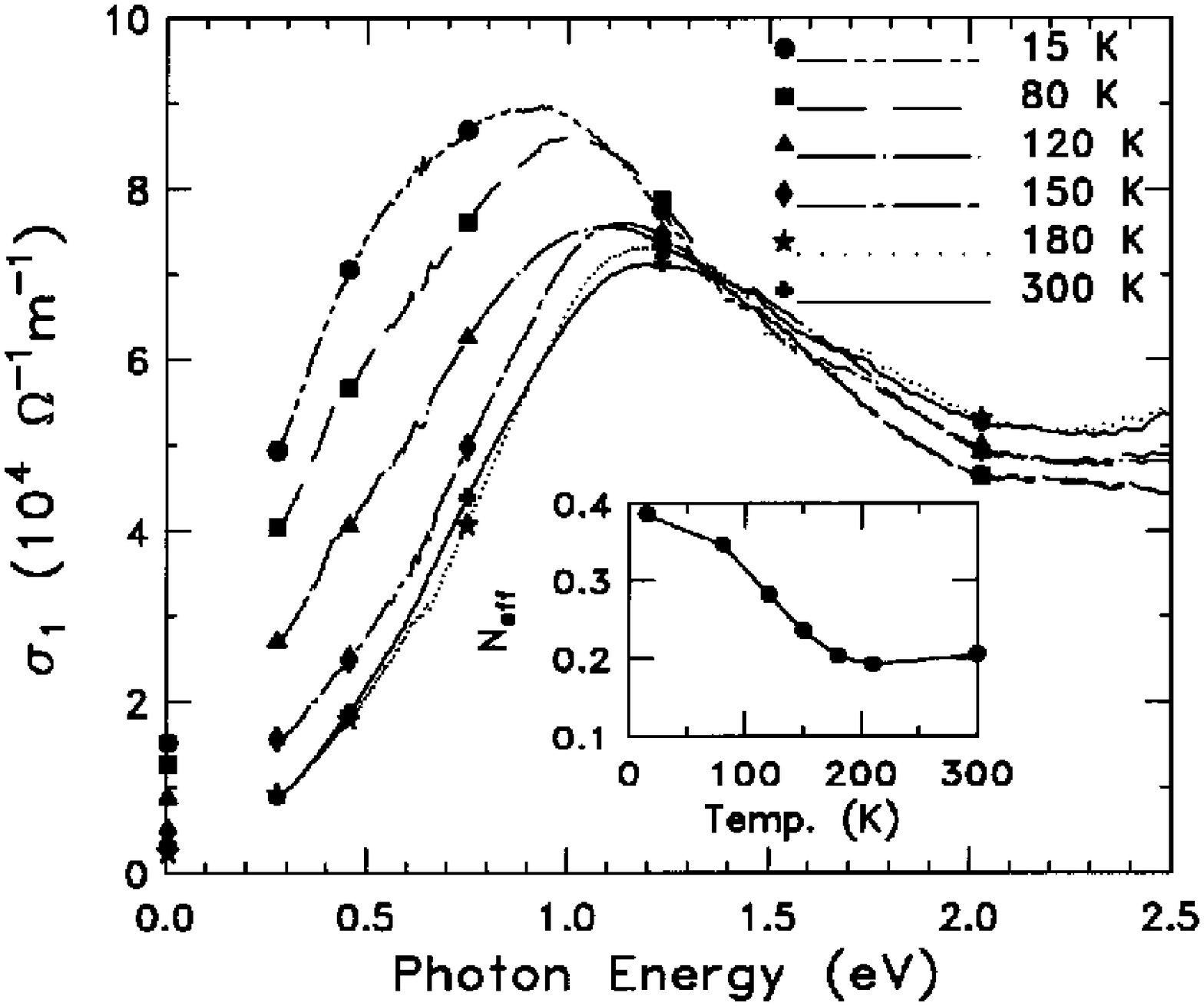}}
  \caption{\label{fig:oc}%
    (a) Calculated optical conductivity for strong electron-phonon coupling
    $g/W=0.16$ in the ferromagnetic state at $T=0$ (full line), the
    paramagnetic state at $T=\Tc$ (dotted line) and the paramagnetic state at
    $T=1.5\Tc$ (dashed line). The plot is for $J=S=\infty$,
    $n=0.5$ and $a=5$ {\AA} \cite{HoEd01}. (b) optical conductivity of
    \chem{Nd_{0.7}Sr_{0.3}MnO_3} at different temperatures, reproduced from
    Kaplan \etal \cite{Ka96}.}
\end{figure}

Using the notation of section~\ref{sec:resist-param-state} we can write the
optical conductivity $\si(\nu)$, corresponding to the elliptic density of
states in equation~(\ref{eq:ell_dos}), in the form
\cite{ChFr98,ChMiDa00,HoEd01}
\begin{equation}\label{eq:optical-cond}
  \si(\nu)=\frac{2\pi e^2}{3a^3\hbar}\int\,\rmd\en\int\,\rmd E
  M_0(E)A_E(\en)A_E(\en+\nu)\frac{f(\en)-f(\en+\nu)}{\nu}\,.
\end{equation}
This expression satisfies the correct one-band sum rule
\cite{ChMiDa00,QuCeSi98} that
$(2/\pi)\int_0^\infty\si(\nu)\rmd\nu=-Ke^2/(3a\hbar)$, where the `kinetic
energy' $K$ is the thermal average per lattice site of the first term in the
Hamiltonian (\ref{eq:h_hde}). The energy $K$ appeared previously in
equation~(\ref{eg:D0b}) for the spin-wave stiffness constant. For $\nu=0$,
equation~(\ref{eq:optical-cond}) yields the dc conductivity given by
equation~(\ref{eq:4.19}). In the ferromagnetic state of the pure DE model
($g/W=0$) at $T=0$ the spectral function $A_E(\en)=\delta(\en-E)$ so that,
from equation~(\ref{eq:optical-cond}), $\si(\nu)\propto\delta(\nu)$. However,
in LSMO with $x=0.3$ the observed intraband transitions spread up to 1 eV
\cite{OkKaIsAr97} which suggests that phonons are important for producing
broadened spectral functions. A sharp Drude peak below 0.04 eV, arising from
$\delta$-like parts of $A_E(\en)$ for $E$ near the Fermi level, accounts only
for half of the total intraband spectral weight \cite{OkKaIsAr97}. An
alternative interpretation is that orbital degrees of freedom in the
doubly-degenerate \chem{e_g} band, combined with strong correlation (no
doubly-occupied sites), lead to incoherent motion of carriers
\cite{IsYaNa97,MaHoPom}. Here we pursue the comparison of results of the
many-body CPA treatment of the Holstein-DE model with experiment. A defect of
Green's \cite{Gr01} CPA treatment is that the sharp quasi-particle peak in
the spectral function, which should exist in the ferromagnetic state at
$T=0$, is missing. This is due to incoherent scattering by the phonons which
gives a spurious residual resistivity. Thus the Drude peak in $\si(\nu)$ is
absent, as in the classical phonon treatment of Millis \etal
\cite{MiMuSh96II}. Also in the one-band model with $J=\infty$ interband
transitions between Hund's rule split bands, and p$\rightarrow$d charge
transfer transitions do not feature. These are observed at photon energy
$\nu>3$ eV and $\si(\nu)$ calculated in the one-band model is only non-zero
in the region $\nu<2.5$ eV with in general only one peak. Hohenadler and
Edwards \cite{HoEd01} calculated $\si(\nu)$ in the Holstein-DE model with
$g/W=0.16$ and compared with experimental data on an unannealed film of
\chem{Nd_{0.7}Sr_{0.3}MnO_3} (NSMO) \cite{Ka96}. The comparison is shown in
figure~\ref{fig:oc}. This data was chosen because the low temperature
incoherent scattering in the sample matches closely the incoherent scattering
introduced by CPA. In annealed NSMO films \cite{QuCeSi98}, and in single
crystals \cite{LeJuLe99}, $\si(\nu)$ in the low temperature ferromagnetic
state continues to rise with decreasing $\nu$ down to much lower photon
energy, and $\si(0)\approx3$ \con. Clearly in our calculation, and those of
Millis \etal \cite{MiMuSh96II}, the shift of the peak to lower energy is held
up due to spurious incoherent scattering in the ground state, which limits
the low-temperature dc conductivity. There seems little doubt that a more
correct treatment of the Holstein-DE model at low temperatures would lead to
something more like the shift observed in good samples. Very recently Perroni
\etal \cite{PeFiCaIa01} seem to have achieved this together with a Drude peak
at low temperatures in their variational Lang-Firsov treatment. It is to be
hoped that these features survive even if their predicted phase separation is
suppressed by Coulomb interaction. In any case one does
not need to invoke a change from unbinding bipolarons to unbinding polarons
to explain the shift, as is done by Alexandrov and Bratkovsky
\cite{AlBrat99}. It is also not clear why in the bipolaron regime these
authors find a Gaussian form for $\si(\nu)$ without a threshold photon
energy. In the bipolaron regime there should be a gap in the single-particle
spectrum, analogous to that in a superconductor, but with magnitude of the
order of the bipolaron binding energy. There is no such gap in the metallic
`unpaired' single polaron regime.

In the paramagnetic state above $\Tc$, $\si(\nu)$ is much less sample
dependent and the CPA calculations are much more reliable. Many authors
\cite{MiMuSh96II,QuCeSi98,Ka96,HoEd01} agree on the following interpretation
of the peak at about 1 eV. During the absorption process an electron moves
from one site to a neighbouring one which was previously unoccupied. The
electron motion is accompanied by a lattice distortion, of Jahn-Teller type,
which corresponds to a displacement of the local phonon oscillator coordinate
in the Holstein-DE model. When an electron enters (leaves) a site the final
displaced (undisplaced) oscillator is generally in an excited state with
typical excitation energy $g^2/\om$. This is the atomic-limit polaron binding
energy and for the parameters assumed here is about 0.5 eV. Thus the peak in
$\si(\nu)$ (figure~\ref{fig:oc:a}) occurs at about twice the polaron binding
energy just as in the standard small-polaron theory \cite{Ma90}. However for the present
intermediate electron-phonon coupling $g/W=0.16$ polaron bands near the Fermi
level are largely washed out above $\Tc$ \cite{Gr01}, so standard
small-polaron theory is not expected to apply to $\si(\nu)$ for low photon
energies. This is certainly the case because small-polaron theory would predict an
activation energy in the dc conductivity $\si(0)$ of 0.25 eV, half the
polaron binding energy. However in NSMO the activation energy is observed to
be about 0.08 eV \cite{ZhKacond-mat} and Green's \cite{Gr01} calculations
(see figure~\ref{fig:rho_T_g0.16}) are in good agreement with this.

We conclude that the energy of the peak in $\si(\nu)$ for $T>\Tc$ is a
reliable measure of twice the polaron binding energy, but that the activation
energy for dc conductivity gives no such direct information. The most extreme
illustration of this is LSMO with $x=0.3$. There is no activation energy for
$\si(0)=\rho^{-1}$ which slowly decreases with temperature above $\Tc$ (see
figure~\ref{fig:fig4}). But the temperature-dependent part of $\si(\nu)$ has
a clear peak at about 0.6 eV for $T>\Tc$ \cite{OkKaIsAr97}. This indicates a
polaron binding energy of 0.3 eV, which corresponds to $g/W\approx0.12$ for
the other parameters assumed in this section. The estimate of $g/W\simeq0.1$
we have sometimes suggested for this material is probably too low. However
the difference between $g/W=0.16$ for LCMO and NSMO and $g/W=0.12$ for LSMO,
all with $x=0.3$, is sufficient to explain their very different behaviour.

\subsection{Spin waves}\label{sec:spinwaves}

\begin{figure}[htbp]
  \centering \includegraphics[width=0.65\textwidth]{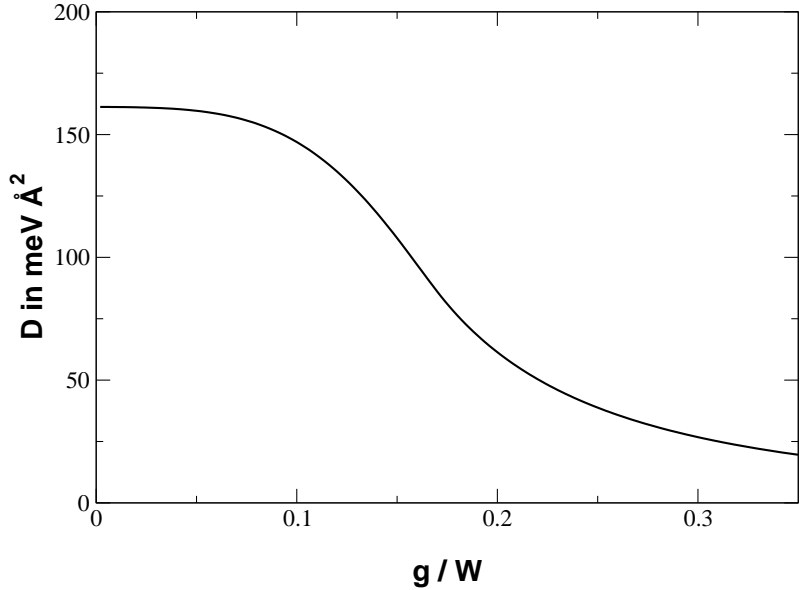}
\caption{\label{fig:spin_stiffness}%
  The spin-wave stiffness $D$ versus electron-phonon coupling $g$ in the
  saturated ferromagnetic state at $T=0$. The plot is for $S=J=\infty$, $W=1$
  eV, $n=0.5$ and $a=4$ {\AA} \cite{HoEd01}.}
\end{figure}
In section~\ref{sec:wider-view} we pointed out that the simple result of
equation~(\ref{eg:D0b}) for the spin-wave stiffness constant $D$ applies even
when local electron-phonon coupling is included. In particular it applies to
the Holstein-DE model and a simple variational argument shows that is is an
upper bound to the true value of $D$ \cite{HoEd01}. The complete spin-wave
dispersion curve takes the form of that for a simple nearest-neighbour
Heisenberg model \cite{Fu96,HoEd01}, as observed in \chem{La_{0.7}Pb_{0.3}MnO_3}
\cite{PeAe96}. Deviations from this simple result are found near the zone
boundary in manganites with lower $\Tc$ than the Pb doped material
\cite{FeDaHw98,HwDaChAe98,DaHwZhFB00}. Hohenadler and Edwards \cite{HoEd01}
calculated $D$ as a function of the electron-phonon coupling $g/W$ and the
result is shown in figure~\ref{fig:spin_stiffness}.
This behaviour of $D$ in the Holstein-DE model is very similar to that of
$\Tc$, as calculated by Green \cite{Gr01} (see
figure~\ref{fig:Tc_coupling}). The main difference is in the
extreme strong-coupling limit where $\Tc$ becomes very small at
$g/W\approx0.35$ whereas $D$ is decreasing quite slowly. The slow decrease of
$D$ is exactly what one expects from equation~(\ref{eg:D0b}) and small-polaron
theory, where the kinetic energy $K\sim g^{-2}$
\cite{AlMo94,FiKaKuAl99}. $\Tc$ seems to be determined more by the width of
the narrow polaron band around the Fermi level, which decreases exponentially
with $g$. This is the result found by R\"oder \etal \cite{RoZaBi96} which is
essentially $\Tc\propto D(g=0)\exp(-\gamma E_{\rm p}/\om)$, in the notation
of section~\ref{sec:polarons-bipolarons}. Here $D(g=0)$ is given by
equation~(\ref{eg:D0b}) for the pure DE model. For manganites with higher $\Tc$,
where the spin-wave dispersion curve is Heisenberg-like, one expects that the
ratio $\delta=D/(k_{\rm B}\Tc a^2)$, where $a$ is the lattice constant,
should be approximately 0.286 which is the value for the $S=3/2$ Heisenberg
model. In fact Hohenadler and Edwards \cite{HoEd01} find, combining their
results with those of Green for $\Tc$ (see figure~\ref{fig:Tc_coupling}),
that $\delta\approx0.24$ for $g/W=0.1$ which is appropriate to a high $\Tc$
manganite like LSMO ($x\approx0.3$). Clearly, from
figures~\ref{fig:Tc_coupling} and~\ref{fig:spin_stiffness}, $\delta$
increases as $g/W$ becomes larger, that is as $\Tc$ becomes smaller. This
agrees with experiment; for example in \chem{Nd_{0.7}Sr_{0.3}MnO_3}
($\Tc=197.9$ K), $\delta\approx0.64$ \cite{FeDaHw98}. In this system, and in
LCMO \cite{LyEr96}, the spin-wave stiffness constant does not collapse to
zero at $T=\Tc$ so the behaviour is certainly not Heisenberg-like. The
properties of the Holstein-DE model in this parameter regime need further
investigation. The phase separation for $T>0.65\Tc$ predicted by Perroni
\etal \cite{PeFiCaIa01} should be examined in the presence of Coulomb
interaction.



\section{Conclusions}

A brief review inevitably involves a personal choice of topic and emphasis,
Here we have concentrated on the physics of manganites such as
\chem{La_{1-x}Ca_xMnO_3} in the ferromagnetic regime with $x \sim~0.2$--0.4. We
have not discussed interesting phenomena such as charge and orbital ordering
which occur elsewhere in the phase diagram.

The main conclusion is that the Holstein-DE model is capable of describing
semi-quantitatively a wide range of experimental data on these materials. The
model combines Zener's double exchange mechanism for ferromagnetism with the
possibility of polaronic effects associated with strong electron-phonon
coupling. These are the essential ingredients proposed by Millis \etal
\cite{MiLiSh95,MiMuSh96II}.  The double-exchange mechanism relies on the
mobile carriers in doped systems being Mn \chem{e_g} electrons rather than O
p holes. The theoretical and experimental evidence that this is indeed the
case is reviewed in section~\ref{sec:electronic-structure}. In
section~\ref{sec:itin-electr-ferr} the concept of double exchange is set
within the general scheme of itinerant electron magnetism. One ingredient
omitted in the Holstein-DE model is A-site disorder and the discussion of
section~\ref{sec:role-disorder} shows that this omission is not serious for
systems with $x\sim0.3$.

The most complete theory of the Holstein-DE model, capable of dealing with
quantum spins and phonons, is that of Green \cite{Gr01}; it is developed in a
concise way in section~\ref{sec:many-body-cpa-hde}. The theory makes use of
the many-body CPA method introduced earlier \cite{EdGrKu99,GrEd99} for
treating the pure double-exchange model, as described in
section~\ref{sec:magn-transp-de}. For the pure DE model this method is
equivalent to Furukawa's dynamical mean-field theory in his limit of
classical spins.  Theoretical results for the Holstein-DE model are compared
with experiment in section~\ref{sec:theory-experiment}. One striking
conclusion is that the manganites fall in the regime of intermediate
electron-phonon coupling, just on the verge of small-polaron formation. Thus
in general the standard results of small-polaron theory do not apply. However
even in the intermediate-coupling regime the atomic-limit polaron binding
energy $E_{\rm p}$ is still a significant parameter and a peak in the optical
conductivity $\si(\nu)$ occurs at photon energy $2E_{\rm p}$, just as in
standard small-polaron theory. But the activation energy for electrical
conduction above $\Tc$ turns out to be much smaller than that predicted by
small-polaron theory, in agreement with experiment. The key parameter in the
theory is $E_{\rm p}/W$, the ratio of the polaron binding energy to the
half-width of the \chem{e_g} band. This determines the Curie temperature
$\Tc$ and the form of the resistivity $\rho(T)$; there is essentially no
explicit dependence on the phonon frequency $\om$ in the
intermediate-coupling regime. Thus the oxygen isotope effect in $\Tc$ and
$\rho(T)$ is not a consequence of the change in $\om$ associated with a
change of isotopic mass. The observed proportionality between the isotope
exponent and the pressure-coefficient indicates that the shift of $\Tc$
arises from a change of $E_{\rm p}/W$ in both cases. The primary effect may
be regarded as a change in $W$ due to a change in Mn--O--Mn bond angle, and
this effect is strongly enhanced by a change in $E_{\rm p}$ due to modified
screening of the electron-phonon coupling arising from a changed density of
states at the Fermi level. It is proposed, following Babushkina \etal
\cite{BaBeOzGo98}, that a change in Mn--O--Mn bond angle arising from
isotopic substitution is due to a modified vibrational amplitude of the O
ions owing to the mass change. This proposal, and that concerning the
screening of the electron-phonon coupling, requires further theoretical and
experimental investigation. The large isotope effect, pressure effect,
colossal magnetoresistance and the striking difference between
\chem{La_{0.7}Sr_{0.3}MnO_3} and \chem{La_{0.7}Ca_{0.3}MnO_3} are all due to
extreme sensitivity to the parameter $E_{\rm p}/W$ in the intermediate
coupling regime ($E_{\rm p}/W\sim0.5$).

Another point which requires investigation is the nature of the ferromagnetic
transition at $\Tc$. It is found, both theoretically and experimentally, that
in manganites with low $\Tc$ the ratio of spin-wave stiffness constant $D$ to
$\Tc$ is much larger than the Heisenberg-like value occurring in those with
high $\Tc$. Experimentally, in the low $\Tc$ systems, $D$ does not collapse
to zero at $\Tc$. Nanoscale phase separation near and above $\Tc$ is a
possible explanation \cite{LyEr96} and further theoretical work along these
lines is required.  

The physics of the manganites involves a subtle interplay
between magnetism and electron-phonon coupling. The same might be said, with
less certainty, of high temperature superconductivity in the cuprates. It
seems clear from this review that a firm theoretical understanding of the
manganites is within reach. Methods of handling electron-phonon coupling in
the intermediate coupling regime have been developed. An intriguing prospect
is that this experience gained with the manganites may lead to important
advances in our understanding of the cuprates.


\section*{Acknowledgments}

This review began life as a conference paper ({\it cond-mat/0109266}) which
was rejected on grounds of length. That paper had a coauthor, Alex Green, and
section~\ref{sec:many-body-cpa-hde} draws extensively on his published work.
However the critical opinions expressed in this much-expanded review are my
own and it was decided, with Alex Green's agreement, that I should take the
responsibility of sole authorship. I am grateful to him for providing two
previously unpublished figures (figures~\ref{fig:rho_T_g0.1}
and~\ref{fig:isotope_LCMO}) and for several years of fruitful collaboration.
More recently I have enjoyed collaborating with Martin Hohenadler who has
also given invaluable assistance in preparing this review for publication. I
am also grateful to Kenn Kubo for inspiring our work on double-exchange in
the early stages. I also thank Lesley Cohen, D. M. Eagles, J. K. Freericks,
N. Furukawa, K. Kamenev, D. Meyer and K. A. M\"uller for helpful discussion.
I am also grateful to the UK Engineering and Physical Sciences Research
Council (EPSRC) for financial support. The EPSRC Magnetic Oxide Network,
coordinated by Gillian Gehring, has also been a stimulating influence.



\clearpage
\bibliography{../bibtex/bibliography}

\begin{thebibliography}{100}

\bibitem{Coey99}
{\sc Coey, J. M.~D.}, {\sc Viret, M.}, and {\sc von Moln\'{a}r, S.}, 1999, {\em
  Advances in Physics\/}, 48, 167

\bibitem{ZaSaAl85}
{\sc Zaanen, J.}, {\sc Sawatzky, G.~A.}, and {\sc Allen, J.~W.}, 1985, {\em
  Phys. Rev. Lett.\/}, 55, 418

\bibitem{Ze51}
{\sc Zener, C.}, 1951, {\em Phys. Rev.\/}, 82, 403

\bibitem{SaPoVu96}
{\sc Satpathy, S.}, {\sc Popovi\'{c}, Z.~S.}, and {\sc Vukajlovi\`{c}, F.~R.},
  1996, {\em Phys. Rev. Lett.\/}, 76, 960

\bibitem{MuHiGi98}
{\sc Murakami, Y.}, {\sc Hill, J.~P.}, {\sc Gibbs, D.}, {\sc Blume, M.}, {\sc
  Koyama, I.}, {\sc Tanaka, M.}, {\sc Kawata, H.}, {\sc Arima, T.}, {\sc
  Tokura, Y.}, {\sc Hirota, K.}, and {\sc Endoh, Y.}, 1998, {\em Phys. Rev.
  Lett.\/}, 81, 582

\bibitem{MaEn96}
{\sc Martin, M.~C.}, {\sc Shirane, G.}, {\sc Endoh, Y.}, {\sc Hirota, K.}, {\sc
  Moritomo, Y.}, and {\sc Tokura, Y.}, 1996, {\em Phys. Rev. B\/}, 53, 14285

\bibitem{MoreoYuDa99}
{\sc Moreo, A.}, {\sc Yunoki, S.}, and {\sc Dagotto, E.}, 1999, {\em
  Science\/}, 283, 2034

\bibitem{Na96}
{\sc Nagaev, E.~L.}, 1996, {\em Phys. Uspekhi\/}, 39, 781

\bibitem{ToAsKu96}
{\sc Tomioka, Y.}, {\sc Asamitsu, A.}, {\sc Kuwuhara, H.}, {\sc Moritomo, Y.},
  and {\sc Tokura, Y.}, 1996, {\em Phys. Rev. B\/}, 53, R1689

\bibitem{ScRaBaCh95}
{\sc Schiffer, P.}, {\sc Ramirez, A.~P.}, {\sc Bao, W.}, and {\sc {Cheong},
  S.-W.}, 1995, {\em Phys. Rev. Lett.\/}, 75, 3336

\bibitem{RoAt96}
{\sc {Rodriguez-Martinez}, L.~M.} and {\sc Attfield, J.~P.}, 1996, {\em Phys.
  Rev. B\/}, 54, 15622

\bibitem{Hw96}
{\sc Hwang, H.~Y.}, {\sc {Cheong}, S.-W.}, {\sc Ong, N.~P.}, and {\sc Batlogg,
  B.}, 1996, {\em Phys. Rev. Lett.\/}, 77, 2041

\bibitem{MaBuIs97}
{\sc Mathur, N.~D.}, {\sc Burnell, G.}, {\sc Isaac, S.~P.}, {\sc Jackson,
  T.~J.}, {\sc {Teo}, B.-S.}, {\sc {MacManus-Driscoll}, J.~L.}, {\sc Cohen,
  L.~F.}, {\sc Evetts, J.~E.}, and {\sc Blamire, M.~G.}, 1997, {\em Nature\/},
  387, 266

\bibitem{Hw95}
{\sc Hwang, H.~Y.}, {\sc Palstra, T. T.~M.}, {\sc {Cheong}, S.-W.}, and {\sc
  Batlogg, B.}, 1995, {\em Phys. Rev. B\/}, 52, 15046

\bibitem{UrMoAr95}
{\sc Urushibara, A.}, {\sc Moritomo, Y.}, {\sc Arima, T.}, {\sc Asamitsu, A.},
  {\sc Kido, G.}, and {\sc Tokura, Y.}, 1995, {\em Phys. Rev. B\/}, 51, 14103

\bibitem{McWaZh96}
{\sc McIlroy, D.~N.}, {\sc Waldfried, C.}, {\sc Zhang, J.}, {\sc {Choi},
  J.-W.}, {\sc Foong, F.}, {\sc Liou, S.~H.}, and {\sc Dowben, P.~A.}, 1996,
  {\em Phys. Rev. B\/}, 54, 17438

\bibitem{PiSi96}
{\sc Pickett, W.~E.} and {\sc Singh, D.~J.}, 1996, {\em Phys. Rev. B\/}, 53,
  1146

\bibitem{SaShBa95}
{\sc Sarma, D.~D.}, {\sc Shanti, N.}, {\sc Barman, S.~R.}, {\sc Hamada, N.},
  {\sc Sawada, H.}, and {\sc Terakura, K.}, 1995, {\em Phys. Rev. Lett.\/}, 75,
  1126

\bibitem{ZaSaAl86}
{\sc Zaanen, J.}, {\sc Sawatzky, G.~A.}, and {\sc Allen, J.~W.}, 1986, {\em J.
  Magn. Magn. Mat.\/}, 54, 607

\bibitem{TeWiOgKu84}
{\sc Terakura, K.}, {\sc Williams, A.~R.}, {\sc Oguchi, T.}, and {\sc K\"ubler,
  J.}, 1984, {\em Phys. Rev. Lett.\/}, 52, 1830

\bibitem{SuKaMaHa00}
{\sc {Su}, Y.-S.}, {\sc Kaplan, T.~A.}, {\sc Mahanti, S.~D.}, and {\sc
  Harrison, J.~F.}, 2000, {\em Phys. Rev. B\/}, 61, 1324

\bibitem{BaChSa94}
{\sc Barman, S.~R.}, {\sc Chainani, A.}, and {\sc Sarma, D.~D.}, 1994, {\em
  Phys. Rev. B\/}, 49, 8475

\bibitem{SiPi98}
{\sc Singh, D.~J.} and {\sc Pickett, W.~E.}, 1998, {\em Phys. Rev. B\/}, 57, 88

\bibitem{ChMaSa93}
{\sc Chainani, A.}, {\sc Mathew, M.}, and {\sc Sarma, D.~D.}, 1993, {\em Phys.
  Rev. B\/}, 47, 15397

\bibitem{SaBoMiNa95}
{\sc Saitoh, T.}, {\sc Bocquet, A.~E.}, {\sc Mizokawa, T.}, {\sc Namatame, H.},
  {\sc Fujimori, A.}, {\sc Abbata, M.}, {\sc Takeda, Y.}, and {\sc Takano, M.},
  1995, {\em Phys. Rev. B\/}, 51, 13942

\bibitem{PaChCh96}
{\sc {Park}, J.-H.}, {\sc Chen, C.~T.}, {\sc Cheong, S.~W.}, {\sc Bao, W.},
  {\sc Meigs, G.}, {\sc Chakarian, V.}, and {\sc Idzerda, Y.~U.}, 1996, {\em
  Phys. Rev. Lett.\/}, 76, 4215

\bibitem{PaChCh98II}
{\sc {Park}, J.-H.}, {\sc Vescovo, E.}, {\sc {Kim}, H.-J.}, {\sc Kwon, C.},
  {\sc Ramesh, R.}, and {\sc Venkatesan, T.}, 1998, {\em Phys. Rev. Lett.\/},
  81, 1953

\bibitem{SouByOsNaAm98}
{\sc {Soulen Jr.}, R.~J.}, {\sc Byers, J.~M.}, {\sc Osofsky, M.~S.}, {\sc
  Nadgorny, B.}, {\sc Ambrose, T.}, {\sc Cheong, S.~F.}, {\sc Broussard,
  P.~R.}, {\sc Tanaka, C.~T.}, {\sc Nowak, J.}, {\sc Moodera, J.~S.}, {\sc
  Barry, A.}, and {\sc Coey, J. M.~D.}, 1998, {\em Science\/}, 282, 85

\bibitem{OsNaSo99}
{\sc Osofsky, M.~S.}, {\sc Nadgorny, B.}, {\sc {Soulen Jr.}, R.~J.}, {\sc
  Broussard, P.}, {\sc Rubinstein, M.}, {\sc Byers, J.}, {\sc Laprade, G.},
  {\sc Mukovskii, Y.~M.}, {\sc Shulyatev, D.}, and {\sc Arsenov, A.}, 1999,
  {\em J. Appl. Phys.\/}, 85, 5567

\bibitem{Lu96}
{\sc Lu, Y.}, {\sc Li, X.~W.}, {\sc Gong, G.~Q.}, {\sc Xiao, G.}, {\sc Gupta,
  A.}, {\sc Lecoeur, P.}, {\sc Sun, J.~Z.}, {\sc Wang, Y.~Y.}, and {\sc Dravid,
  V.~P.}, 1996, {\em Phys. Rev. B\/}, 54, R8357

\bibitem{SuKrDu97}
{\sc Sun, J.~L.}, {\sc {Krusin-Elbaum}, L.}, {\sc Duncombe, P.~R.}, {\sc Gupta,
  A.}, and {\sc Laibowwitz, R.~B.}, 1997, {\em App. Phys. Lett.\/}, 70, 1769

\bibitem{PaChCh98}
{\sc {Park}, J.-H.}, {\sc Vescovo, E.}, {\sc {Kim}, H.-J.}, {\sc Kwon, C.},
  {\sc Ramesh, R.}, and {\sc Venkatesan, T.}, 1998, {\em Nature\/}, 392, 794

\bibitem{AlEiMu76}
{\sc Alvarado, S.~F.}, {\sc Eib, W.}, {\sc Munz, P.}, {\sc Siegmann, H.~C.},
  {\sc Campagna, M.}, and {\sc Remeika, J.~P.}, 1976, {\em Phys. Rev. Lett.\/},
  13, 4918

\bibitem{Li99}
{\sc Livesay, E.~A.}, {\sc West, R.~N.}, {\sc Dugdale, S.~B.}, {\sc Santi, G.},
  and {\sc Jarlborg, T.}, 1999, {\em J. Phys.: Condens. Matter\/}, 11, L279

\bibitem{PiSi97}
{\sc Pickett, W.~E.} and {\sc Singh, D.~J.}, 1997, {\em Phys. Rev. B\/}, 55,
  R8642

\bibitem{De98}
{\sc Dessau, D.~S.}, {\sc Saitoh, T.}, {\sc {Park}, C.-H.}, {\sc {Shen},
  Z.-X.}, {\sc Villella, P.}, {\sc Hamada, N.}, {\sc Moritomo, Y.}, and {\sc
  Tokura, Y.}, 1998, {\em Phys. Rev. Lett.\/}, 81, 192

\bibitem{JuSoKr97}
{\sc Ju, H.~L.}, {\sc {Sohn}, H.-C.}, and {\sc Krishnan, K.~M.}, 1997, {\em
  Phys. Rev. Lett.\/}, 79, 3230

\bibitem{He93}
{\sc Hewson, A.~C.}, 1993, {\em The Kondo problem to Heavy Fermions\/}
  (Cambridge University Press)

\bibitem{AnHa55}
{\sc Anderson, P.~W.} and {\sc Hasegawa, H.}, 1955, {\em Phys. Rev.\/}, 100,
  675

\bibitem{Ge60}
{\sc de~Gennes, P.~G.}, 1960, {\em Phys. Rev.\/}, 118, 141

\bibitem{KuOh72}
{\sc Kubo, K.} and {\sc Ohata, N.}, 1972, {\em J. Phys. Soc. Jpn.\/}, 33, 21

\bibitem{Gr01}
{\sc Green, A. C.~M.}, 2001, {\em Phys. Rev. B\/}, 63, 205110

\bibitem{HoEd01}
{\sc Hohenadler, M.} and {\sc Edwards, D.~M.}, 2001, {\em cond-mat/0111175\/},
  submitted to {\it J. Phys.: Condens. Matter}

\bibitem{EdMu85}
{\sc Edwards, D.~M.} and {\sc Muniz, R.~B.}, 1985, {\em J. Phys. F: Met.
  Phys.\/}, 15, 2339

\bibitem{MuCoEd85}
{\sc Muniz, R.~B.}, {\sc Cooke, J.~F.}, and {\sc Edwards, D.~M.}, 1985, {\em J.
  Phys. F: Met. Phys.\/}, 15, 2357

\bibitem{SoTe99}
{\sc Solovyev, I.~V.} and {\sc Terakura, K.}, 1999, {\em Phys. Rev. Lett.\/},
  82, 2959

\bibitem{GeKoKrRo96}
{\sc Georges, A.}, {\sc Kotliar, G.}, {\sc Krauth, W.}, and {\sc Rosenberg,
  M.~J.}, 1996, {\em Rev. Mod. Phys.\/}, 68, 13

\bibitem{pruschke_review}
{\sc Pruschke, T.}, {\sc Jarrell, M.}, and {\sc Freericks, J.~K.}, 1995, {\em
  Advances in Physics\/}, 44, 187

\bibitem{EdFi71}
{\sc Edwards, D.~M.} and {\sc Fisher, B.}, 1971, {\em J. Physique Coll. C1\/},
  32, 697

\bibitem{QuCeSi98}
{\sc Quijada, M.}, {\sc \v{C}erne, J.}, {\sc Simpson, J.~R.}, {\sc Drew,
  H.~D.}, {\sc Ahn, K.~H.}, {\sc Millis, A.~J.}, {\sc Shreekala, R.}, {\sc
  Ramesh, R.}, {\sc Rajeswari, M.}, and {\sc Venkatesan, T.}, 1998, {\em Phys.
  Rev. B\/}, 58, 16093

\bibitem{EdRi92}
{\sc Edwards, D.~M.} and {\sc Rigby, A.~M.}, unpublished, {Rigby}, {A. M.},
  1992, Ph.D. thesis (University of London)

\bibitem{Zh00}
{\sc {Zhao}, G.-m.}, 2000, {\em Phys. Rev. B\/}, 62, 11639

\bibitem{LiKa01}
{\sc Lichtenstein, A.~I.} and {\sc Katsnelson, M.~I.}, 2001, {\em
  Band-ferromagnetism\/}, edited by K.~Baberschke, M.~Donath, and W.~Nolting
  (Springer), p~75

\bibitem{EnHi97}
{\sc Endoh, Y.} and {\sc Hirota, K.}, 1997, {\em J. Phys. Soc. Japan\/}, 66,
  2264

\bibitem{Ve69}
{\sc Velicky, B.}, 1969, {\em Phys. Rev.\/}, 184, 614

\bibitem{HiEd73}
{\sc Hill, D.~J.} and {\sc Edwards, D.~M.}, 1973, {\em J. Phys. F: Met.
  Phys.\/}, 3, L162

\bibitem{Fu73}
{\sc Fukuyama, H.}, 1973, {\em AIP Conf. Proc.\/}, 10, 1127

\bibitem{EdHi76}
{\sc Edwards, D.~M.} and {\sc Hill, D.~J.}, 1976, {\em J. Phys. F: Met.
  Phys.\/}, 6, 607

\bibitem{WaEdWo71}
{\sc Wakoh, S.}, {\sc Edwards, D.~M.}, and {\sc Wohlfarth, E.~P.}, 1971, {\em
  J. Physique C1\/}, 32, 1073

\bibitem{Wa98}
{\sc Wang, X.}, 1998, {\em Phys. Rev. B\/}, 57, 7427

\bibitem{FuPom}
{\sc Furukawa, N.}, 1999, {\em Physics of Manganites\/}, edited by T.~A. Kaplan
  and S.~D. Mahanti (Kluwer, New York), p~1

\bibitem{IzKiKu63}
{\sc Izuyama, T.}, {\sc Kim, D.~J.}, and {\sc Kubo, R.}, 1963, {\em J. Phys.
  Soc. Japan\/}, 18, 1025

\bibitem{BrEd98}
{\sc Brunton, R.~E.} and {\sc Edwards, D.~M.}, 1998, {\em J. Phys.: Condens.
  Matter\/}, 10, 5421

\bibitem{Ok97}
{\sc Okabe, T.}, 1997, {\em Prog. Theor. Phys.\/}, 97, 21

\bibitem{WuMu98}
{\sc Wurth, P.} and {\sc {M\"uller-Hartmann}, E.}, 1998, {\em Eur. Phys. J.
  B\/}, 5, 403

\bibitem{Ed68}
{\sc Edwards, D.~M.}, 1968, {\em J. Appl. Phys.\/}, 39, 481

\bibitem{Ro69}
{\sc Roth, L.~M.}, 1969, {\em Phys. Rev.\/}, 186, 428

\bibitem{HeEd73}
{\sc Hertz, J.~A.} and {\sc Edwards, D.~M.}, 1973, {\em J. Phys. F: met.
  Phys.\/}, 3, 2174

\bibitem{EdHe73}
{\sc Edwards, D.~M.} and {\sc Hertz, J.~A.}, 1973, {\em J. Phys. F: Met.
  Phys.\/}, 3, 2191

\bibitem{KaMaSu01}
{\sc Kaplan, T.~A.}, {\sc Mahanti, S.~D.}, and {\sc {Su}, Y.-S.}, 2001, {\em
  Phys. Rev. Lett.\/}, 86, 3634

\bibitem{Ed67}
{\sc Edwards, D.~M.}, 1967, {\em Prog. Roy. Soc. A\/}, 300, 373

\bibitem{Go00}
{\sc Golosov, D.~I.}, 2000, {\em Phys. Rev. Lett.\/}, 84, 3974

\bibitem{Th65}
{\sc Thompson, E.~D.}, 1965, {\em J. Appl. Phys.\/}, 36, 1133

\bibitem{EdGrKu99}
{\sc Edwards, D.~M.}, {\sc Green, A. C.~M.}, and {\sc Kubo, K.}, 1999, {\em J.
  Phys.: Condens. Matter\/}, 11, 2791

\bibitem{GrEd99}
{\sc Green, A. C.~M.} and {\sc Edwards, D.~M.}, 1999, {\em J. Phys.: Condens.
  Matter\/}, 11, 10511, {\it erratum}, 2000, 12, 9107

\bibitem{Hu64}
{\sc Hubbard, J.}, 1964, {\em Proc. Roy. Soc.\/}, 281, 401

\bibitem{ElKrLe74}
{\sc Elliot, R.~J.}, {\sc Krumhansl, J.~A.}, and {\sc Leath, P.~L.}, 1974, {\em
  Rev. Mod. Phys.\/}, 46, 465

\bibitem{EdHe68}
{\sc Edwards, D.~M.} and {\sc Hewson, A.~C.}, 1968, {\em Rev. Mod. Phys.\/},
  40, 810

\bibitem{EdHe90}
{\sc Edwards, D.~M.} and {\sc Hertz, J.~A.}, 1990, {\em Physica B\/}, 163, 527

\bibitem{LuEd95}
{\sc Luchini, M.~U.} and {\sc Edwards, D.~M.}, 1995, {\em J. Low. Temp.
  Phys.\/}, 99, 305

\bibitem{PoHeNo98}
{\sc Potthoff, M.}, {\sc Herrmann, T.}, and {\sc Nolting, W.}, 1998, {\em Eur.
  Phys. J. B\/}, 4, 485

\bibitem{BuHePr98}
{\sc Bulla, R.}, {\sc Hewson, A.~C.}, and {\sc Pruschke, T.}, 1998, {\em J.
  Phys.: Condens. Matter\/}, 10, 8365

\bibitem{Fu94}
{\sc Furukawa, N.}, 1994, {\em J. Phys. Soc. Jpn.\/}, 63, 3214

\bibitem{Fu96}
{\sc Furukawa, N.}, 1996, {\em J. Phys. Soc. Jpn.\/}, 65, 1174

\bibitem{MeSaNo01}
{\sc Meyer, D.}, {\sc Santos, C.}, and {\sc Nolting, W.}, 2001, {\em J. Phys.:
  Condens. Matter\/}, 13, 2531

\bibitem{Ku74}
{\sc Kubo, K.}, 1974, {\em J. Phys. Soc. Jpn.\/}, 36, 32

\bibitem{MiLiSh95}
{\sc Millis, A.~J.}, {\sc Littlewood, P.~B.}, and {\sc Shraiman, B.~I.}, 1995,
  {\em Phys. Rev. Lett.\/}, 74, 5144

\bibitem{MiMuSh96II}
{\sc Millis, A.~J.}, {\sc M\"uller, R.}, and {\sc Shraiman, B.~I.}, 1996, {\em
  Phys. Rev. B\/}, 54, 5405

\bibitem{RoZaBi96}
{\sc R\"oder, H.}, {\sc Zang, J.}, and {\sc Bishop, A.~R.}, 1996, {\em Phys.
  Rev. Lett.\/}, 76, 1356

\bibitem{Va96}
{\sc Varma, C.~M.}, 1996, {\em Phys. Rev. B\/}, 54, 7308

\bibitem{MuDa96}
{\sc {M\"uller-Hartmann}, E.} and {\sc Dagotto, E.}, 1996, {\em Phys. Rev.
  B\/}, 54, R6819

\bibitem{Naga96}
{\sc Nagaev, E.~L.}, 1996, {\em Phys. Rev. B\/}, 54, 16608

\bibitem{ShXiShTi97}
{\sc Sheng, L.}, {\sc Xing, D.~Y.}, {\sc Sheng, D.~N.}, and {\sc Ting, C.~S.},
  1997, {\em Phys. Rev. B\/}, 56, R7053

\bibitem{LiZaBiSo97}
{\sc Li, Q.}, {\sc Zang, J.}, {\sc Bishop, A.~R.}, and {\sc Soukoulis, C.~M.},
  1997, {\em Phys. Rev. B\/}, 56, 4541

\bibitem{SmXiZhRa00}
{\sc Smolyaninova, V.~N.}, {\sc Xie, X.~C.}, {\sc Zhang, F.~C.}, {\sc
  Rajeswari, M.}, {\sc Greene, R.~L.}, and {\sc {Das Sarma}, S.}, 2000, {\em
  Phys. Rev. B\/}, 62, 3010

\bibitem{AuKo00}
{\sc Auslender, M.} and {\sc Kogan, E.}, 2000, {\em cond-mat/0006184\/}

\bibitem{ZhDoWa98}
{\sc Zhong, F.}, {\sc Dong, J.}, and {\sc Wang, Z.~D.}, 1998, {\em Phys. Rev.
  B\/}, 58, 15310

\bibitem{LeFr01}
{\sc Letfulov, B.~M.} and {\sc Freericks, J.~K.}, 2001, {\em
  cond-mat/0103471\/}

\bibitem{Ma90}
{\sc Mahan, G.~D.}, 1990, {\em Many-particle Physics\/} (Plenum Press), 2nd ed.

\bibitem{Eagl66}
{\sc Eagles, D.~M.}, 1966, {\em Phys. Rev.\/}, 145, 645

\bibitem{ZaBiRo96}
{\sc Zang, J.}, {\sc Bishop, A.~R.}, and {\sc R\"oder, H.}, 1996, {\em Phys.
  Rev. B\/}, 53, R8840

\bibitem{BeZe99}
{\sc Benedetti, P.} and {\sc Zeyher, R.}, 1999, {\em Phys. Rev. B\/}, 59, 9923

\bibitem{HeVo00}
{\sc Held, K.} and {\sc Vollhardt, D.}, 2000, {\em Phys. Rev. Lett.\/}, 84,
  5168

\bibitem{Ho59a}
{\sc Holstein, T.}, 1959, {\em Ann. Phys. (N.Y.)\/}, 8, 325

\bibitem{Ho59b}
{\sc Holstein, T.}, 1959, {\em Ann. Phys. (N.Y.)\/}, 8, 343

\bibitem{Su72}
{\sc Sumi, H.}, 1972, {\em J. Phys. Soc. Jpn.\/}, 33, 327

\bibitem{Su74}
{\sc Sumi, H.}, 1974, {\em J. Phys. Soc. Jpn.\/}, 36, 770

\bibitem{CiPaFrFe97}
{\sc Ciuchi, S.}, {\sc de~Pasquale, F.}, {\sc Fratini, S.}, and {\sc Feinberg,
  D.}, 1997, {\em Phys. Rev. B\/}, 56, 4494

\bibitem{FrJaSc93}
{\sc Freericks, J.~K.}, {\sc Jarrell, M.}, and {\sc Scalapino, D.~J.}, 1993,
  {\em Phys. Rev. B\/}, 48, 6302

\bibitem{MiMuShI}
{\sc Millis, A.~J.}, {\sc M\"uller, R.}, and {\sc Shraiman, B.~I.}, 1996, {\em
  Phys. Rev. B\/}, 54, 5389

\bibitem{EmHo69}
{\sc Emin, D.} and {\sc Holstein, T.}, 1969, {\em Ann. Phys. (N.Y.)\/}, 53, 439

\bibitem{WoMiGe98}
{\sc Worledge, D.~C.}, {\sc Mi\'{e}ville, L.}, and {\sc Geballe, T.~H.}, 1998,
  {\em Phys. Rev. B\/}, 57, 15267

\bibitem{ZhKacond-mat}
{\sc {Zhao}, G.-m.}, {\sc Kang, D.~J.}, {\sc Prellier, W.}, {\sc Rajeswari,
  M.}, and {\sc Keller, H.}, 1999, {\em cond-mat/9912355\/}

\bibitem{AlBrat99}
{\sc Alexandrov, A.~S.} and {\sc Bratkovsky, A.~M.}, 1999, {\em J. Phys.:
  Condens. Matter\/}, 11, L531

\bibitem{AlBr99}
{\sc Alexandrov, A.~S.} and {\sc Bratkovsky, A.~M.}, 1999, {\em Phys. Rev.
  Lett.\/}, 82, 141

\bibitem{AlexBr99}
{\sc Alexandrov, A.~S.} and {\sc Bratkovsky, A.~M.}, 1999, {\em J. Phys.:
  Condens. Matter\/}, 11, 1989

\bibitem{AlKo99}
{\sc Alexandrov, A.~S.} and {\sc Kornilovich, P.~E.}, 1999, {\em Phys. Rev.
  Lett.\/}, 82, 807

\bibitem{Fr54}
{\sc Fr\"ohlich, H.}, 1954, {\em Adv. Phys.\/}, 3, 325

\bibitem{FeLoWe99}
{\sc Fehske, H.}, {\sc Loos, J.}, and {\sc Wellein, G.}, 1999, {\em
  cond-mat/9911218\/}

\bibitem{ChRaFe98}
{\sc Chakraverty, B.~K.}, {\sc Ranninger, J.}, and {\sc Feinberg, D.}, 1998,
  {\em Phys. Rev. Lett.\/}, 81, 433

\bibitem{ChRaFe99}
{\sc Chakraverty, B.~K.}, {\sc Ranninger, J.}, and {\sc Feinberg, D.}, 1999,
  {\em Phys. Rev. Lett.\/}, 82, 2621

\bibitem{ZhSmPrKe00}
{\sc {Zhao}, G.-m.}, {\sc Smolyaninova, V.}, {\sc Prellier, W.}, and {\sc
  Keller, H.}, 2000, {\em Phys. Rev. Lett.\/}, 84, 6086

\bibitem{Comment49}
2000, Comment and reply on ref.~\cite{AlBr99}, {\it Phys. Rev. Lett.}, 84, 2042

\bibitem{PeFiCaIa01}
{\sc Perroni, C.~A.}, {\sc {De Filippis}, G.}, {\sc Cataudella, V.}, and {\sc
  Iadonisi, G.}, 2001, {\em cond-mat/0106588\/}

\bibitem{CaFiIa01}
{\sc Cataudella, V.}, {\sc {De Filippis}, G.}, and {\sc Iadonisi, G.}, 2001,
  {\em Phys. Rev. B\/}, 63, 52406

\bibitem{KiGuChPa96}
{\sc Kim, K.~H.}, {\sc Gu, J.~Y.}, {\sc Choi, H.~S.}, {\sc Park, G.~W.}, and
  {\sc Noh, T.~W.}, 1996, {\em Phys. Rev. Lett.\/}, 77, 1877

\bibitem{Sa96}
{\sc Sarma, D.~D.}, {\sc Shanti, N.}, {\sc Krishnakumar, S.~R.}, {\sc Saitoh,
  T.}, {\sc Mizokawa, T.}, {\sc Sekiyama, A.}, {\sc Kobayashi, K.}, {\sc
  Fujimori, A.}, {\sc Weschke, E.}, {\sc Meier, R.}, {\sc Kaindl, G.}, {\sc
  Takeda, Y.}, and {\sc Takano, M.}, 1996, {\em Phys. Rev. B\/}, 53, 6873

\bibitem{DaZhMo96}
{\sc Dai, P.}, {\sc Zhang, J.}, {\sc Mook, H.~A.}, {\sc {Liou}, S.-H.}, {\sc
  Dowben, P.~A.}, and {\sc Plummer, E.~W.}, 1996, {\em Phys. Rev. B\/}, 54,
  R3694

\bibitem{LoEgBr97}
{\sc Louca, D.}, {\sc Egami, T.}, {\sc Brosha, E.~L.}, {\sc R\"oder, H.}, and
  {\sc Bishop, A.~R.}, 1997, {\em Phys. Rev. B\/}, 56, R8475

\bibitem{Naga99}
{\sc Nagaev, E.~L.}, 1999, {\em Phys. Lett. A\/}, 258, 65

\bibitem{ZhaoPom}
{\sc {Zhao}, G.-m.}, {\sc Keller, H.}, {\sc Greene, R.~L.}, and {\sc M\"uller,
  K.~A.}, 1999, {\em Physics of Manganites\/}, edited by T.~A. Kaplan and S.~D.
  Mahanti (Kluwer, New York), p~221

\bibitem{BaBeOzGo98}
{\sc Babushkina, N.~A.}, {\sc Belova, L.~M.}, {\sc Ozhogin, V.~I.}, {\sc
  {Gorbenko}, O.~Y.}, {\sc Kaul, A.~R.}, {\sc Bosak, A.~A.}, {\sc Khomskii,
  D.~I.}, and {\sc Kugel, K.~I.}, 1998, {\em J. Appl. Phys.\/}, 83, 7369

\bibitem{ZhKeHo97}
{\sc {Zhao}, G.-m.}, {\sc Keller, H.}, {\sc Hofer, J.}, {\sc Shengelaya, A.},
  and {\sc M\"uller, K.~A.}, 1997, {\em Solid State Commun.\/}, 104, 57

\bibitem{ZhHuKe97}
{\sc {Zhao}, G.-m.}, {\sc Hunt, M.~B.}, and {\sc Keller, H.}, 1997, {\em Phys.
  Rev. Lett.\/}, 78, 955

\bibitem{Naga98}
{\sc Nagaev, E.~L.}, 1998, {\em Phys. Rev. B\/}, 58, 12242

\bibitem{FrIsCh98}
{\sc Franck, J.~P.}, {\sc Isaac, I.}, {\sc Chen, W.}, {\sc Chrzanowski, J.},
  and {\sc Irwin, J.~C.}, 1998, {\em Phys. Rev. B\/}, 58, 5189

\bibitem{ZhCoKeMu00}
{\sc {Zhao}, G.-m.}, {\sc Conder, K.}, {\sc Keller, H.}, and {\sc M\"uller,
  K.~A.}, 2000, {\em Phys. Rev. B\/}, 62, 5334

\bibitem{WeYeVa97}
{\sc Wei, J. Y.~T.}, {\sc {Yeh}, N.-C.}, and {\sc Vasquez, R.~P.}, 1997, {\em
  Phys. Rev. Lett.\/}, 79, 5150

\bibitem{BiElRaBh99}
{\sc Biswas, A.}, {\sc Elizabeth, S.}, {\sc Raychaudhuri, A.~K.}, and {\sc
  Bhat, H.~L.}, 1999, {\em Phys. Rev. B\/}, 59, 5368

\bibitem{MoYu99}
{\sc Moreo, A.}, {\sc Yunoki, S.}, and {\sc Dagotto, E.}, 1999, {\em Phys. Rev.
  Lett.\/}, 83, 2773

\bibitem{ChFr98}
{\sc Chung, W.} and {\sc Freericks, J.~K.}, 1998, {\em Phys. Rev. B\/}, 57,
  11955

\bibitem{ChMiDa00}
{\sc Chattopadhyay, A.}, {\sc Millis, A.~J.}, and {\sc {Das Sarma}, S.}, 2000,
  {\em Phys. Rev. B\/}, 61, 10738

\bibitem{OkKaIsAr97}
{\sc Okimoto, Y.}, {\sc Katsufuji, T.}, {\sc Ishikawa, T.}, {\sc Arima, T.},
  and {\sc Tokura, Y.}, 1997, {\em Phys. Rev. B\/}, 55, 4206

\bibitem{IsYaNa97}
{\sc Ishihara, S.}, {\sc Yamanaka, M.}, and {\sc Nagaosa, N.}, 1997, {\em Phys.
  Rev. B\/}, 56, 686

\bibitem{MaHoPom}
{\sc Mack, F.} and {\sc Horsch, P.}, 1999, {\em Physics of Manganites\/},
  edited by T.~A. Kaplan and S.~D. Mahanti (Kluwer, New York), p~103

\bibitem{Ka96}
{\sc Kaplan, S.~G.}, {\sc Quijada, M.}, {\sc Drew, H.~D.}, {\sc Tanner, D.~B.},
  {\sc Xiong, G.~C.}, {\sc Ramesh, R.}, {\sc Kwon, C.}, and {\sc Venkatesan,
  T.}, 1996, {\em Phys. Rev. Lett.\/}, 77, 2081

\bibitem{LeJuLe99}
{\sc Lee, H.~J.}, {\sc Jung, J.~H.}, {\sc Lee, Y.~S.}, {\sc Ahn, J.~S.}, {\sc
  Noh, T.~W.}, {\sc Kim, K.~H.}, and {\sc {Cheong}, S.-W.}, 1999, {\em Phys.
  Rev. B\/}, 60, 5251

\bibitem{PeAe96}
{\sc Perring, T.~G.}, {\sc Aeppli, G.}, {\sc Hayden, S.~M.}, {\sc Carter,
  S.~A.}, {\sc Remeika, J.~P.}, and {\sc {Cheong}, S.-W.}, 1996, {\em Phys.
  Rev. Lett.\/}, 77, 711

\bibitem{FeDaHw98}
{\sc {Fernandez-Baca}, J.~A.}, {\sc Dai, P.}, {\sc Hwang, H.~Y.}, {\sc Kloc,
  C.}, and {\sc {Cheong}, S.-W.}, 1998, {\em Phys. Rev. Lett.\/}, 80, 4012

\bibitem{HwDaChAe98}
{\sc Hwang, H.~Y.}, {\sc Dai, P.}, {\sc {Cheong}, S.-W.}, {\sc Aeppli, G.},
  {\sc Tennant, D.~A.}, and {\sc Mook, H.~A.}, 1998, {\em Phys. Rev. Lett.\/},
  80, 1316

\bibitem{DaHwZhFB00}
{\sc Dai, P.}, {\sc Hwang, H.~Y.}, {\sc Zhang, J.}, {\sc {Fernandez-Baca},
  J.~A.}, {\sc {Cheong}, S.-W.}, {\sc Kloc, C.}, {\sc Tomioka, Y.}, and {\sc
  Tokura, Y.}, 2000, {\em Phys. Rev. B\/}, 61, 9553

\bibitem{AlMo94}
{\sc Alexandrov, A.~S.} and {\sc Mott, N.~F.}, 1994, {\em Rep. Prog. Phys.\/},
  57, 1197

\bibitem{FiKaKuAl99}
{\sc {Firsov}, Y.~A.}, {\sc Kabanov, V.~V.}, {\sc Kudinov, E.~K.}, and {\sc
  Alexandrov, A.~S.}, 1999, {\em Phys. Rev. B\/}, 59, 12132

\bibitem{LyEr96}
{\sc Lynn, J.~W.}, {\sc Erwin, R.~W.}, {\sc Borchers, J.~A.}, {\sc Huang, Q.},
  {\sc Santoro, A.}, {\sc {Peng}, J.-L.}, and {\sc Li, Z.~Y.}, 1996, {\em Phys.
  Rev. Lett.\/}, 76, 4046

\end{thebibliography}
\bibliographystyle{./aip}


\end{document}